\documentclass[11pt]{article}
\usepackage{amsfonts, amsthm, amssymb, amsmath, mathrsfs, soul, comment}
\usepackage{graphicx, subcaption, array}
\usepackage{outlines}
\usepackage{float, wrapfig}
\usepackage[subnum]{cases}

\usepackage[colorlinks=true,linkcolor=blue,citecolor=red]{hyperref}

\newtheorem{Lemma}{Lemma}[section]
\newtheorem{Theorem}{Theorem}
\newtheorem{Proposition}[Lemma]{Proposition}

\newtheorem{Remark}[Lemma]{Remark}
\newtheorem{Def}[Lemma]{Definition}

\def\bR{\mathbb{R}}
\def\bZ{\mathbb{Z}}
\def\bT{\mathbb{T}}
\def\bZ{\mathbb{Z}}
\def\bN{\mathbb{N}}
\def\bx{\mathbf{x}}
\def\br{\mathbf{\bar{r}}}
\def\eqx{\mathbf{x}_{\rm e}}

\def\acDi{{\mathcal{D}}}
\def\PD{{\mathbb{P}}}

\def\by{\mathbf{y}}
\def\be{\mathbf{e}}
\def\ve{\vec{e}}
\def\bv{\mathbf{v}}
\def\VEE{\overline{\mathrm{E}}}
\def\EHV{\mathrm{E}_{\mathrm HV}}
\def\beq{\begin{equation}}
\def\eeq{\end{equation}}
\def\mrd{\mathrm d}
\def\mrD{\mathrm D}

\def\mrH{\mathrm H}
\def\mrE{\mathrm E}
\def\mrF{\mathrm F}
\def\a{\alpha}

\def\cF{\mathcal{F}}

\def\cA{\mathcal{A}}
\def\cB{\mathcal{B}}
\def\cC{\mathcal{C}}
\def\cH{\mathcal{H}}
\def\cP{\mathcal{P}}

\def\olr{\overline{r}}
\def\bpart{\bar{\partial}}

\def\tf{\tilde{f}}
\def\tc{\tilde{c}}

\DeclareMathOperator\arctanh{arctanh}

\renewcommand{\vert}[1]{{\vec{v}_{#1}}}
\newcommand{\site}[1]{{\vec{x}_{#1}}}
\newcommand{\region}[1]{V_{#1}}

\newcommand{\vl}[1]{{\vec{l}_{#1}}}

\newcommand{\vd}[2]{{\vec{d}_{#1,#2}}}
\newcommand{\vy}[2]{{\vec{y}_{#1,#2}}}
\newcommand{\vA}[2]{{A_{#1,#2}}}

\newcommand{\vpsi}[2]{{\psi_{#1,#2}}}
\newcommand{\vr}[2]{{r_{#1,#2}(\theta)}}

\numberwithin{equation}{section}
\newcommand{\tcb}[1]{\textcolor{black}{#1}}

\title{Defects and Frustration in the Packing of Soft Balls}
\author{Kenneth Jao\footnote{Michigan State University, jaokenne@msu.edu}, Keith Promislow\footnote{Michigan State University, promislo@msu.edu}, Samuel Sottile\footnote{Michigan State University, sottile1@msu.edu}}
\date{\today }
\begin{document}

\maketitle
%\tableofcontents
%\pagebreak
\begin{abstract}
This work introduces the Hookean-Voronoi energy, a minimal model for the packing of soft, deformable balls. This is motivated by recent studies of quasi-periodic equilibria arising from dense packings of diblock and star polymers.  Restricting to the planar case, we investigate the equilibrium packings of identical, deformable objects whose shapes are determined by an $N$-site Voronoi tessellation of a periodic rectangle. We derive a reduced formulation of the system showing at equilibria each site must reside at the ``max-center'' of its associated Voronoi region and construct a family of ordered ``single-string'' minimizers whose cardinality is $O(N^2)$. We identify sharp conditions under which the system admits a regular hexagonal tessellation and establish that in all cases the average energy per site is bounded below by that of a regular hexagon of unit size. 
However, numerical investigation of gradient flow of random initial data, reveals that for modest values of $N$ the system preponderantly equilibrates to quasi-ordered states with low energy and large basins of attraction. For larger $N$ the distribution of equilibria energies appears to approach a $\delta$-function limit, whose energy is significantly higher than the ground state hexagon. This limit is possibly shaped by two mechanisms: a proliferation of moderate-energy disordered equilibria that block access of the gradient flow to lower energy quasi-ordered states and a rigid threshold on the maximum energy of stable states. 
%The system frustration, measuring the gap between the average equilibrium energy of the gradient flow and the system ground state, increases with $N$ but saturates.
\end{abstract}

\section{Introduction}

There has been substantial recent interest in space-filling packings by soft objects that are spherical when unconstrained. These arise naturally from phase separation of classes of amphiphilic diblock polymers whose composition lends an energetic preference to forming spherical micelles. Computational studies, largely based upon models derived from self-consistent mean field theory, have identified a wide range of quasi-periodic structures and periodic structures with large periodicity.  This  work started with the experimental observation of Frank Kasper $\sigma$ phases in sphere forming block copolymers, \cite{Bates-Sci10}. In the space-filling arrangement the micelles form dodecagonal quasicrystals, and the authors attributed  the disorder to frustration in the macromolecular packing. This was followed by work examining the roles of symmetry breaking and exchange of mass, \cite{Bates-PNAS14},  issues of stability \cite{Shi-18}, effective descriptions from Voronoi type models \cite{Reddy-18},  and identification of periodic Laves phases with large unit cells, \cite{Dorfman-M21}.
The study of complex packing phases in diblock copolymers is complicated by the possibility of micelles exchanging mass via transport of the diblock polymer chains. This facilitates symmetry breaking since a diversity of sphere volumes can arise dynamically from variations induced by initial distributions. 

More recently complex packing phases have been investigated in Miktoarm star polymers, \cite{Fred-ACS20}.   Star-shaped polymers have a central core with arms that radiate out symmetrically, Figure\,\ref{f:star_voronoi} (left).  These molecules form domains from a single polymer molecule, eliminating the possibility of mass exchange. In a dilute setting in the presence of a good solvent they form soft spheres of prescribed radius. When $N$ star polymers are packed into a domain whose volume is inferior to the natural volume of the $N$ star-polymers, they compress and roughly form a Voronoi tessellation of the domain. In $\bR^3$
the only  Voronoi tessellation that can be formed from identical shapes is the lattice cube which has degenerate vertices and is not a good candidate for energy minimization. Thus it is not surprising that minimizers are quasi-periodic, or have large-period configurations identified computationally by tuning the aspect ratio of a periodic box. 

We present a systematic analytical and computational study of a simplified energy, Hookean-Voronoi energy, defined for $N$ polymer sites placed in a periodic rectangle within the plane.  We show that the system posses a large collection of ordered equilibria, establish sharp criteria that identify when these ordered equilibria include the regular hexagonal tiling, and show that in all cases the area-scaled regular hexagon provides a lower bound for the system energy. \tcb{Using a computational investigation, we investigate the structure of the inherent states -- the stable local minima of the system,  \cite{Weber-83}. Despite the existence of ordered local minima, for large values of $N$ gradient flows of the system originating from randomly distributed initial data generically converge to inherent states that have a significant measure of disorder and an associated energy that is above the ground state.  The system ground state may be ordered or disordered, but the generic limit of the gradient flow is disordered and energetically frustrated. While the motivation for this problem arises from three dimensional packing problems, we simplify the analysis and the computations by focusing on a two dimensional rectangular domain. There are applications in which the two dimensional reduction is common, for example the vertex model of \cite{Manning15} in which arrays of cells in biological tissues are assigned energy based upon the area of their Voronoi regions, or the self-propelled Voronoi model in  \cite{Manning16} which includes a sum of region area and perimeter.} 

\tcb{The Hookean-Voronoi energy combines a Voronoi partition of the domain with a minimal Hookean formulation of elastic energy for the packing of $N\in\bZ_+$ identical star polymers. We consider a rectangular domain $\Omega_{\alpha,N}\subset\bR^2$ with aspect ratio $\alpha\in(0,1]$ and area $N$ and construct classes of ordered, irregular hexagonal tessellations that are equilibria of the gradient flow. For each $N$ and $\alpha$ there are $O(N^2)$ such ordered tessellations. 
%We identify sharp conditions on $N$ and $\alpha$ under which the class of ordered equilibria contains a \emph{regular} hexagon tiling.  
In Theorem\,\ref{t:PosVEE} we establish that the per-site energy of any tessellation is bounded from below by the energy of the regular hexagon of unit area. This result holds even if the system admits no regular hexagonal tessellation. Nevertheless, numerical evidence shows  that the minimum energy of the ordered tessellations is very sensitive to both site number $N$ and aspect ratio $\alpha$ and can be significantly higher than the global minimum energy for those values.  Significantly, under the gradient flow the fraction of the total phase space that is contained in the basins of attraction of the ordered equilibria depends sensitively upon the values of $N$ and $\alpha$. For large values of $N$, the probability of random initial data converging to an ordered Voronoi tessellation becomes vanishing small. Indeed the limiting equilibrium generically have defects, defined as Voronoi regions that are not six sided, and the numerical investigation suggests that the expected number of defects scales linearly with system size $N$ while the expected system energy converges to a $\delta$-function distribution with non-zero average energy.}

Most surprisingly, the system frustration grows with size. Specifically the per-site energy of the equilibrium obtained from the gradient flows \emph{increases} with $N$, and apparently saturates as the probability distribution of equilibrium energies converges to a limiting distribution.  
The simulations suggest that there is sharp cut-off on the maximum energy that an inherent state can possess. However the vanishing of the basins of attraction of the ordered equilibria suggest that the number of moderately defect-filled inherent states grows significantly faster than the number of ordered and even weakly-disordered ones. %\tcb{These local minima are called the inherent states of the system, \cite{Weber-83}.}
With large $N$ the basins of attraction of moderately disordered equilibria fill space. The ordered equilibrium and low-energy states with isolated defects are rendered inaccessible to the gradient flow and the system is frustrated.  Indeed, this simple system shows that ordered equilibria may not be good predictors of bulk (large domain) behavior.

This paper is organized as follows. In section 2 the Hookean-Voronoi energy is presented and its equilibria are characterized in terms of max-center points that are analogous to, but different than, centroids of the Voronoi regions. In section 3 we construct the ``single-string'' families of  ordered equilibria,  characterize the site number $N$ and aspect ratio $\alpha$ which admit tessellations by regular hexagons, and establish the unit-hexagon as the a lower bound on the per-site energy of all tessellations.  In section 4 we present the results of the numerical simulations and address the large $N$ behaviour of the system.

\section{The Hookean-Voronoi Energy}
\begin{figure}[H]
    \centering
    \begin{subfigure}{.4\textwidth}
        \centering
        \includegraphics[width=.8\textwidth]{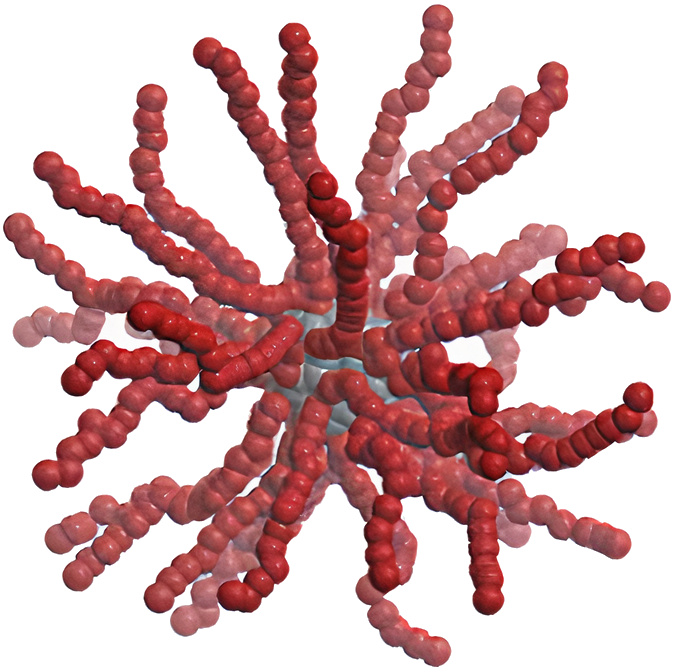}
    \end{subfigure}
    \begin{subfigure}{.4\textwidth}
        \centering
        \includegraphics[width=.8\textwidth]{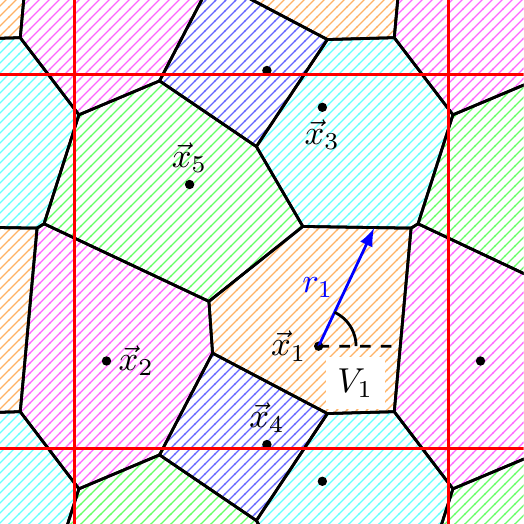}
    \end{subfigure}
\caption{(left) A depiction of a ``miktoarm'' star polymer, with arms radiating from a central core, reprinted with permission from \cite{SP-09}. (right) Voronoi diagram generated from $N=5$ sites $\{\site{1},\ldots, \site{5}\}$ on a periodic domain. The radius $r_1$ to $\partial\region{1}$ at angle $\theta$ relative to site $\site{1}$ is depicted by a blue arrow. }
\label{f:star_voronoi}
\end{figure}

For simplicity of analysis we restrict our attention to subsets of $\bR^2$ which contain a collection of $N$ sites, each representing a star polymer or other deformable object with radial symmetry. The sites are used to form a Voronoi partition of the rectangular domain $\Omega_{\alpha,N}\subset \bR^2$ with aspect ratio $\a\in(0,1]$. The domain is taken to rectangular $\Omega_{\a,N}=[0,\sqrt{N\a}]\times [0,\sqrt{N/\a}]$, normalized to have unit area per site. We impose periodic boundary conditions so that the domain is equivalent to the torus, $\bT_\alpha := \bR^2 / \Omega_{\alpha,N}.$ \tcb{This choice best approximates the ``infinite'' bulk systems we seek to recover in the large $N$ limit.}   We assume a simple Hookean spring energy in which the arms have linear resistance to compression and extension.  The $N$ sites are denoted 
$\bx:=(\site{1}, \site{2}, \ldots, \site{N}) \in \bR^{2N}$, 
with each $\site{i}\in\Omega_{\alpha,N}$.
The Voronoi partition of $\Omega_{\alpha,N}$ divides it into regions  $\{\region{j}\}_{j=1}^N$ with disjoint interior where $\region{j}$ is defined to be the subset of $\Omega_{\alpha,N}$ whose points are closer, in toroidal distance, to $\site{j}$ than to any other any other site. 
With this notation the Hookean-Voronoi energy takes form
\beq\label{e:energy-def}
    \EHV(\bx) = \sum\limits_{j=1}^N \int_0^{2\pi} \big|r_j(\theta)-r_*\big|^2 \mrd \theta, 
\eeq
where $r_j$ is the distance on $\bT_\alpha$ from $\site{j}$ to $\partial V_j$ along the ray making angle $\theta$ to the positive $x_1$ axis, Figure\,\ref{f:star_voronoi} (right). The quantity $r_*$ denotes the equilibrium length of each polymer arm of the star polymer. \tcb{The energy models the compression of each polymer arm as a spring with unit Hooke's law constant and assumes a sufficiently high density of polymer arms that the energy of compression dominates the splay between arms.} Since Voronoi regions are convex the radii $\{r_j\}_{j=1}^N$ are well defined  and the Hookean-Voronoi energy is well-posed.  

It is instructive to reformulate the Hookean-Voronoi energy in terms of the average radius of the Voronoi region, $\region{j}$, with respect to its site $\site{}$, defined as
\beq\label{e:rbar}
\olr_j(\bx):=\frac{1}{2\pi} \int_0^{2\pi} r_j(\theta)\ \mrd\theta.
\eeq
Expanding the square in \eqref{e:energy-def} and using the polar-coordinate formula for region area 
\[
\begin{aligned}
    \EHV(\bx) &= \sum\limits_{j=1}^N \int_0^{2\pi}\left( r_j^2(\theta)-2r_*r_j(\theta)+r_*^2\right) \mrd \theta, \\
    &= \sum\limits_{j=1}^N \left( 2 |V_j|-4\pi r_* \olr_j +2\pi r_*^2\right),
    %\\
    %&= 2\pi r_0^2 
    %\sum\limits_{j=1}^N \left(\frac{A_j}{\pi r_0^2}-2 \olr_j +1\right),
\end{aligned}
\]
where $|V_j|$  denotes the area of $V_j$.  Since the areas  of the Voronoi regions sum to the domain size $|\Omega_{\alpha,N}|=N$, we arrive at the expression
\beq\label{e:energy-def-2}
 \EHV(\bx) = 2N\left(1 +\pi r_*^2\right) - 4\pi r_* \sum_{j=1}^N \olr_j.
\eeq
Up to a constant, the Hookean Voronoi energy depends only upon the sum of the average radii of the Voronoi sets. \tcb{The average radius has an inverse relation to the perimeter of the set. Indeed Jensen's inequality provides an \emph{upper} bound on the average radius in terms of the domain area.}
\begin{Proposition}
Let $V$ be a convex domain with a rectifiable boundary. Then for any $\site{}\in V$ the average radius of $V$ with respect to center $\site{}$ is well posed and satisfies
\beq\label{e:isopar}
\pi \olr^2 \leq |V|.
\eeq
\end{Proposition}
\begin{proof}
Since $V$ is convex the ray leaving $\site{}$ at angle $\theta$ has a unique intersection with $\partial V$ and hence $r=r(\theta;\site{})$ is well-defined. 
Jensen's inequality applied to convex function $s\mapsto s^2$ implies
\beq\label{e:Jensen}
\left(\frac{1}{2\pi} \int_0^{2\pi} r(\theta)\ \mrd \theta\right)^2 \leq \frac{1}{2\pi} \int_0^{2\pi} r^2(\theta)\ \mrd \theta=\frac{|V|}{\pi}.
\eeq
From the definition of $\bar{r}$, this is equivalent to \eqref{e:isopar}.
\end{proof}

\begin{Remark}
The result \eqref{e:isopar} is a form of ``inverse'' isoparametric inequality. For any convex domain which encloses a specified area, the average radius is maximized for the circle with the site $\site{}$ at the circle's center.  This motivates the introduction of the averaged isoparametric reciprocal 
\beq
\rho_\circ(V;\site{}):=\frac{\pi \olr^2}{|V|}\leq 1,
\eeq
which resembles the reciprocal of the standard isoperimetric ratio of area to the square of the perimeter. 
\end{Remark}

\subsection{Voronoi Notation and Symmetry}
\begin{figure}[H]
    \centering
    \begin{subfigure}{.45\textwidth}
        \centering
        \includegraphics{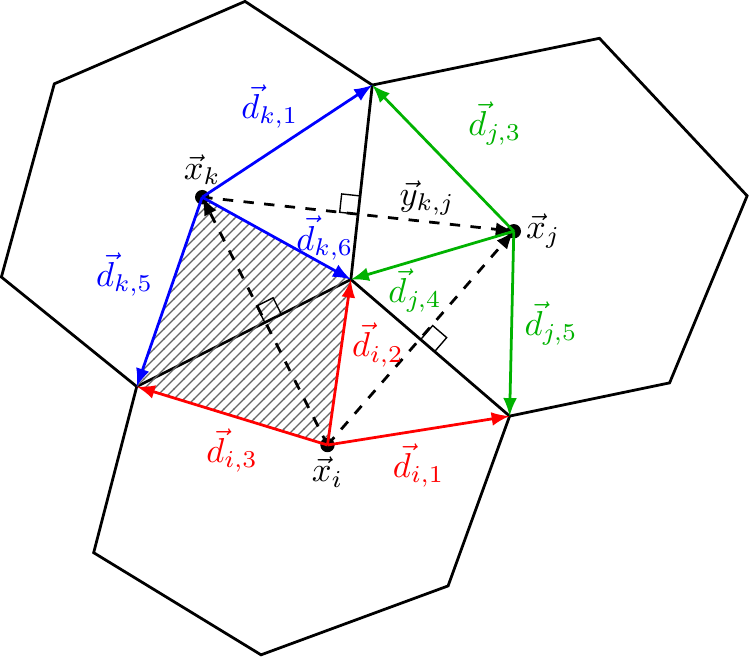}
    \end{subfigure}
    \begin{subfigure}{.45\textwidth}
        \centering
        \includegraphics{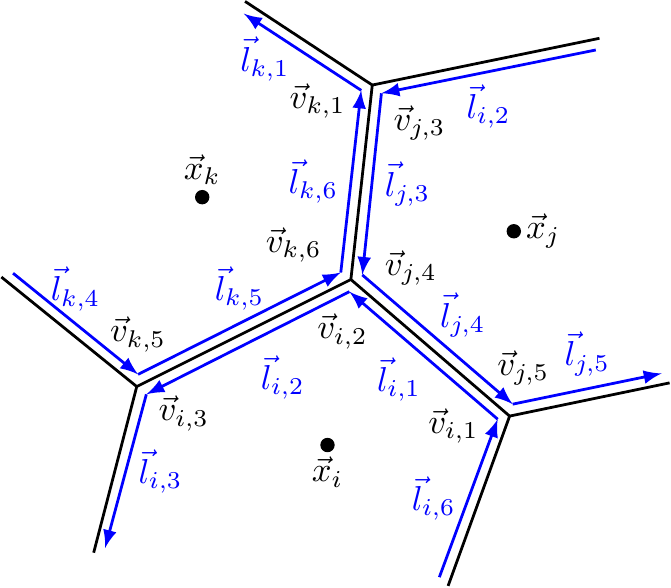}
    \end{subfigure}
\caption{Sample of Voronoi cells with shared vertex $\vert{}$ labeled with key vectors. Relative vertex vectors $\vec{d}$, Voronoi edges as perpendicular bisectors, and the kite region formed between sites $\site{i}$ and $\site{k}$ are shown (left). The edge vectors $\vec{l}$ are drawn adjacent to the edge, showing the opposing directions between neighboring sites due to the counterclockwise orientation (right).
}%close caption
\label{f:voronoi_notation}
\end{figure}

 The gradient of the Hookean-Voronoi energy can be simplified using the symmetries of the Voronoi partition.  For a given Voronoi partition we denote the total number of unique vertices of the regions by $M=M(\bx)$, and enumerate them $\{\vert{j}\}_{j=1}^M.$  However, for each individual  Voronoi region,
$\region{i}$, its $m_i$ vertices are also given a double-subscript notation $\{\vert{i,1}, \ldots, \vert{i,m_i}\}$, enumerated in the second index so that the vertices lie counter-clockwise about $\site{i}$, with $\vert{i,1}$ making the smallest non-negative angle with the positive $x$-axis. 
For $i=1,\ldots, N$ and $j=1,\ldots, m_i$ we denote the angle made by vertex $\vert{i,j}$, site $\site{i}$ and the positive $x$-axis by $\varphi_{i,j}.$ These satisfy $0\leq \varphi_{i,1}< \varphi_{i,2}<\cdots< \varphi_{i,m_i}.$
%\}_{\nu\in\vertindex{i}}$ indexed through the set
%\[
%    \vertindex{i} := \{\nu_{i,1}, \nu_{i,2}, \dots, \nu_{i,m_i} \}.
%\]
%We say $V_i$ has vertices  As an example, $V_5$ might have vertices $\{\vert{3},\vert{5},\vert{7},\vert{11}\}$ such that $\vertindex{5}=\{3, 5, 7, 11\}$. Additionally, fo
%For $\nu \in \vertindex{i}$, the vertex $\vnext{i}{\nu}$/$\vprev{i}{\nu}$ denotes the next/previous vertices from $\vert{\nu}$ moving in the counterclockwise direction about site $\site{i}$. 
We also introduce the vectors
\[
    \vd{i}{j} := \vert{i,j} - \site{i}, \quad \vl{i,j} := \vert{i,j+1} - \vert{i,j}, 
\]
where here and below the vertex $j$ is understood to be taken mod $m_i$. These respectively represent the vertex-site and counterclockwise vertex-vertex vectors,
see Figure\,\ref{f:voronoi_notation}.  The set of near-neighbors of $\site{i}$, $\mathcal{N}(\site{i})$, includes the sites whose Voronoi sets share an edge with $\region{i}$, or more precisely the sites whose Voronoi regions share precisely two distinct vertices with $\region{i}.$ Specifically two sites whose Voronoi regions intersect at a single vertex are not near neighbors. 

Voronoi tessellations possess a key symmetry. The vector between two near-neighbor sites is perpendicularly bisected by the edge that the two regions share.  The vector from $\site{i}$ to the near neighbor $\site{j}$ that lies on the perpendicular to edge with vector $\vl{i,j}$ is denoted
$\vy{i}{j} := \site{j} - \site{i},$. The  site vector $\vy{i}{j}$
is perpendicularly  bisected by the corresponding edge. This implies that the quadrilateral region with vertices given by the two neighboring sites $\site{i}$ and $\site{k}$, and their vertices is a kite, with area given by the cross product
\beq\label{e:kite-area}
  |\vA{i}{j}|=  \left|\vy{i}{j} \times \vl{i,j}\right| = %\frac{1}{2}\vy{i}{j}^\bot \cdot \vl{i,j} =
    \left|\vl{i,j}\right
    |\Bigl|\vy{i}{j}\Bigr|.
\eeq
Indeed the triangle with vertices $\site{i}$, $\vert{i,j}$, $\vert{i,j+1}$ is equivalent to the triangle formed by the same two vertices and associated near-neighbor $\site{k}$, 
see the shaded region in Figure\,\ref{f:voronoi_notation} (left). We call this the kite-symmetry of the Voronoi regions.
%where here and below $\vec{x}^\bot := R\vec{x}$, and $R$ is rotation by $\pi/2.$ 
%The defining property of the Voronoi tessellation is that each vertex is the circumcenter of the three closest site. We denote by $\mathcal{X}(\vert{i})$ the indices of the sites whose Voronoi region have $\vert{i}$ as a vertex.

\subsection{Calculation of the Average Radius}
The average radius admits a closed form expression. For each region $\region{i}$ the integrand $r_i=r_i(\theta)$ in \eqref{e:rbar} is continuous on $[0,2\pi]$ and smooth in $\theta$ on the sub-interval $\{[\varphi_{i,j},\varphi_{i,j+1}]\}_{i=1}^{m_i},$
where for $j=m_i$ we denote $\varphi_{i,m_i+1}=\varphi_{i,1}+2\pi.$
For each $i=1,\ldots N$ we decompose the average radius $\bar{r}_i$ defined in \eqref{e:rbar} into the sum of $\bar{r}_{ij}$ over $j=1,\ldots, m_i$ defined through
\beq
\label{e:rbar-ij}
%\frac{1}{2\pi}\sum_{j=1}^{m_i} \bar{r}_{i,j} = 
\bar{r}_{i,j}:=\frac{1}{2\pi}\int_{\varphi_{i,j}}^{\varphi_{i,j+1}} r_{i,j}(\theta)\  \mrd\theta.
\eeq
The terms $\bar{r}_{i,j}=\bar{r}_{i,j}(\site{i},\vert{i,j},\vert{i,j+1})$ denote the contribution to the average radius of region $\region{i}$ from the triangle formed by site $\site{i}$ and vertices $\vert{i,j}$ and $\vert{i,j+1}.$ For $\theta\in[\varphi_{i,j},\varphi_{i,j+1}]$, the radius takes the form
\beq\label{e:rij-def}
    \vr{i}{j} = - \frac{2\vA{i}{j}}{(\vl{i,j})^\bot \cdot (\cos\theta, \sin\theta)^T},
\eeq
where the triangle area $A_{i,j}$ is given by
\beq 
\label{e:Aij-def}
    \vA{i}{j} := \frac12 \big|\vd{i}{j} \times \vl{i,j} \Big| = \frac12\vd{i}{j}^\bot \cdot \vl{i,j}=\frac12(\site{i}-\vert{i,j})\cdot \vl{i,j}^\bot.
\eeq
Here and below $\vec{x}^\bot := R\vec{x}$, where $R$ is rotation by $\pi/2.$  Notably the sub-parts $r_{i,j}$ of $r_i$ dependent linearly upon $\site{i}$ through $A_{i,j}$. We re-write the denominator of $\vr{i}{j}$ as
$$ (\vl{i,j})^\bot \cdot (\cos\theta, \sin\theta)^T= |\vl{i,j}|\sin(\theta + \vpsi{i}{j}),$$ 
where the phase is given by
\beq\label{e:psi}
\psi_{i,j}=-\arctan\left(\frac{l_{i,j}^{(1)}}{l_{i,j}^{(2)}}\right).
\eeq
The integral can be evaluated through the substitution $u=\tan((\theta+\vpsi{i}{j})/2)$ yielding the expression
\beq\label{e:r_ij-exp}
    \bar{r}_{i,j}=
    %\frac{1}{2\pi} \int_{\varphi_{i,j}}^{\varphi_{i,j+1}}
%    \frac{1}{\sin(\theta + \vpsi{i}{j})} \mrd \theta =
 \frac{A_{i,j}}{\pi|\vl{i,j}|}\left(   \arctanh\left(\frac{\vd{i}{j+1} \cdot \vl{i,j}}{|\vd{i}{j+1}||\vl{i,j}|}\right) - \arctanh\left(\frac{\vd{i}{j} \cdot \vl{i,j}}{|\vd{i}{j}||\vl{i,j}|}\right)\right).
\eeq
Denoting the angle between the edge $\vl{i,j}$ and the site-vertex vectors $\vd{i}{j}$ and $\vd{i}{j+1}$ as $\alpha_{i,j}$ and $\beta_{i,j}$ respectively, measured counter-clockwise, then the average radius has the expression
\beq\label{e:barr-exact}
\bar{r}_i:=\frac{1}{2\pi}\sum_{j=1}^{m_i}
%\frac{\vd{i}{j}^\bot\cdot\vl{i,j}}{|\vl{i,j}|}
|\vd{i}{j}|\sin(\alpha_{i,j})\left( \arctanh (\cos\beta_{i,j}) -\arctanh(\cos\alpha_{i,j})\right).
\eeq

\subsection{The Gradient Flow}
\label{ss:GF}
Despite the closed from expression for the average radius, it is informative to calculate the gradient of the Hookean-Voronoi energy directly. The vertices of the Voronoi regions, collectively enumerated as $\bv=\bv(\bx)$, depend upon the choice of sites. In all cases the vertices are a continuous function of the sites, \cite{Du-1999}. A vertex $\vert{}$ is non-degenerate if the three sites that are closest to it are at the same distance, $\rho(\vert{})>0$, and all other sites are strictly further away. In this case the tessellation is said to be Delaunay and vertices depend smoothly upon the site locations.
\begin{Lemma}
A non-degenerate vertex depends smoothly upon the sites.
\end{Lemma}
\begin{proof}
A non-degenerate vertex $\vert{}$ and its distance $\rho>0$ are defined as the solution of the system
$$ \|\vert{}-\site{j}\|_2^2=\rho^2,$$
for $j=j_1,j_2,j_3$ which denote the indices of the three sites closest to $\vert{}$. This system can be written as 
$F(\vert{},\rho; \site{j_1},\site{j_2},\site{j_3})=0,$
with the Jacobian of $F$ at $(\vert{},\rho)$ taking the form
$$\nabla_{\vert{},\rho}F = 
2
\left(\begin{matrix} (\vert{}- \site{j_1})^T & \rho \cr
                    (\vert{}- \site{j_2})^T & \rho \cr
                     (\vert{}- \site{j_3})^T & \rho
\end{matrix}\right)\in \mathbb R^{3\times3}.
$$
So long as $\rho>0$ and $\{\site{j_1},\site{j_2},\site{j_3}\}$ are not collinear, then det$\nabla_{\vert{},\rho}F\neq 0.$ If $\vert{}$ is non-degenerate then the three sites are distinct and equal distance from $\vert{}$. Hence they cannot be collinear and the Jacobian is invertible. Since $F$ is smooth function of all of its arguments, the implicit function theorem implies that $\vert{}$ and $\rho$ depend smoothly upon the sites in some neighborhood.
\end{proof}

This gives us the following result.
\begin{Proposition}
The Hookean-Voronoi energy is a continuous function of the sites, $\bx\in\mathbb R^{2N}.$ If each of the sites $\bv=\bv(\bx)$ is non-degenerate, then the Hookean-Voronoi energy lies in 
${\mathcal C}^3\left([\Omega_{\alpha,N}]^{2N}\right).$
\end{Proposition}
\begin{proof}
The Hookean-Voronoi energy is a smooth function of the sites $\bx$ and the vertices $\bv.$ If each vertex is non-degenerate, then the vertices are smooth functions of the sites, and the result follows.
\end{proof}

The gradient of the energy naturally arises from the variation of the average radius due two effects: the variation of site $\site{i}= (\site{i}^{(1)}, \site{i}^{(2)})\in\bR^2$ within a fixed domain $\region{i}$, and the variation of the domain $\region{i}$ due to motion of the vertices, $\bv=\bv(\bx)$.  Correspondingly, for any function $f:\bR^{2N}\times\bR^M\mapsto\bR$ of the sites $\bx$ and the vertices $\bv=\bv(\bx)$ we express the derivative with respect to $\site{i}$ in the form 
\beq\label{e:energy-grad-form}
    \underbrace{D_{\site{i}} f(\bx, \bv) }_{2\times 1} =  \underbrace{\bpart_{\site{i}} f(\bx, \bv)}_{2 \times 1} + \sum_{\vert{} \in \bv} \underbrace{(\partial_{\site{i}} \vert{})^T}_{2 \times 2} \underbrace{\partial_{\vert{}}f(\bx,\bv)}_{2 \times 1}.
\eeq
Here $D_{\site{i}}$ denotes the partial Jacobian of $f$ with respect to $\site{i}$ with $f$ viewed as function only of $\bx$ with $\bv$ eliminated through the relation $\bv=\bv(\bx).$ The ``fixed-vertex'' Jacobian $\bpart_\site{i}$ denotes a partial Jacobian of $f$ with respect to $\site{i}$ with $\bv$ fixed, $\partial_{\vert{}}f$ denotes the usual partial Jacobian of $f$ with respect to $\bv$ with $\bx$ fixed, and $\partial_\site{i}$ is the usual partial Jacobian of a function of $\bx$ only.
With this notation the gradient flow takes the form
\beq
\label{e:HV-gradflow}
    \frac{d\bx}{dt} = -D_\bx \EHV(\bx) = -\begin{bmatrix}D_{\site{1}} \EHV(\bx) \\
    % D_{\site{2}} \EHV(\bx) \\
     \vdots\\D_{\site{N}} \EHV(\bx) \end{bmatrix},
\eeq
subject to initial data.

%\subsection{Simplification of the Gradient}
The kite-symmetry of the Voronoi diagram leads to a simplification of the gradient due to a cancellation of terms arising from the variation of the vertices.   We take $D_{\site{k}}$ of equation \eqref{e:energy-def-2} and bring the $\partial_{\vert{}}$ derivatives into the integral to obtain
\beq 
D_{\site{k}} \EHV = -4\pi r_*\sum_{i=1}^N\left(  \bpart_{\site{k}}\bar{r}_i+ 
\frac{1}{2\pi}\int_0^{2\pi} \sum_{j=1}^{m_i} (\partial_{\site{k}}\vert{i,j})^T \partial_{\vert{i,j}} r(\theta)\mrd \theta\right).
\eeq
The function $\bar{r}_i$ depends upon $\site{k}$ only if $i=k$, and the  sum over $r_i$ reduces to a single term.  Using the kite-wise formulas 
$r_{i,j}=r_{i,j}(\theta,\vert{i,j},\vert{i,j+1},\site{i})$ for 
$r_i$ in \eqref{e:rbar-ij} the integrals can be broken into sums over edges of $\region{i},$ for which the $\partial_{\vert{}}$ terms reduce to two per side,
\[
    D_{\site{k}} \EHV =-4\pi r_* \left(\bpart_{\site{k}}r_k  +\frac{1}{2\pi}\sum_{i=1}^N \sum_{j=1}^{m_i}  \int_{\varphi_{i,j}}^{\varphi_{i,j+1}} \left(
      (\partial_{\site{k}} \vert{i,j})^T\partial_{\vert{i,j}}+
 (\partial_{\site{k}}\vert{i,j+1})^T\partial_{\vert{i,j+1}}\right) 
      r_{i,j} (\theta) \mrd \theta\right). 
\]

 The double sum vanishes due to the kite-symmetry. To see this 
we relabel the $K$ edges in the Voronoi diagram as $\{\vl{\ell}\}_{\ell=1}^K$. The two sites in the associated kite are labeled $\site{\ell+}$ and $\site{\ell-}$ with site $\site{\ell+}$ having the larger $x$-coordinate or larger $y$-coordinate if the $x$-coordinates are equal. The vertices that terminate the edges of $\vl{\ell}$ are labeled $\vert{\ell\pm}$ respecting the counter-clockwise orientation about $\site{\ell+}.$ The radius functions $r_{i,j}$ on each kite are relabeled $r_{\ell\pm}$ to align with their site $\site{\ell\pm}$. The angle end-points of $r_{\ell+}$ are labeled $\varphi_{\ell\pm}$ while the $\theta$ dependence of $r_{\ell-}$ is linearly translated so that its end-points are also $\varphi_{\ell_\pm}$. This is possible as the two halves of the kite are isomorphic. The double sum is regrouped into a single sum over the Voronoi domain edges,
\[
    D_{\site{k}} \EHV = -4\pi r_* \Bigg(\bpart_{\site{k}}\bar{r}_k
    %\sum_{j=1}^{m_k} \int_{\varphi_{k,j}}^{\varphi_{k,j+1}} \partial_{\site{k}} \vr{k}{j} \mrd \theta\\
    + \frac{1}{2\pi}\sum_{\ell=1}^K \int_{\varphi_{\ell-}}^{\varphi_{\ell^+}} \left((\partial_{\site{k}}\vert{\ell^+})^T \partial_{\vert{\ell+}}  + 
    (\partial_{\site{k}}\vert{\ell^-})^T \partial_{\vert{\ell-}}\right)
    (r_{\ell+}+r_{\ell-})(\theta)
    \mrd \theta \Bigg).
\]
Here we have introduced the function 
$$ r_\ell:= 
(r_{\ell+}+r_{\ell-})(\theta,\vert{\ell+},\vert{\ell-},\site{i+},\site{i-}),$$
which inherits properties from the kite symmetry. In particular $$\partial_{\vert{\ell\pm}} r_\ell \cdot \vl{\ell}=0,$$
since moving either vertex $\vert{\ell\pm}$ in the direction parallel to $\vl{\ell}$ does not change the distances $r_{\ell\pm}$ of $\vl{\ell}$ to $\site{\pm}$, respectively. 
Moreover, as the two triangles in the kite are isomorphic, the function 
$$
t\in\mathbb R\mapsto r_\ell(\theta;\vert{\ell+}+t \vl{\ell}^\bot,\vert{\ell-},\site{\ell+},\site{\ell-}),
$$
has even parity about $t=0,$ hence its derivative at $t=0$ is zero. Similar arguments with perturbations to $\vert{\ell-}$ allow us to deduce that
$$ \partial_{\vert{\ell\pm}} r_\ell\cdot  \vl{\ell}^\bot=0.$$
Since $\vl{\ell}$ and $\vl{\ell}^\bot$ span $\bR^2$, the vectors $\partial_\vert{\ell\pm}r_\ell\in\mathbb R^2$ are zero, and the integrands in $D_{\site{k}}\EHV$ are identically zero. The $\site{k}$-gradient of the Hookean-Voronoi energy reduces to
\beq\label{e:HV-grad}
    D_{\site{k}} \EHV =-4\pi r_* \bpart_{\site{k}} \bar{r}_k,
\eeq
for $k=1,\ldots, N.$ The Hookean-Voronoi gradient flow reduces to
\beq
\label{e:HV-gradflow2}
    \frac{d\bx}{dt} = 
    %-D_\bx \EHV(\bx) = 
     4\pi r_*\begin{bmatrix}
     \bpart_ {\site{1}} \bar r_1(\bx;\bv) \\
%     \partial_ {\site{2}} \bar r_2(\bx) \\
             \vdots\\
         \bpart_{\site{N}} \bar r_N(\bx;\bv)
    \end{bmatrix}.
\eeq
The gradient depends only upon the constant-vertex derivatives of the average radii. The flow remains fully coupled since the motion of each site shifts the vertices, thereby impacting the neighboring site's motion. This formulation shows that $\bx$ is an equilibrium if and only if each site $\site{i}$ is a critical point of the average radius of its domain $\region{i}$. We examine this in the following sub-section.

\subsection{Max-centers and Equilibrium}
The reduced form of the gradient of $\EHV$ motivates the introduction of the max-center of a convex domain. 

\begin{Def}

For a fixed, bounded, convex region $V \subset \bR^2$ with a piece-wise smooth boundary. The max-centers of $V$ are the elements of 
\beq
    \underset{\vec{x} \in V}{\arg \max{\bar{r}(\vec{x})}}.
\eeq
These are the points $\site{}\in V$ that maximize the average radius of region $V.$
\end{Def}

For a domain $\region{}$ with a piecewise linear boundary, the max-center is unique. 
\begin{Proposition}
\label{t:MCU}
Let $V$ be convex with a piecewise linear boundary, then  the average radius $\bar{r}(\site{};V)$ is strictly concave function of $\site{}.$ In particular $\bar{r}$ has a unique critical point which is a maximum and $V$ has a unique max-center.
\end{Proposition}

\begin{proof}
We show that $\bar{r}:\region{}\mapsto\mathbb R_+$ is a strictly concave function of $\site{}$ and hence has a unique maximum.
The integrand of $\bar{r}$ is continuous with respect to $\site{}\in\region{}$ but is not ${\mathcal C^1}(\region{}).$
We adopt the notation of sub-section\,\ref{ss:GF}, by setting $\region{}=\region{i}$ for some $i$. 
Taking the fixed-vertex partial $\bpart_{\site{i}}$ of \eqref{e:rbar}, we may exchange the order of one derivative with the integral, yielding
\[
 \bpart_{\site{i}}\bar{r}_i(\site{i}) = 
 \frac{1}{2\pi}\int_0^{2\pi}
 \bpart_{\site{i}}r_i(\theta)\mrd\theta =
 \frac{1}{2\pi}\sum_{j=1}^{m_i} \int_{\varphi_{i,j}}^{\varphi_{i,j+1}} \bpart_{\site{i}}r_{i,j}(\theta)\mrd\theta. 
 \]
 The fixed-vertex Hessian, $\bpart^2_{\site{i}}$, of $\bar{r}_i$ does not commute with the full integral and must be taken in the component-wise formulation
\beq\label{e:Hessian-barr}
    \bpart^2_{\site{i}}\bar{r}_i = 
    \frac{1}{2\pi} \sum_{j=1}^{m_i} \left(
    \left[
   \bpart_{\site{i}}r_{i,j}(\varphi_{i,j+1})  (\partial_{\site{i}}\varphi_{i,j+1})^T - 
     \bpart_{\site{i}}r_{i,j}(\varphi_{i,j})    (\partial_{\site{i}}\varphi_{i,j})^T\right]
    + \int_{\varphi_{i,j}}^{\varphi_{i,j+1}} \bpart^2_{\site{i}}r_{i,j}(\theta)\mrd\theta\right).
\eeq
The functions $r_{i,j}$ given in \eqref{e:rij-def} are linear in $\site{i}$ through $A_{i,j}$, hence their $\site{i}$-Hessian is zero and the integral above vanishes. To evaluate the boundary terms we determine that
\beq\label{e:part-rij} 
\bpart_{\site{i}} r_{i,j} \bigl|_{\theta=\varphi_{i,k}}= -\frac{\vl{i,j}^\bot}{\vl{i,j}^\bot\cdot (\vd{i}{k} / |\vd{i}{k}|)}
= \frac{\vl{i,j}^\bot}{\vl{i,j}^\bot\cdot \vd{i}{j}}|\vd{i}{k}|,
\eeq
for $k=j, j+1$ where we used $\vd{i}{j}+\vl{i,j}=\vd{i}{j+1}$ to derive the last equality. From the definition of $\varphi$ we have
\beq\label{e:grad-varphi}
    \bpart_{\site{i}} \varphi_{i,j} =- \frac{\vd{i}{j}^\bot}{|\vd{i}{j}|^2}.
\eeq
Substituting these expressions into \eqref{e:Hessian-barr} we express the fixed-vertex hessian of $\bar{r}_i$ as a sum of $m_i$ rank-one terms,
\beq\label{e:Hessian-barr2}
    \bpart^2_{\site{i}}\bar{r}_i = -\frac{1}{2\pi} \sum_{i=1}^{m_i} \frac{\vl{i,j}}{\vl{i,j}^\bot\cdot \vd{i}{j}} \cdot \left(\frac{(\vd{i}{j+1})^T}{|\vd{i}{j+1}|} - \frac{(\vd{i}{j})^T}{|\vd{i}{j}|} \right).
\eeq
For any term in the summation
\beq
    \begin{aligned}
    \textrm{tr}(\bpart_{\site{i}}^2\bar{r}_{i, j}) &= 
   - \frac{1}{2\pi}\frac{\vl{i,j}}{\vl{i,j}^\bot\cdot \vd{i}{j}} \cdot \left(\frac{\vd{i}{j+1}}{|\vd{i}{j+1}|} - \frac{\vd{i}{j}}{|\vd{i}{j}|} \right),
    \\
    &= - \frac{1}{2\pi}\frac{\vd{i}{j+1}-\vd{i}{j}}{\vl{i,j}^\bot\cdot \vd{i}{j}} \cdot \left(\frac{\vd{i}{j+1}}{|\vd{i}{j+1}|} - \frac{\vd{i}{j}}{|\vd{i}{j}|} \right),    \\
    &= - \frac{1}{2\pi}\frac{|\vd{i}{j+1}| + |\vd{i}{j}|}{\vl{i,j}^\bot\cdot \vd{i}{j}} 
       \left(1-\frac{\vd{i}{j+1}\cdot\vd{i}{j}}{|\vd{i}{j+1}||\vd{i}{j}|}\right) < 0.
    \end{aligned}
\eeq
Each matrix $\partial_{\site{i}}^2\bar{r}_{i, j}$ is rank one and negative semi-definite, and their ranges $\{\vl{i,j}\}_{j=1}^{m_i}$ span $\bR^2$ since they are the edges of $\region{i}$ and cannot all be linearly dependent. For any non-zero vector $\vec{w}\in\bR^2$, the bilinear form is non-positive
$$ \vec{w}^T [\bpart_{\site{i}}^2\bar{r}_{i}] \vec{w}=
\sum_{j=i}^{m_j} \vec{w}^T [\bpart_{\site{i}}^2\bar{r}_{i,j}] \vec{w}\leq 0,$$
and equals zero if and only $\vec{w}$ is in the kernel of each of $\bpart_{\site{i}}^2\bar{r}_{i, j}$. Since their ranges span, this is not possible, hence the matrix $\bpart_{\site{i}}^2\bar{r}_{i}$ is negative definite. Thus the Hessian of $\bar{r}_i$ is uniformly negative on the compact set $\region{i}$ and $\bar{r}_i$ has a unique maximum. 
\end{proof}

%\textcolor{blue}{Keith: Are the Hessians self-adjoint? How smooth is $\bar{r}$ wrt $\site{}$? Might need to work directly with the kernel or range of adjoint.}

\subsection{Max-center and Centroidal energies}
We denote the max-center of each Voronoi region $\region{i}$ by $\site{i,*}.$ From Proposition\,\ref{t:MCU} the max-center is unique and depends upon the sites only through their determination of the vertices, $\site{i,*}=\site{i,*}(\bv(\bx)).$ Moreover the max-center is the unique solution of $\partial_{\site{i}}\bar{r}=0$ for $\site{i}\in\region{i}.$  Consequently $\bx$
is an equilibrium of the Hookean-Voronoi gradient flow \eqref{e:HV-gradflow2} if and only if it solves $\bx=\bx_*(\bx). $ We denote the Hessian of the full energy $\EHV$ by 
$$\mrH=D_\bx^2 \EHV.$$
On the torus the Hookean-Voronoi energy is invariant up to translation of $\bx$ by
$$ \be_j:=(\ve_j, \ldots, \ve_j, \cdots, \ve_j)\in\mathbb R^{2N},$$
for $j=1,2$ where $\ve_1=(1,0)$ and $\ve_2=(0,1)$ are the canonical unit vectors. The Hessian generically has a two-dimensional kernel corresponding to these two translational invariants. An equilibrium of the system with a non-degenerate Voronoi decomposition is stable if the Hessian at $\bx=\bx_*$ is strictly positive
\beq
\label{e:coercivity}
\by^T \mrH (\bx_*)\by \geq \nu(\bx_*) |\by|^2,
\eeq
for all $\by$ orthogonal to the kernel $\{\be_1,\be_2\}^\perp$. When these conditions are satisfied we call $\nu(\bx_*)>0$ the coercivity of $\EHV$ at $\bx_*.$ When $\bx$ is close to $\bx_*(\bx)$, the Hookean-Voronoi energy satisfies the relation
\beq\label{e:HV-quad}
\EHV(\bx) = \frac{1}{2}(\bx-\bx_*(\bx))^T \mrH(\bx_*) (\bx-\bx_*(\bx))+ 
O\left(|\bx-\bx_*|^3\right).
\eeq
A tempting simplification is to replace $\mrH(\bx_*)$ with the identity matrix and drop the error terms. This yields the ``max-center'' energy
\beq\label{e:MC-energy} 
E_{\rm MC}(\bx):= \frac12  |\bx-\bx_*(\bx)|^2,
\eeq
so named in analogy to the centroidal energy frequently associated to Voronoi tessellations. The global minima of the max-center energy is zero, and is achieved precisely when the sites $\bx$  lie at the max-centers  $\bx_*$. In particular all equilibria, including local minima or saddles of $\EHV$, are global minimizers of $E_{\rm MC}.$ However non-global minima and saddle points of the max-center energy cannot be critical points of $\EHV.$  Indeed from its construction the max-center energy is not a Lyapunov functional for the Hookean-Voronoi flow, as it converts saddle points of $\EHV$ to global minima of $E_{\rm MC}$.  
%However the unstable-HV equilibria are a global minima of $E_{\rm MC}$ and hence the value of the max-center energy value will increase, from nearly zero, under the HV-flow. 

In the sequel we argue that the replacement of the Hookean Voronoi Energy with the max-center energy represents a significant loss of information contained in the distribution of the energies of the Hookean Voronoi equilibria. Indeed the max-center energy is a better approximation of the gradient-squared Hookean Voronoi energy, defined by
$$ E_{\rm GS-HV}(\bx):= \frac12 |D_\bx \EHV(\bx)|^2.$$
Energies of this form have been suggested in the context of saddle-point search methods, \cite{Prom-BGS}. For $\bx$ near $\bx_*$ the gradient square energy has the approximation
$$ E_{\rm GS-HV}(\bx)=\frac12 |\mrH(\bx_*)(\bx-\bx_*)|^2 + O\left(|\bx-\bx_*|^3\right).$$
At this level of approximation $E_{\rm GS-HV}$ relates to $E_{\rm HV}$ through the replacement of $\mrH(\bx_*)$ in \eqref{e:HV-quad} with $\mrH^2(\bx_*)$. This has a significant impact on gradient flow dynamics near these critical points as it flips the signs of all negative eigenvalues of the Hessian. However a subsequent replacement of $\mrH^2(\bx_*)$ with the identity does not introduce any sign-flips, making it plausible that the max-center energy and the gradient-squared energy give qualitatively similar gradient flow dynamics when the sites are near their max-centers. 

%$$ E_{\rm EV}(\bx):=\sum_{i=1}^{m_i} \int_{\region{i}}\frac 12 |y-\site{i}|^2\ \mrd y.$$

The relation of the Hookean-Voronoi, max-center, and the gradient squared Hookean-Voronoi energy have analogs with the centroidal energy. Indeed the volumetric Voronoi energy also called the quantizer energy, \cite{Tor10}, defined by
$$ \cF_{\rm VV}(\bx) :=\sum_{i=1}^N \int_{\region{i}} |y-\site{i}|^2 \mrd y,
$$
is minimized at $\bx$ iff the sites $\bx$ lie at the centroids, $\bx_c$ of the Voronoi regions, see Proposition 3.1 \& 3.2 of \cite{Du-1999}. This suggests the introduction of the centroidal energy
$$\cF_{\rm C}(\bx)= \sum_{i=1}^N \frac12 \|\site{i}-\site{c,i}\|^2,$$
where $\site{c,i}$ is the centroid of the Voronoi region $\region{i}.$ The centroidal Voronoi energy relates to the volumetric Voronoi energy in much the same way that the max-center energy relates to the Hookean Voronoi energy. The centroidal energy is more closely related to the gradient-squared volumetric Voronoi energy than to the volumetric Voronoi energy.

Generically the centroid and the max-center of a convex set are not equal, but are often quite close. It is relatively simple calculation to determine the centroid of a domain, and the centroidal energy has attracted significant attention  for their use in mesh generation, \cite{Du-1999} and image processing, \cite{Du-06}. Recently strategies for the efficient computation of centroidal Voronoi gradient flows, including approaches to enhance the rate of convergence to equilibria have been presented, \cite{Choksi-SJSC}. Analysis has shown convergence to hyper-uniform distributions and large $N$ distributions of cell energies in two dimensions, \cite{Tor19}.

\section{Ordered Equilibria}

A Voronoi tessellation said to be ordered if all of the Voronoi regions are identical up to a rigid body motion. From symmetry considerations the collection of sites that produced ordered tessellations is invariant under the gradient flow, since each site is interchangeable, by symmetry they remain so under the flow.
Consequently the gradient-flow induced motion of an ordered tessellation can be reduced to that of a single site moving towards the evolving max-center of its evolving region. The Euler characteristic relates the number of faces $N$, edges $K$, and vertices $M$ of the Voronoi tessellation
\[  N-K+M=\xi,\]
where $\xi$ is the Euler characteristic of the underlying domain. The Euler characteristic of a torus is zero. For an ordered tessellation, the number of vertices per site, $m$ is fixed. Since each edge joins two sites the edge number satisfies $K=mN/2$. Assuming non-degeneracy of vertices,  each vertex joins three edges, hence $M=mN/3$. The Euler characteristic reduces to the relation $N(1-m/6)=0$ and we deduce that a non-degenerate ordered tessellation has hexagonal regions with $6$ edges and $6$ vertices.

 %the $\The vertex collision generically produces four defects, in the form of two pairs of 5- and 7-sided defects. In simulations we find that defects in a Hookean-Voroni system prefer to align in pairs of 5- and 7-sided regions. In particular a single 5-7 pair in a background of 6-sided regions cannot be removed by a vertex collision. It must join another 5-7 pair, form a 5-7 quadripartite structure as depicted in Figure\,\ref{f:pair_collision} (right) and undergo the vertex collision in the reverse direction. This mechanism supports the persistence persistence of isolated $5-7$ pairs within a background of hexagons. The pairs must be brought in order to remove them. \textcolor{red}{This requires a transport phenomena of defects in a network that is evocative of the transport of excess protons through the hydrogen bounded network that occurs within water.}

\subsection{Single-String Voronoi Tessellations}
The single-string tessellations are a subset of the ordered tessellations that are critical points of the Hookean Voronoi energy. These are associated to closed, straight-lines geodesics of the torus $\mathbb T_\alpha.$  In the plane, the geodesic can be extended to a line in $\bR^2$, which may be shifted to pass through $(0,0)$. A closed geodesic corresponds to a line that passes through one of the lattice points
$\{(s_1 \sqrt{N\alpha} , s_2\sqrt{N/\alpha})\}_{s\in{\bZ^2}}$
corresponding to the $\Omega_{\alpha,N}$-periodic images of the upper-right corner of $\Omega_{\alpha,N}$. The line then forms a periodic orbit in $\Omega_{\a,N}.$
The single-string site vector $\bx$ is formed by placing $N$ equally-spaced points along the straight periodic orbit.  Indeed, fixing lattice point index $s=(s_1,s_2) \in \bZ^2$, then without loss of generality $s_1, s_2 \neq 0$ and $\gcd(s_1,s_2) = 1$,  and for $j=1,\ldots, N$  the site $\site{j}$ of $\bx $ is given by
\[
 \site{j} =  \left( j s_1 \frac{\sqrt{N\alpha}}{N} \textrm{ mod } \sqrt{N\alpha}, j s_2 \frac{\sqrt{N/\alpha}}{N} \textrm{ mod } \sqrt{N/\alpha} \right)^T.
\]

\begin{figure}[H]
    \centering
    \begin{subfigure}{.4\textwidth}
        \centering
        \includegraphics[width=0.8\textwidth]{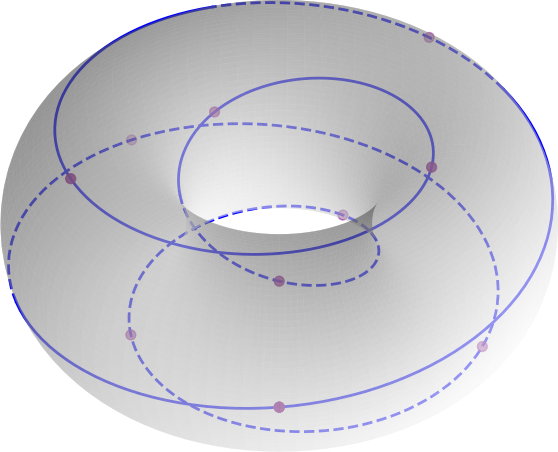}
    \end{subfigure}
    \begin{subfigure}{.4\textwidth}
        \centering
        \includegraphics[width=0.65\textwidth]{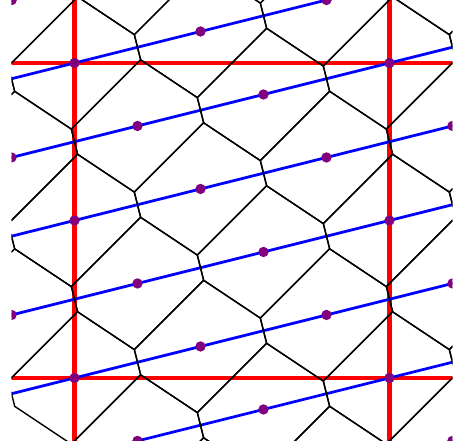}
    \end{subfigure}
\caption{Construction of a single-string tessellation for $\alpha=1$. (left) A closed, straight-line geodesic of the equal-radius torus $\mathbb T_1$ divided into $N=10$ sub-units. (right) The mapping of the ten points onto $\Omega_1$ (red lines), and the associated single-string Voronoi tessellation.}
    \label{f:singlestring}
\end{figure}

% We notice that due to periodicity, a configuration generated by $s=(x+kN,y+kN), k \in \bZ$ are all equivalent to one another. 
By the symmetry of the construction, the collection of vectors  $\{\vy{i}{j}\}$ from site $\site{i}$ to the sites of its near-neighbors are the same. Since these vectors define the Voronoi regions, each Voronoi region is isometric. 
%To see this, we observe that for any two neighboring sites $\site{i}$ and $\site{j}$, it must be that for an arbitrary $\site{k}$, $\site{k} + (\site{j}-\site{i})$ is its neighbor. 
For each $N$ and $\alpha$ the number of single-sting Voronoi tessellations is finite, as it cannot exceed the number of coprime pairs $(s_1,s_2)\in\mathbb Z^2$ with $1 \leq s_1,s_2 \leq N$.

\begin{Proposition}
\label{p:SS-equil}
For each $N\in\mathbb N_+$ and each aspect ratio $\alpha,$ the single-string Voronoi tessellations correspond to sites that are equilibria of the Hookean-Voronoi gradient flow \eqref{e:HV-gradflow2}.
\end{Proposition}
\begin{proof}
We show that $\bx=\bx_*.$  For the single-string construction each Voronoi region is a fixed rigid translation of each-other under any vector that points from site to site. Since these vectors have a $\pi$ rotation symmetry, each region is invariant under rotation by $\pi$ about its defining site.  This symmetry maps the associated max-center from $\site{i*}$ to its $\pi$ rotation image $2\site{i}-\site{i*}.$ Since the max-center is unique, the max-center and its $\pi$-rotation image must be the same, hence $\site{i}=\site{i,*}.$ 
\end{proof}

\subsection{Regular Hexagonal Voronoi Tessellations and Volumetric Excess Energy}
Regular hexagonal tessellations are a special case of single-string tessellations.  
From  \eqref{e:barr-exact} the average radius of the regular $s$-gon of unit area takes the form
\begin{equation}
\label{e:def-sgon}
   % \bar{\mathbf{r}}_s 
   \br_s= \frac{1}{\pi} \sqrt{\frac{s}{\tan(\pi/s)}}\arctanh(\sin(\pi/s)).
\end{equation}
A regular hexagon, $V_{\rm hex},$ of unit area 
%the average radius satisfies
%$$\bar{r}_6=\frac{1}{2\pi}\frac{6\sqrt{2}}{3^{1/4}}\arctanh\left(\frac{1}{2}\right),
%$$
has corresponding energy
\beq\label{e:Ehex}
\mrE_{\rm hex}:= 2(1+\pi r_*^2) -4\pi r_* \br_6.
\eeq
To facilitate comparisons of energy between different $N$ and $\alpha$ we introduce the {\sl volumetric excess energy}, \tcb{denoted $\VEE$}, of a tessellation and of a region within the tessellation. The first is computed by dividing the tessellation energy $\EHV(\bx)$ by $|\Omega_{\a,N}|=N$, to obtain an average energy per unit area, and subtracting the energy of a regular hexagon of unit area, 
\beq
\label{e:VEE-def}
\VEE(\bx):= \frac{\EHV(\bx)}{N} -\mrE_{\rm hex}.
\eeq
The volumetric excess energy of an individual Voronoi region $\region{i}$ with average radius $\bar{r}_i$ denotes the contribution of that region to the tessellation's $\VEE$. From \eqref{e:energy-def-2} this is defined as
\beq
\label{e:VEE-region}
\VEE(\region{i}):= 2(|\region{i}|+\pi r_*^2) -4\pi r_*\bar{r}_i - \mrE_{\rm hex},
\eeq
so that $\VEE(\bx)$ equals the average of the $\{\VEE(\region{i})\}$ of its regions. 

It is natural to ask which pairs $(N,\alpha)$ admit a tessellation by regular hexagons. The sites of a regular hexagonal tessellation lie on a triangular lattice. Fixing a site at the origin, we recall the Eisenstein integers generated by the cube root of unity,
\[ \bZ[\omega] \subset \mathbb{C},\quad \omega = e^{i\frac{2\pi}{3}}, \quad i = \sqrt{-1}. \]
The Eisenstein integers form a lattice in $\mathbb{C}$ corresponding to centers of a tiling by regular hexagons with area $\sqrt{3}/2$. There exists a regular hexagonal tiling of a periodic domain $\Omega_{\alpha,N}$ iff the vertices of $\Omega_{\alpha,N}$ correspond to Eisenstein integers $\{0, z, z', z+z'\} \in \bZ[\omega]$ with $z,z'$ orthogonal.  In this case $\Omega_{\alpha,N}$ must be a conformal transformation of the smallest rectangular domain $\Omega_{\rm hex}$ with vertices $\{0, 1, (1+2\omega), 2+2\omega\}$. The following result yields a constructive enumeration of the possible aspect ratios $\alpha$ of periodic domains that admit a regular hexagonal tiling with $N$ regions.

\begin{Proposition}
\label{p:Hex-tiling}
Let $N\in\bZ_+$ be even. Write $N = PQ$, with $P,Q\in\mathbb Z_+$ where $P$ is the product of all the prime factors $p$ of $N$ that satisfy $p \equiv 2 \textrm{ mod } 3$. 
%and $q \equiv \textrm{ mod } 3$ respectively, 
The aspect ratios of a rectangular domains that have a tessellation by $N$ regular hexagons is precisely
\beq\label{e:alpha-hex}
\cH(N) = \left\{\frac{N}{2a^2b}\sqrt{3}\quad \Biggl |\quad \forall a, b\in \mathbb Z_+: \, a \mid \frac{N}{2b}, \, b \mid Q\right\}.
\eeq
\end{Proposition}

\begin{Remark}
By convention the domains $\Omega_{\alpha,N}$ have aspect ratio $\alpha\in(0,1].$ To respect this convention we replace an element $\alpha\in \cH$ with its reciprocal, $\alpha^{-1}$, if $\alpha>1.$  This corresponds to rotating $\Omega_{\alpha,N}$ by $\pi/2$.
\end{Remark}
\begin{proof}
A general conformal transformation that maps $\bZ[\omega]$ to itself, transforms $\Omega_{\rm hex}$ onto $\Omega_{\alpha,N}$ where the vertices take the form
\[
    \{0,\ g\gamma,\ h\gamma(1+2\omega),\ (g+h+2h\omega)\gamma \},
\]
with $\gamma \in \bZ[\omega]$ and $g,h \in \bZ_+$. To avoid double counting, we require that $\gamma/k \notin \bZ[\omega]$ for any $k \in \bZ, k\geq 2$. The region $\Omega_{\alpha,N}$, has aspect ratio $\alpha=h|1+2\omega|/g=h\sqrt{3}/g.$ Since it is comprised of $N$ hexagons of size $\sqrt{3}/2$ it has area $|\Omega_{\alpha,N}|=N\sqrt{3}/2.$ The transformed domain has area $|g\gamma||h\gamma(1+2\omega)| = gh\sqrt{3}|\gamma|^2$. Equating these two yields the constraint
\[
    gh|\gamma|^2 = \frac{N}{2}.  
\]
In particular, $N$ must be even. 
Without loss of generality $h$ can be chosen in the form $h = N/(2g|\gamma|^2)$, where we require that $(2g|\gamma|^2) \mid N$. The Eisenstein integers,  $\bZ[\omega]$, are a unique factorization domain. The Eisenstein primes enjoy the following dichotomy, \cite{cox_1989}:  an element $\pi \in \bZ[\omega]$ is an Eisenstein prime if and only if one of two mutually exclusive conditions hold,
\begin{enumerate}
    \item $|\pi|^2$ is a prime  and $\pi \not\equiv 2 \textrm{ mod }3$
    \item $\pi$ is the product of a unit $\{\pm 1, \pm \omega, \pm \omega^2\}$ and a prime integer $p \equiv 2 \textrm{ mod }3$.
\end{enumerate}
This motivates the prime factorization of $N$ in the form 
\[
    N = PQ = \Big(p_1^{a_1}p_2^{a_2}\cdots p_i^{a_i}\Big)\left(q_1^{b_1}q_2^{b_2}\cdots q_j^{b_j}\right),
\]
for some exponents $a_k,b_k\in\bZ_+$ where the prime numbers $p_k \equiv 2 \textrm{ mod } 3$ for $k=1,\ldots i,$ and $q_k\not\equiv 2 \textrm{ mod } 3$ for $k=1,\ldots, j$. The constraint $\gamma/k \notin \bZ[\omega]$ implies that the Eisenstein primes that factor $\gamma$ cannot satisfy condition 2. We deduce that
\[ 
    |\gamma|^2 = |\pi_1^{\phi_1}|^2|\pi_2^{\phi_2}|^2\cdots|\pi_k^{\phi_k}|^2  = \left(q_1^{c_1}q_2^{c_2} \cdots q_j^{c_j}\right),
\]
for exponents $c_k\in\bZ_+$ that satisfy $0\leq c_k \leq b_k$ for $k=1, \ldots, j.$ In particular we may choose $g$ to be any divisor of $N/(2|\gamma|^2)$. Eliminating $h$ from the expression for aspect ratio we find the aspect ratios that admit a tiling by regular hexagons take the form  
\[
    \cH(N) = 
    \left\{\frac{N \sqrt{3}} {2g^2|\gamma|^2}\quad \Bigg|\, \forall g \mid \frac{N}{2|\gamma|^2}, \forall |\gamma|^2 \mid Q\right\}.
\]
Relabeling, we choose $b=|\gamma|^2$ that divides $Q$ and $a=g$ that $a$ divides $N/(2b)$, which yields \eqref{e:alpha-hex}.

\end{proof}
\begin{Remark}
For $N=36$, then $N=2^2\cdot3^2$, so that $Q = 3^2$, with divisors $b = \{1, 3, 9\}$. For each of the three choices of $b$ the divisors of $N/(2b)$ are
\[
    a_1 = \{1, 2, 3, 6, 9, 18\}, a_3 = \{1, 2, 3, 6\}, a_9 = \{1, 2\}.
\]
From \eqref{e:alpha-hex}, the set of admissible aspect ratios reduces to $6$ unique elements, which after taking reciprocals where necessary, take the form
\[
    \cH(36) = \sqrt{3}\left\{\frac{1}{54}, \frac{1}{18}, \frac{2}{27}, \frac{1}{6}, \frac{2}{9}, \frac{1}{2}\right\}.
\]
\end{Remark}

\begin{figure}[H]
    \begin{center}
    \begin{tabular}{cc}
    \includegraphics[width=.4\textwidth]{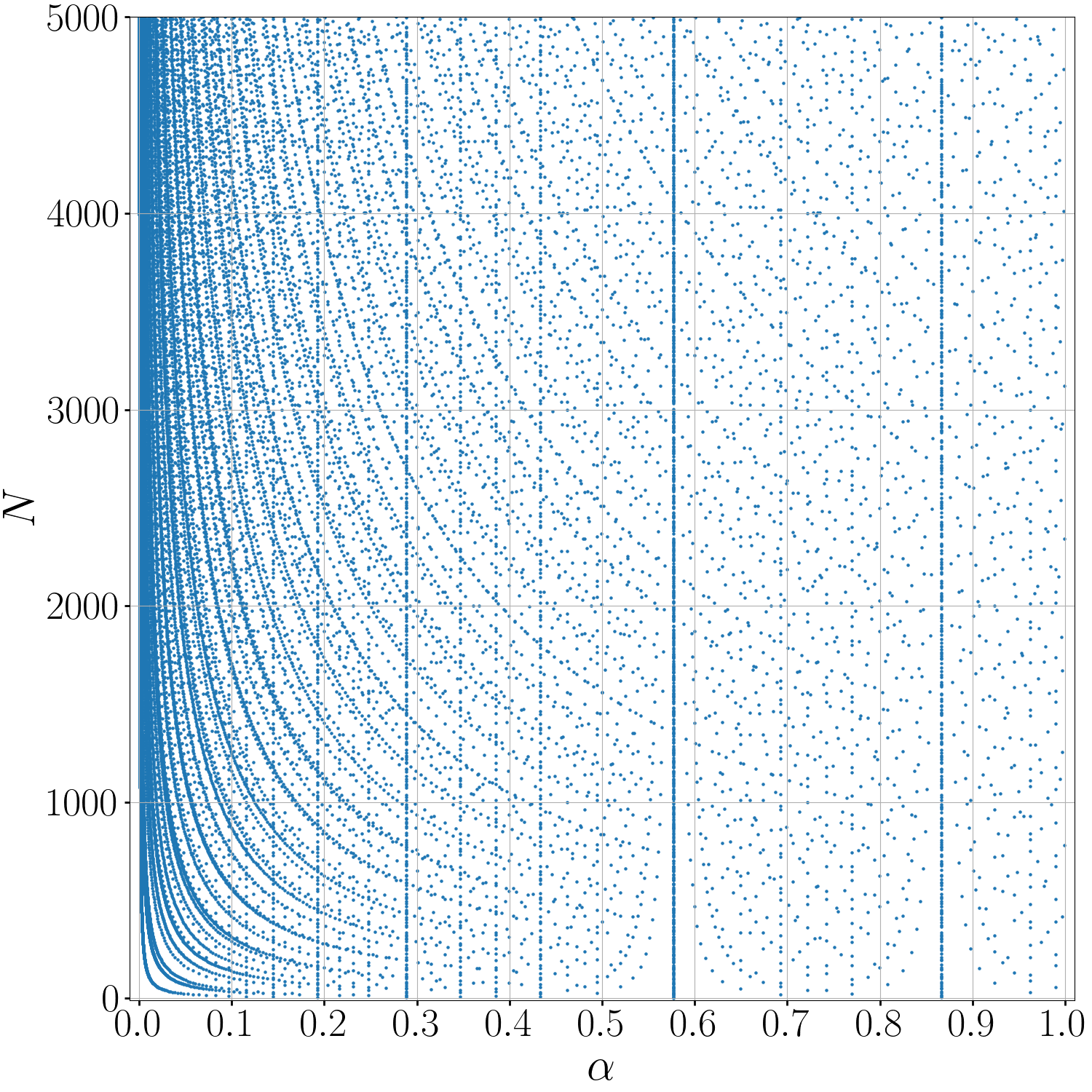}
    \includegraphics[width=.4\textwidth]{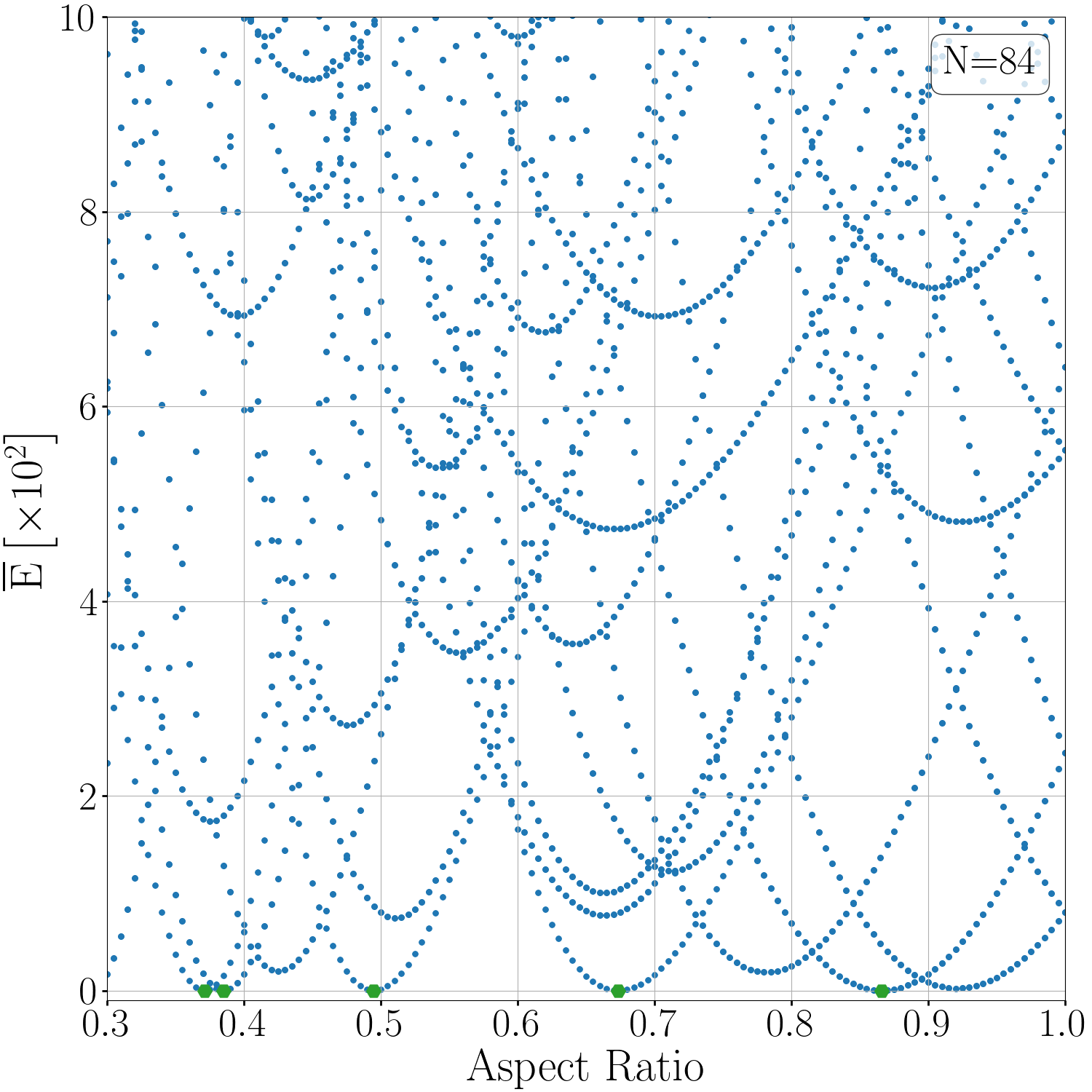}
    \end{tabular}
    \end{center}
\vspace{-0.2in}
    \caption{ (left) The aspect ratios $\alpha$ for which $\Omega_{\alpha,N}$ supports a tiling by $N$ regular hexagons. (right) $\VEE$ of all single-string equilibria for $N=84$ and $\alpha$ at intervals of $10^{-2}$ in $[0.3,1]$. The values $(0.3,1]\cap \cH(84)=
\Bigl\{
%\frac{{1}{126}, \frac{1}{42}, \frac{2}{63}, \frac{1}{ 18}, \frac{1}{14}, \frac{2}{21}, \frac{1}{ 6}, 
\frac{3\sqrt{3}}{14}, \frac{2\sqrt{3}}{9}, \frac{2\sqrt{3}}{7}, \frac{7\sqrt{3}}{ 18}, \frac{\sqrt{3}}{2}\Bigr\},$ corresponding to regular hexagonal tessellations are indicated by a green dot.}
    \label{f:Hex}
\end{figure}

As depicted in Figure\,\ref{f:Hex} (left), for $N\in[0,5000]$ sets $\cH(N)$ have a rich structure. Figure\,\ref{f:Hex} (right) presents the values  of $\VEE$ for all the single-string tessellations of $\Omega_{\alpha,N}$ for $N=82$ and $\alpha\in[0.3,1].$ The minimum ordered $\VEE$ at a given value of $\alpha$ is achieved as the minimum of the $\VEE$ over these discrete families of single-string equilibria which vary smoothly in $\alpha.$  For fixed $N$, the single-string equilibria can be decomposed into families parameterized by aspect ratio $\alpha.$ Each family has a $\VEE$ that is roughly parabolic in $\alpha$, and minimized at a particular value $\alpha=\alpha_*$. The green dots indicate the $\alpha$  for which the $\VEE$ of zero is attained at a regular hexagonal tessellation. This figure suggests that $\VEE$ is always positive, equivalently that the per-site energy of a tessellation is always greater than $\mrE_{\rm hex}.$ This is established in the next sub-section.

\subsection{Maximizing the Average Radius}
%From  \eqref{e:barr-exact} the average radius of the regular $s$-gon of unit area takes the form
%\begin{equation}
%\label{e:def-sgon}
   % \bar{\mathbf{r}}_s 
%   \br_s= \frac{1}{\pi} \sqrt{\frac{s}{\tan(\pi/s)}}\arctanh(\sin(\pi/s)).
%\end{equation}
The average radius of a  tessellation $\{\region{i}\}_{i=1}^N$ generated by sites $\bx\in\bR^{2N}$ is defined by the formula,
$$\bar{r}(\bx) :=\frac{1}{N}\sum_{i=1}^N \bar{r}(\region{i}).$$
We establish that $\VEE(\bx)\geq0$ by showing that any collection $\bx$ of sites satisfies $\br(\bx)\leq\br_6.$
We approach this by replacing the constraint that the regions of a tessellation form a Voronoi partition with a much weaker constraint, leading to a larger collection of regions. In this weaker formulation the average radius is a convex function of the collection, and we show that this convex problem is optimized at the regular hexagon. 

\begin{figure}[H]
    \centering
    \includegraphics[width=.7\textwidth]{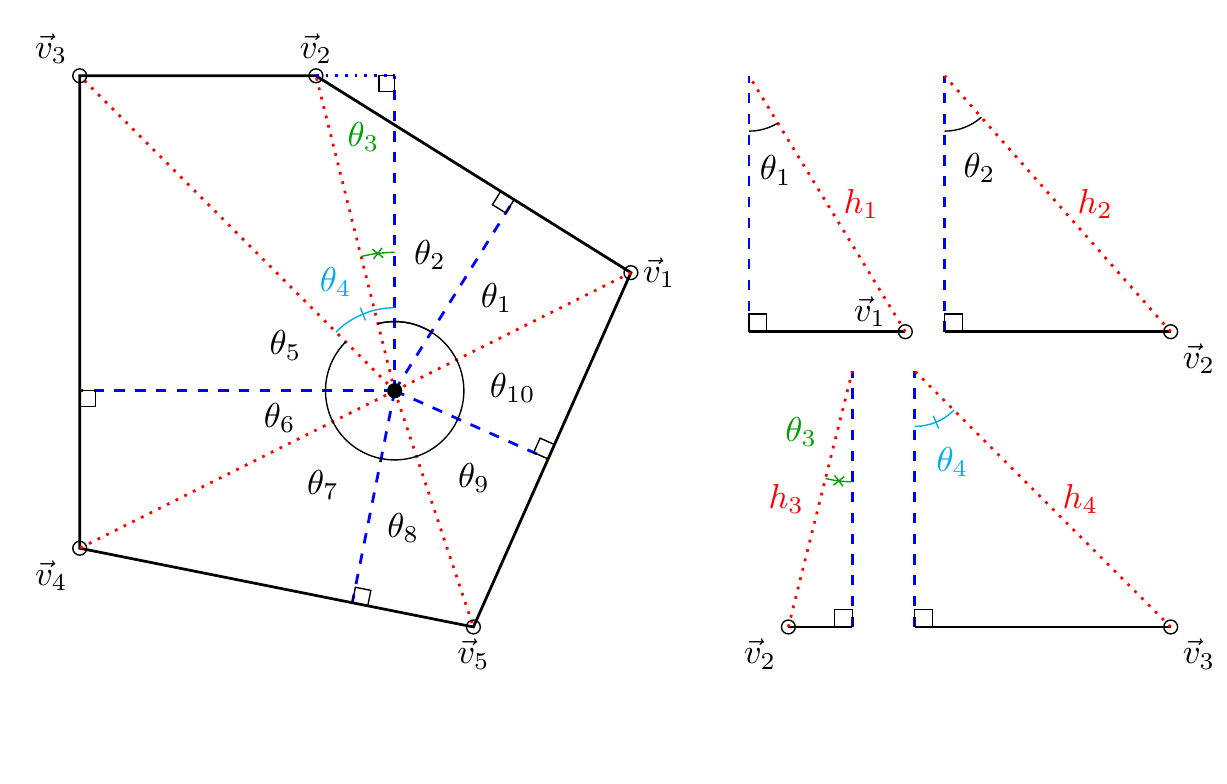}
    \vspace{-0.4in}
    \caption{(left) Right triangle subdivision of a five-sided Voronoi region with perpendiculars (blue dashed) and hypotenuse (red dotted). The angle $\theta_3$ is negative and overlaps with $\theta_2$ and $\theta_4$ so that the sum of all 10 angles in $2\pi.$ (right) Four of the right triangles formed from the subdivision of the Voronoi region. As generated the hypotenuses of the triangles satisfy $h_{2j}=h_{2j+1}$, however this and other geometric restrictions are relaxed in the admissible set $\cA(K).$}
    \label{f:polygon_triangle}
\end{figure}

The weaker formulation is obtained by replacing the collections of regions formed by the 
Voronoi tessellation with geometrically-uncoupled collections of right triangles. These are formed by subdividing each Voronoi region $\region{i}$ in a given tessellation into two right triangles by extending perpendiculars from the site $\site{i}$ to the edge of $\region{i},$ or to the extension of that edge that intersects the perpendicular. For a Voronoi region with $m_i$ edges, this generates $2m_i$ right triangles, each described by two parameters: the length $h$ of the perpendicular and the angle $\theta$ that is adjacent to the site $\site{i}.$ 
Collectively the subdivision generates                  
$$K=2\sum_{i=1}^N m_i$$ 
right triangles, parameterized by  $Y=(h_1,\theta_1, \ldots, h_K,\theta_K)\in\bR^{2K}$.  In the case that a  pair of right triangles are constructed by extending the edge of the Voronoi region beyond its two vertices, then the triangles overlap with their adjacent neighbors. To accommodate this overlap one of the pair is assigned a positive angle and the other a negative angle, with associated positive and negative contributions to area and average radius balance, see equations \eqref{e:fdef} and \eqref{e:c1def} and Figure\,\ref{f:polygon_triangle}.

\subsubsection{Positive Angle Case}
Each triangle makes a contribution to the average radius of the tessellation. Since the tessellation has $2N-2<2K$ degrees of freedom, there are many hidden constraints within these parameters that reflect the geometry of the tessellation. In the weak formulation these geometric constraints are dropped and we maximize an extension of the average radius over the collections of $K$ right triangles that satisfy only two constraints: that the total area of the triangles sums to $N$ and that the interior angles sum to $2\pi N$. We first restrict attention to the case of positive angles only, seeking to maximize the average radius the collections of $K$ triangles over the convex admissible set
 \beq \label{PO:def-Admiss}
 \cA(K):=\left\{Y\,\Bigl |  h_i\in[0,R], \theta_i\in\left[0,\pi/2\right],\, i=1,\dots, K \right\}.
\eeq
The upper bound $R$ on the hypotenuse of the triangles will be selected sufficiently large, depending only upon $N.$ 

For a non-degenerate Voronoi tessellation, Euler's equation with characteristic zero implies that the tessellation has $3N$ edges. Each tessellation edge abuts two right triangles in each tessellation region, so that a non-degenerate tessellation yields $K=4\cdot3N=12N$ right triangles via the subdivision process. This is the maximum value of $K$. Conversely the constraint on the sum of the angles and their individual range requires that $K\geq 4N$. This yields the range $K\in[4N,12N].$ The set $\cA(K)$ contains the set of right triangles $Y=Y(\bx)$ produced from subdividing the Voronoi tessellation generated by $\bx$, but is much larger. Generically the triangles corresponding to an arbitrary $Y\in\cA(K)$ cannot be reassembled into a Voronoi tessellation of $\Omega_{\alpha,N}.$

We generalize the constrained maximization problem to illuminate its structure. Given $K\in\bN_+$ geometric regions, each parameterized by $y_k\in\bR^\ell$ for $k=1,\ldots, K$. We form $Y=(y_1, \ldots, y_K)\in \bR^{\ell\times K}$ and introduce a compact set $\cA\subset \bR^{\ell\times K}$ that defines the collections of $K$ admissible regions. We seek to maximize a function $F:\cA\mapsto \bR$  subject to $\ell$ constraints $C(Y)=\cC$, where $C:\cA\mapsto \mathbb{R}^\ell$ and $\cC\in\bR^\ell$. More specifically we consider a function and constraints that treat each geometric region independently and additively, so that 
\beq\label{F-def}
F(Y)=\sum_{k=1}^Kf(y_k),
\eeq
and 
\beq\label{C-def}
C(Y)=\sum_{k=1}^K c(y_k),
\eeq
for smooth functions $f:\bR^\ell\mapsto\bR$ and $c:\bR^\ell\mapsto\bR^\ell.$

\begin{Lemma}
\label{l:diagMin}
Consider the function $F:\cA\mapsto \bR$ and the constraints $C: \cA\mapsto \bR^\ell$ defined on the admissible set $\cA \subset \bR^{\ell\times K}$ as given above. If the function $g$ introduced in \eqref{g-def} is well-defined and one-to-one on $\cA$, then the interior critical points of the problem
\beq \max\limits_{Y\in \cA} F(Y),
\eeq
subject to the $\ell$ constraints
\beq C(Y)=\cC\in\bR^\ell,
\eeq
occur on the diagonal of $A$. That is the  interior critical points are of the form $Y_*=(y_*, \ldots, y_*)$ for some $y_*\in\bR^\ell.$
\end{Lemma}
\begin{proof}
From the form of $F$ and $C$ in \eqref{F-def} and \eqref{C-def} the gradients $\nabla_Y$ of $F$ and $C$ are block diagonal with the k'th sub-block given in terms of $\nabla_y f\bigl|_{y_k}$ and $\nabla_y c\bigl|_{y_k}$.  For a critical point that is not on the boundary of $\cA,$ then each column vector of the minimizer $Y_*=(y_{*1}, \ldots, y_{*K})$ satisfies an identical Lagrange multiplier problem. Specifically,
$$ \nabla_y f(y_{*k}) = \nabla_y c(y_{*k}) \Lambda,$$
for each $k=1,\ldots K,$ where $\Lambda=(\lambda_1,\ldots, \lambda_\ell)$ are the $k$-independent Lagrange multipliers. By assumption the $\ell\times\ell$ matrix $\nabla_y c$ is invertible on $\cA$. We introduce the map $g:\bR^\ell\mapsto\bR^\ell,$
\beq\label{g-def} 
g(y):= [\nabla_y c(y)]^{-1}\nabla_y f(y).
\eeq
The columns $\{y_{*,k}\}_{k=1}^K$ of an interior critical point $Y_*$ solve $g(y_{*k})=\Lambda$ for $k=1,\ldots, K.$ By assumption $g$ is one-to-one, so if $Y_*$ is a critical point then each column is in the range of $g$, in particular $y_{*k}=y_*:=g^{-1}(\Lambda)$ is independent of $k$ and $Y_*=(y_*,\ldots, y_*)$.
\end{proof}

We apply this Lemma inductively to produce a sharp upper bound on the average radius of a tessellation.
A key point is that in the limit that $R\to\infty$, the boundary of $\cA(K)$ maps onto $\cA(K')$ for some $K'<K$. The proof is presented for the case of positive angles only, Voronoi tessellations including negative angles are eliminated as maximizers of the average radius in subsection\,\ref{s:neg}.
\begin{Theorem}
\label{t:PosVEE}
For any $(N,\alpha)$ and any collection of sites, $\bx$, the average radius of the associated Voronoi tessellation is bounded from above by the average radius of the unit hexagon,
\beq 
\br(\bx)\leq\br_6.
\eeq 
In particular $\VEE(\bx)\geq 0,$ with equality achieved by the regular hexagonal tessellation when $\alpha\in\cH(N).$ 
\end{Theorem}
\begin{proof}

We fix $N$ and $\alpha$ and consider the set of all right triangle subdivisions of tessellations of $\Omega_{\alpha,N}$ by $\bx\in\Omega_{\alpha,N}^N.$ For the positive-angle case, each subdivision forms $K$ right-triangles for some $K\in[4N,12N]$.
To bound the average radius of these tessellations we introduce the convexified family of constrained problems, $\cP(K,\cC)$ for $\cC\in\bR^2.$ These problems seek to maximize
\beq\label{e:Feq}
F(Y):= \sum_{k=1}^K f(h_k,\theta_k),
\eeq
over $Y\in\cA(K)$, subject to the two constraints,
\beq
\label{e:Ci}
C_i(Y):= \sum_{k=1}^K c_i(h_k,\theta_k) = \cC_i,
%|\Omega_{N,\alpha}|=N,
\eeq
for $i=1,2.$ We make the choices
\begin{align}
    f(h,\theta) &= \frac{1}{2\pi} h \cos\theta \arctanh(\sin \theta)=:h \tilde{f}(\theta),
    \label{e:fdef}\\
    %\sin\theta \arctanh(\cos \theta)=h \tilde{f}(\theta), \\
    c_1(h,\theta) &= \frac14 h^2 \sin(2\theta)=:h^2\tilde{c}_1(\theta), \label{e:c1def} \\
    c_2(h,\theta) &= \theta.
    \label{e:c2def}
\end{align}
so that if $Y= Y(\bx)$ arises from the partitioning of the Voronoi tessellation of $\bx$ into right triangles then $\br(\bx)=\frac{1}{N} F(Y(\bx)).$ The natural choices for $\cC$ is $\cC_1=|\Omega_{\alpha,N}|=N$ and $\cC_2=2\pi N$ so that constraint on $C_1$ enforces total triangle area $N$ and the constraint on $C_2$ requires that the site angles sum to $2\pi N$. These are the values of $\cC$ when $Y=Y(\bx)$ arises as a subdivision of an $N$-site tessellation of $\Omega_{\alpha, N}$. Other choices arise from considering maxima that can occur on $\partial\cA(K).$

To apply Lemma\,\ref{l:diagMin} for the choices \eqref{e:fdef}-\eqref{e:c2def} we establish that $g$ defined in \eqref{g-def} is one-to-one on $\cA(K)$. We determine that
\beq \label{e:nabla-c} \nabla_y c = 
\begin{pmatrix} 
 2h \tilde{c}_1(\theta)  & 0 \cr
h^2 \tilde{c}_1'(\theta) & 1 
\end{pmatrix},
\eeq
where a prime denotes the derivative of a function of one variable with respect to that variable. The function $g$ takes the form
\beq\label{e:geq} g := 
\begin{pmatrix} \frac{1}{2 h \tilde{c}_1} & 0 \cr
-\frac{h \tilde{c}_1'}{2 \tilde{c}_1} & 1 
\end{pmatrix} 
\begin{pmatrix} \tilde{f} \cr
 h\tilde{f}' 
\end{pmatrix} =
\begin{pmatrix} 
\frac{\tilde{f}}{2 h \tilde{c}_1} \cr
\frac{h}{2} \frac{\tilde{f}'\tilde{c}_1-\tilde{f}\tilde{c}_1'/2}
{\tilde{c}_1} \end{pmatrix} .
%= \begin{pmatrix} \frac{\tilde{f}}{2 h \tilde{c}_1} \cr
%\frac{h\sqrt{|\tilde{c}_1|}}{2} \left(\frac{\tilde{f}}{\sqrt{|\tilde{c}_1|}}\right)'
%\end{pmatrix}.
\eeq
The product  $g_1g_2$ depends only upon $\theta$, so that for Lagrange multipliers $\Lambda=(\lambda_1,\lambda_2)$ the critical point angle $\theta_*$ satisfies 
\beq
\lambda_1\lambda_2=g_1g_2= \frac{1}{8}\left(\frac{\tilde{f}^2}{\tilde{c}_1}\right)'. %=:\tilde{g}(\theta).
\eeq
%\sign(\theta) \frac{\tilde{f}}{4\sqrt{|\tilde{c}_1|}}\left(\frac{\tilde{f}}{\sqrt{|\tilde{c}_1|}}\right)'\!\!(\theta),
The function $\tilde{g}:=\tilde{f}^2/\tilde{c}_1$ is concave on $(0,\pi/2)$ and $\tilde{g}''$ is strictly negative on this set. This establishes that $\tilde{g}'$ is 1-1 on $(0,\pi/2)$ and from the simple form of the $h$ dependence in $g$ it easily follows that $g$ is 1-1 on $[0,R]\times[0,\pi/2]\mapsto \bR^2$. From Lemma\,\ref{l:diagMin} the critical points of $F$ that are interior to $\cA(K)$ occur on the diagonal of $\cA(K)$, and are of the form $Y_*(K,\cC)=(h_*,\theta_*, \ldots, h_*,\theta_*)\in \bR^{2K}$ for some values $(h_*,\theta_*)\in [0,R]\times (0,\pi/2)$. 
The angle constraint $C_2(Y)=\cC_2$ imposes the condition 
\beq \label{e:theta*}
\theta_*=\frac{\cC_2}{K},
\eeq
and the area constraint $C_1(Y)=\cC_1$ forces $$h_*=2\sqrt{\frac{\cC_1}{K\sin(2\theta_*)}}.
%=2\sqrt{\frac{N}{K\sin(4\pi N/K)}}.
$$ 
In particular, defining the  critical values $F_*(K,\cC):= F(Y_*(K,\cC))$ we have the suggestive form
\beq\label{e:F*}
%\begin{aligned}
F_* = \sqrt{\cC_1\cC_2} \frac{\cos(\frac{\cC_2}{K})\arctanh(\sin(\frac{\cC_2}{K}))}
{\pi\sqrt{\frac{\cC_2}{K}\sin(\frac{2\cC_2}{K})}}.
%= \frac{\cos(\frac{2\pi N}{K})\arctanh(\sin(\frac{2\pi N}{K}))} {\pi\sqrt{\frac{N}{K}\sin(\frac{4\pi N}{K})}}.
\eeq
For fixed $\cC,$ the function $F_*$ is monotonically increasing in $K$, and achieves its maximum value at $K=12N.$ For the natural choice of constraint values $\cC=(N,2\pi N)$ we have
\beq 
\label{PO-hexmax}
\frac{1}{N} F_*(K,N,2\pi N)\leq \frac{1}{N} F_*(K,N,2\pi N)= \sqrt{12} \frac{\cos(\frac{\pi}{6})\arctanh(\sin(\frac{\pi}{6}))}
{\pi\sqrt{\sin(\frac{\pi}{3})}}=\br_6.
\eeq
This establishes that the hexagonal lattice, for which $K=12N$, provides the largest interior critical point of the constrained problem for any value of $K\in[4N,12N].$

To establish that the constrained maximum does not occur on the boundary of $\cA(K)$ we recast the search of the maximum over the boundary as a problem $\cP$ with modified values for $K$ and $\cC.$ Indeed a point $Y$ lies on $\partial\cA(K)$ if and only if there are one or more $i=1, \ldots K$ for which either $\theta_i\in\{0,\pi/2\}$ or $h_i\in\{0,R\}$. If $n_1$ of the angles equal $0$ then the associated triangles makes no contribution to $F$, have no area, and make no contribution to the total angle constraint. The interior critical points of $F$ over this are equivalent to the those of $\cP(K-n_1,\cC).$  
If $n_2$ angles take the value $\pi/2$ then the equivalent problem is $\cP(K-n_2,\cC_1,\cC_2-\alpha_2n_2\pi/2)$ for some $\alpha_2\in[0,1].$
If $n_3$ hypotenuses are $0$ then $Y$ is a competitor for $\cP(K-n_3,\cC_1,\cC_2-\alpha_3 n_3 \pi/2)$.  If a triangle has hypotenuse $h_i=R\gg1$, then the area $A_i\leq N$, and angle $\theta_i=O(N/R^2)$ so that $f(y_i)=O(N/R).$ 
For fixed $N$, up to terms that are small in $1/R$, the maximum of $F$ over this set is equivalent to $\cP(K-n_4,\cC_1-n_4\alpha_4,\cC_2)$. Combining all these boundary cases we are brought to the problem $\cP(K',\cC')$ with $K'=K-n_1-n_2-n_3-n_4$, and $\cC'=(\cC_1',\cC_2')$ where $\cC_1'=\cC_1-\alpha_3n_3,$ and $\cC_2'=\cC_2-\frac{\pi}{2}(\alpha_2n_2+\alpha_3n_3)-\alpha_4 n_4.$ 
Since $F_*$ is increasing in $K$ for fixed $\cC_2$ and is clearly increasing in $\cC_1',$ the interior critical points of these boundary problems satisfy the bounds
$$F_*(K',\cC_1',\cC_2')<F_*(K,\cC_1,\cC_2')\leq F(12N,\cC_1,\cC_2'),$$
 Since $\cC_2'\leq 2\pi N$ the ratio $\cC_2'/12N \leq \frac{\pi}{6}.$ On the range $\cC_2'\in (0,\pi/6)$ the function $F_*$ is strictly increasing in $\cC_2$ for fixed $K,$ hence
 $$ F_*(K',\cC')\leq F_*(12N,N,2\pi N).$$ 
Taking $R$ sufficiently large that 
$$F_*(12N,N,2\pi N)-F_*(12N-1,N,2\pi N)\gg O(N/R),$$ 
we deduce that the maximum of $F$ over $\cA(K)$ for any $K\in [4N,12N]$ occurs as the interior critical point of $\cA(12N)$. In particular for all collections of $N$ sites
\beq
\br(\bx)\leq \max\limits_{\bx}\frac{1}{N}F(Y(\bx)) \leq
\max\limits_{K\leq 12N} \max\limits_{Y\in A(K)}\frac{1}{N}F(Y) = \frac{1}{N} F_*(12N,N,2\pi N)= \br_6.
\eeq
This is equivalent to the statement that $\VEE(\bx)\geq 0.$ 
\end{proof}

\subsubsection{Negative Angles}
\label{s:neg}
If negative angles arise in the right triangle subdivision of a Voronoi tessellation then the naive extension of the approach outlined above fails when large values of the hypotenuse are associated to negative angles. On the other hand, the inclusion of negative angles as  wholly independent parameter leads to a loss of convexity. To prevent this the right triangles are kept as pairs, grouping each negative angle triangle with a positive angle triangle with a shared side that denotes the height of the perpendicular to their common Voronoi edge.  Each pair of triangles can be parameterized by 3 variables: the height $d \in [0,\infty]$, and the two angles $\theta_+, \theta_- \in [-\pi/2, \pi/2]$, with $\theta_+ + \theta_- > 0$.  Crucially, the Lemma below shows that for a single triangle pair their contribution to the average radius is maximized when both angles are positive.

\begin{Lemma}
\label{l:negative}
A single right triangle pair with area $A$ and angle sum $\phi \in [0, \pi]$ maximizes its average radius when $\theta_+ = \theta_-=\phi/2$. This maximum average radius is given by
\beq
\label{e:rmax}
    \bar{r}_{\rm max}=\frac{1}{\pi}\sqrt{A}\frac{\arctanh(\sin(\phi/2))}{\sqrt{\tan(\phi/2)}}.
\eeq
\end{Lemma}
\begin{proof}
A right triangle pair with fixed area $A$, and angle sum $\phi \in [0,\pi]$, is parameterized by height $d \in [0, \infty)$ and angle $\theta \in [\phi-\pi/2, \pi/2]$. The average radius is given by $f$ under constraint $c_1=A$,
\begin{align}
    f(d, \theta) &= \frac{1}{2\pi}d(\arctanh(\sin\theta) + \arctanh(\sin(\phi-\theta))) = d(\tf(\theta)+\tf(\phi-\theta))\\
    c_1(d, \theta) &= \frac{1}{2}d^2(\tan(\theta) + \tan((\phi-\theta))) = d^2(\tc_1(\theta) + \tc_1(\phi-\theta)).
\end{align}
Here we have introduced $\tf(\theta)=\arctanh(\sin\theta),$ and $\tc_1(\theta)=\tan(\theta). $ The internal critical point equations of the system subject to the area constraint are expressed in terms of the Lagrange multiplier $\lambda_1$,
\beq
\begin{pmatrix}
    \tf(\theta) + \tf(\phi-\theta)\\
    d\tf'(\theta)\\
    -d\tf'(\phi-\theta)
\end{pmatrix}
= \lambda_1
\begin{pmatrix}
    2d(\tc_1(\theta) + \tc_1(\phi-\theta))\\
    d^2\tc'_1(\theta)\\
    -d^2\tc'_1(\phi-\theta)
\end{pmatrix}.
\eeq
Eliminating $\lambda_1$ from the second and third equations yields the relation
\beq
    \frac{\tf'(\theta)}{\tc'_1(\theta)} = \frac{\tf'(\phi-\theta)}{\tc'_1(\phi-\theta)}.
\eeq
The function $\tf'/\tc'_1$ is even and monotonic on $\theta>0$ and $\theta<0$. We deduce that $\theta = \pm (\phi-\theta)$ and see that this system only has a solution on the ``plus'' branch, for which $\theta = \phi/2$. From the area condition we see that $d\to0$ and $f\to 0$ as $\theta\to \pi/2$ or $\theta\to\phi-\pi/2.$ Moreover $d$ must remain bounded due to the area constraint, so the maximum is attained at the internal critical point. At this point $\theta=\phi/2$ for which
\beq
    d=\sqrt{\frac{A}{\tan(\phi/2)}},
\eeq
and maximum of $f$ is given by \eqref{e:rmax}.
\end{proof}

Cutting a tessellation into $K$ pairs of right triangle, we can bound the average radius from above by maximizing
\beq\label{e:Feq-negative}
F(Y):= \sum_{k=1}^K f(d_k,\theta_{k,+},\theta_{k,-}),
\eeq
over $Y\in\cA(K)$, subject to the area and angle constraints,
\beq
\label{e:Ci-negative}
C_i(Y):= \sum_{k=1}^K c_i(h_k,\theta_{k,+},\theta_{k,-}) = \cC_i,
%|\Omega_{N,\alpha}|=N,
\eeq
for some $\cC_i\in\bR$ for $i=1,2.$ Here the functions $f$ and $c_i$ are as in Lemma\,\ref{l:negative} while 
\beq
    c_2(d,\theta_+, \theta_-):= \theta_+ + \theta_-.
    \label{e:c2def_neg}
\eeq
From Lemma\,\ref{l:negative} for each pair of triangles the maximum average radius contribution occurs when $\theta_-=\theta_+\geq0$. Consequently negative angles are incompatible with local minima of the energy and the maximization over the triangle pairs can be embedded within the larger positive angle maximization problem.  We deduce that $\bar{r}(\bx)\leq \bar{r}_6$ for all tessellations. This addresses the negative angle case and completes the proof of Theorem\,\ref{t:PosVEE}.

\section{Defects and Frustration in Disordered Equilibria}

We investigate the role of site number $N$ and aspect ratio $\alpha$ in determining the probability that a random initial configuration of sites converges to an ordered or a disordered equilibrium  through the gradient flow of the Hookean Voronoi energy. \tcb{While the existence of ordered states and the ground state status of the regular hexagonal tessellation established in Section 3 suggest that ordered structures should play a dominant role in long time behavior of the gradient flow of the Hookean-Voronoi energy, numerical simulations show that in fact the system is frustrated, and suggest that in the large $N$ limit the system tends to a non-zero per-site average frustration.}

\begin{figure}[H]
    \centering
    \begin{subfigure}{.3\textwidth}
        \centering
        \includegraphics{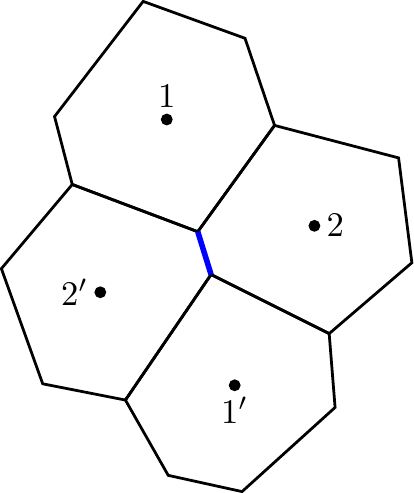}
    \end{subfigure}
    \begin{subfigure}{.3\textwidth}
        \centering
        \includegraphics{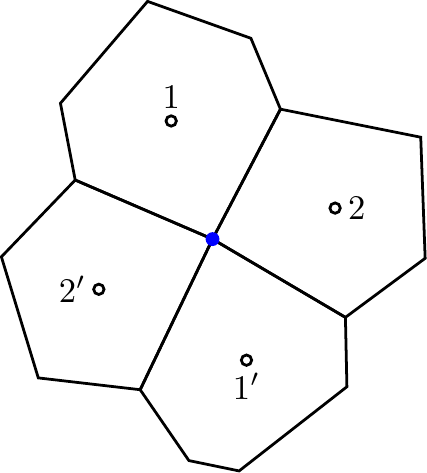}
    \end{subfigure}
    \begin{subfigure}{.3\textwidth}
        \centering
        \includegraphics{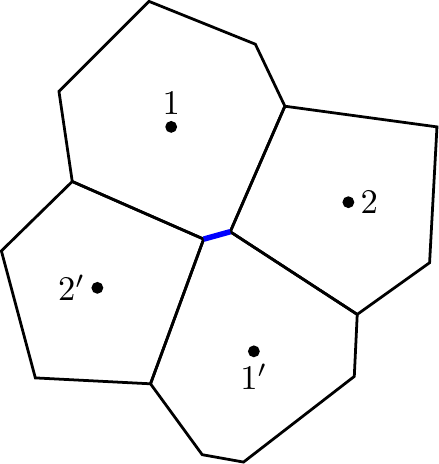}
    \end{subfigure}
    \caption{A vertex collision that generates two pair of 5- and 7-sided defect regions from a tessellation of 6-sided regions. The sites of 5-, 6-, and 7- sided regions are indicated by pentagons, circles, and stars respectively, while sites with 6 edges and a degenerate vertex are indicated by open circles. The energy $\VEE(\bx)$ is continuous, piece-wise smooth, but not differential at a vertex collision.
    }
    \label{f:pair_collision}
\end{figure}

The motion of the sites under the gradient flow can induce changes in the edge count $m_i$ of a Voronoi region via the formation of a degenerate vertex with more than three edges. This generically arises through a ``vertex collision'' in which an edge shrinks to zero length, bringing two vertices together, eliminating one vertex and one edge from the graph. \tcb{The energy $\mrE(\bx)$ is continuous, but not differentiable at these non-generic degenerate configurations.} Given a tessellation that is locally composed of 6-sided regions, the generic vertex collision arises when deformation of the site vector $\bx$ creates two pair of 5- and 7-sided regions, as presented in Figure\,\ref{f:pair_collision}. Under deformation an edge separating two six-sided sites, labeled $\site{1}$ and $\site{1'}$, shortens to bring its two vertices together, forming a single degenerate vertex with four edges that is the corner of sites $\site{1}, \site{1'}, \site{2}, \site{2'}$. 
 %with the $1-1'$ and $2-2'$ regions sharing one vertex and no edge. 
 Continuing with the deformation of the sites, the degenerate vertex may split into two vertices by forming a new edge that separates the regions of $\site{2}$ and $\site{2'}.$ This process changes the four six-sided regions into two five-sided and and two seven-sided.
 
Vertex collisions can occur among regions with arbitrary numbers of sides, generically adding one side to each of the $\site{1}-\site{1'}$ regions and subtracting one vertex from each of the $\site{2}-\site{2'}$ regions. 

 \begin{Def}
 We say that  a site and its region are a defect if each of the region's vertices are non-degenerate and the region does not have 6 sides.
 \end{Def}  
 
In computations we take $r_*=4$. Although this value is immaterial to the outcomes of the simulations, this choice scales the relaxation time of the system and the values of $\VEE$ of the intrinsic states.

\subsection{Computational equilibria}
The gradient flow is implemented through a second order predictor-corrector scheme. The SciPy library is employed to compute the Voronoi diagram from a set of points. The gradient is computed via \eqref{e:HV-gradflow2}  using a second-order midpoint method and an embedded first-order (Euler's) method to obtain an error estimate for the adaptive step size, with initial step size $\delta_{\mathrm{step}}$. Writing the gradient flow in the form $y'(t) = f(t,y)$, we have increments
\[
    k_1 = f(t_n, y_n), \quad k_2 = f(t_n+\delta_{\mathrm{step}}, y_n+\delta_{\mathrm{step}}k_1).
\]
The second order update $y_{n+1}$ and error estimate $\epsilon$ take the form
\[
    y_{n+1} = y_n + \frac{\delta_{\mathrm{step}}}{2}(k_1 + k_2) \quad \epsilon = \frac{\delta_{\mathrm{step}}}{2}(k_1 + k_2) - k_1.
\]
The step size is updated using the rule 
\[
    \delta_{\mathrm{step}} \rightarrow \delta_{\mathrm{step}} \sqrt{\frac{10^s}{\|\epsilon\|_{2}}}, \quad s = \min \{ -3, -2 + \log_{10}\|f(t_n, y_n) \|_{2} \}
\]

%\begin{Remark}
%Present gradient flow. Use coercivity of Hessian and small residual to establish that computational equilibria are legitimate. Include perturbation and relaxation back to equilibria.
%\end{Remark}

We establish a stopping criteria that determines if a simulation of the 
gradient flow is sufficiently close to a stable equilibrium. For each putative computational equilibrium  $\eqx$ we compute the hessian
$$ \mrH(\eqx):= D_\bx^2(\eqx),$$
and determine its spectrum. We remove the two dimensional kernel, spanned by the translational vectors $\{\be_1,  \be_2\}$, and determine the minimum eigenvalue on the space perpendicular to the kernel. When this quantity is positive, it is labeled $\nu(\eqx),$ motivated by the coercivity defined in \eqref{e:coercivity}. The stopping condition is that
\beq\label{e:eq-cond}
\frac{\|D_\bx(\eqx)\|_{2}}{\nu(\eqx)} \leq \delta_{\rm eq},
\eeq
for $\delta_{\rm eq}\ll1$ chosen suitably small.
The motivation is that for $\bx$ near $\eqx$ we have the expansion
$$ D_\bx (\bx) = D_\bx(\eqx) + H(\eqx)(\bx-\eqx) + R,$$
where the remainder $R\sim \|\bx-\eqx\|_2^2.$ If $\bx$ is an exact equilibrium then
\beq\label{e:eq-exp}
\bx= \eqx - H^{-1}\left(D_\bx(\eqx)+R\right)=:F(\bx),
\eeq
where we may adjust $\bx$ so that $\bx-\eqx$ is orthogonal to the translational modes. Then for some $c>0$ depending upon a bound of $D_{\site{}}^3$ in a neighborhood of $\eqx$, we have the estimate
$$ \|H^{-1}\left(D_\bx(\eqx)+R\right)\|_{2} \leq \delta_{\rm eq} + \frac{c}{\nu}\|\bx-\eqx\|_2^2.$$
Returning this result to \eqref{e:eq-exp} we have the bound
$$\|\bx-\eqx\|_2\leq \delta_{\rm eq} +\frac{c}{\nu}\|\bx-\eqx\|_2^2.$$
This shows that any equilibrium of the system that lies inside the ball of radius $\alpha/c$ centered at $\eqx$ in fact lies within the smaller ball of radius $\delta_{\rm eq}.$ Rigorous statements on this problem can be made, see Theorem 2.2 of \cite{Day07}, including existence of exact equilibrium.  We treat this issue informally in this work, taking $\delta_{\rm eq}=10^{-5}.$ We say that a site-vector $\bx$ is a \emph{computational equilibrium}, and denote it by $\eqx$, if it satisfies this stopping condition for the gradient flow.

\begin{figure}[H]
    \vspace{-0.2in}
    \begin{center}
    \includegraphics[width=0.8\textwidth]{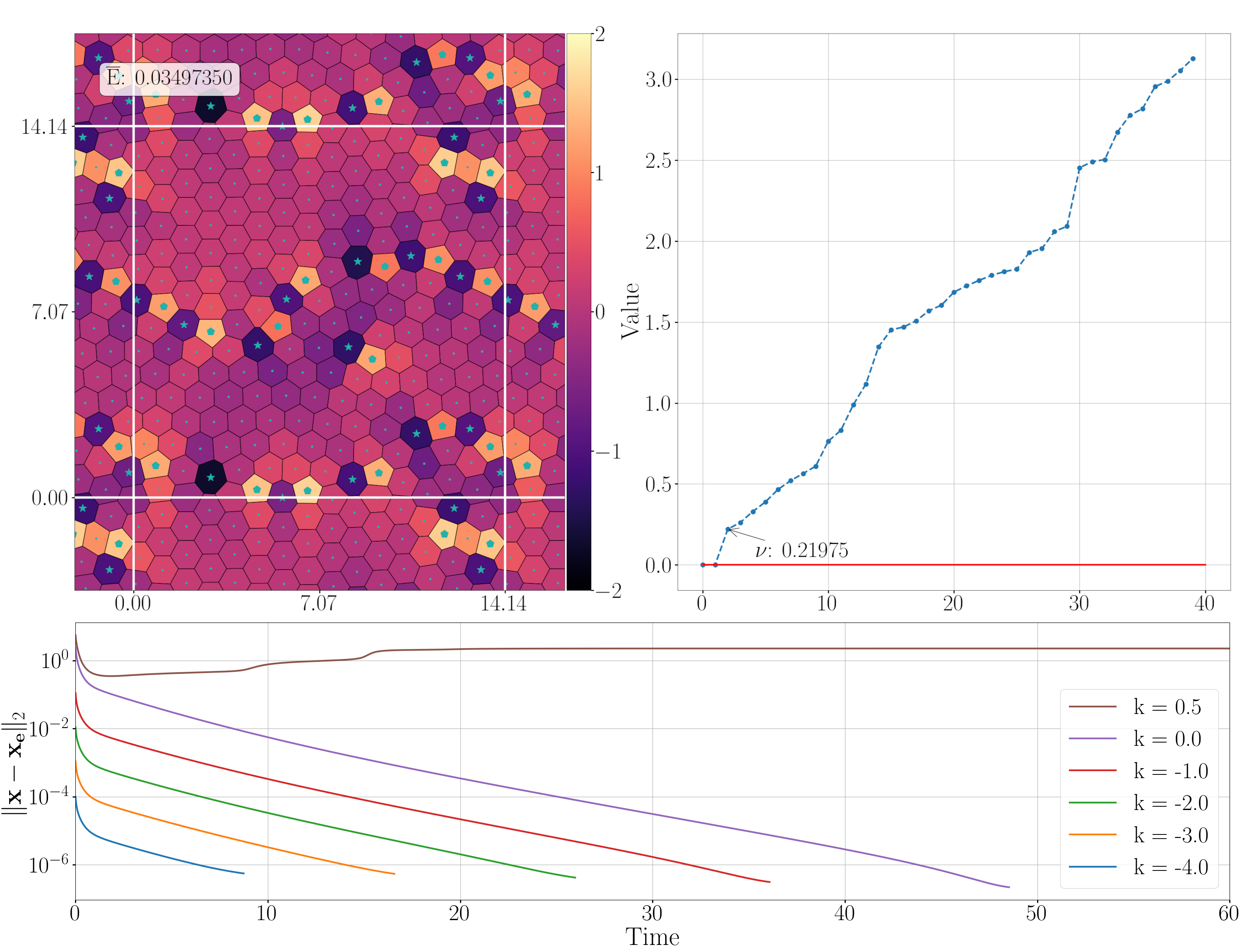}
    \end{center}
    \vspace{-0.2in}
    \caption{(top-left) The tessellation generated from an equilibrium $\eqx$ with $N=200$ sites in the $\alpha=1$ domain containing 16 pairs of $5-7$ defects. The regions with 5, 6, and 7 sides have sites indicated with pentagon, dot, or star, respectively. Solid white lines depict periodic domain $\Omega_{\alpha,N}$ with adjacent domains shown for continuity.  (top-right) The 40 smallest eigenvalues of the hessian, including the two dimensional kernel, and the coercivity $\nu=0.21975$. (bottom) The evolution of $\|\bx(t)-\eqx\|_2$ verses $\VEE(\bx(t))$ for six simulations of the gradient flow starting from initial sites $\bx(0)$ that are random perturbations of $\eqx$ with $\|\bx(0)-\eqx\|_2=10^k$ for $k=-4, -3, -2, -1, 0,$ and $0.5.$ }
    \label{f:Equilibium}
\end{figure}

As an example, we consider the computational equilibrium $\eqx$ corresponding to $N=200$ and $\alpha=1$ which has 16 \emph{pairs} of 5-7 defects. 
The associated Voronoi tessellation is presented in Figure\,\ref{f:Equilibium} (top-left). The equilibrium has $\VEE(\eqx)=3.497\times10^{-2}$, and its constituent Voronoi regions are colored according to their $\VEE$, which range from $1.65$ (light) achieved at a 5-sided defect to $-1.68$ (dark) achieved at a 7-sided defect. This range of values is typical of a disordered equilibria. The spectrum of the Hessian of $\EHV$ at $\eqx$ is computed numerically, and the smallest 40 eigenvalues are shown in Figure\,\ref{f:Equilibium} (top-right). The two zero eigenvalues correspond to the kernel of the Hessian and are spanned by the translational modes $\{\be_1,\be_2\}.$ The coercivity $\nu(\eqx)=0.21975$ is defined by the smallest non-zero eigenvalue. The existence of an exact equilibrium within $O(\delta_{\rm eq})$ of $\eqx$ is supported by six simulations of the gradient flow corresponding to initial data that are random perturbations of $\eqx$ with $\|\bx(0)-\eqx\|_2$ taking the value $10^{-k}$ for $k=-4, -3, -2, -1, 0,$ and $0.5.$ As depicted in Figure\,\ref{f:Equilibium} (bottom), each of the orbits corresponding to the smallest 5 perturbations relax back to $\eqx$ with identical exponential rates. The orbit corresponding to the largest perturbation diverges to a different equilibrium.

\subsection{Moderate \texorpdfstring{$N$}{N}: Ordered verses Disordered Equilibrium}
\begin{figure}[H]
    \begin{center}
    \begin{tabular}{ccc}
    \includegraphics[width=.309\textwidth]{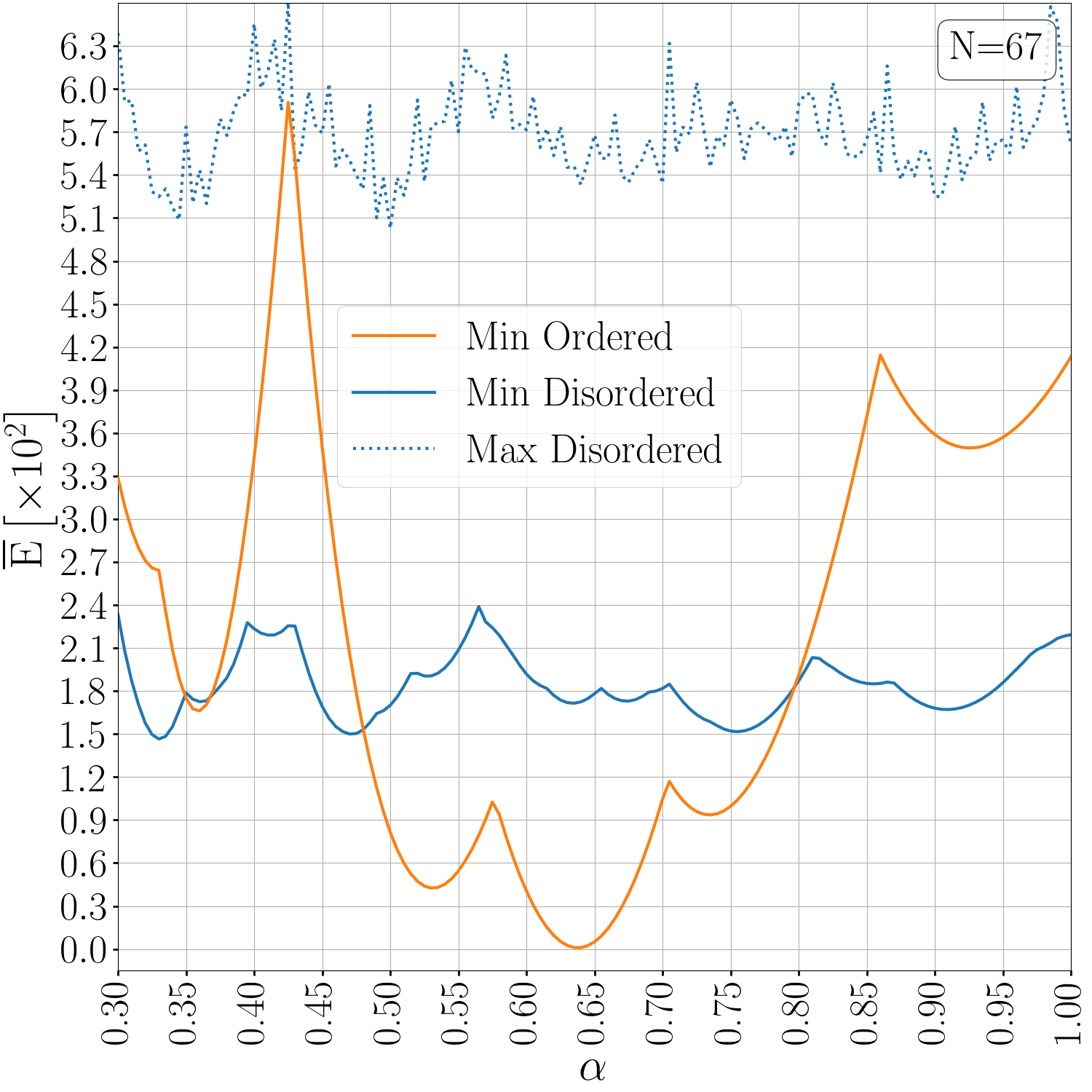} & 
    \includegraphics[width=.309\textwidth]{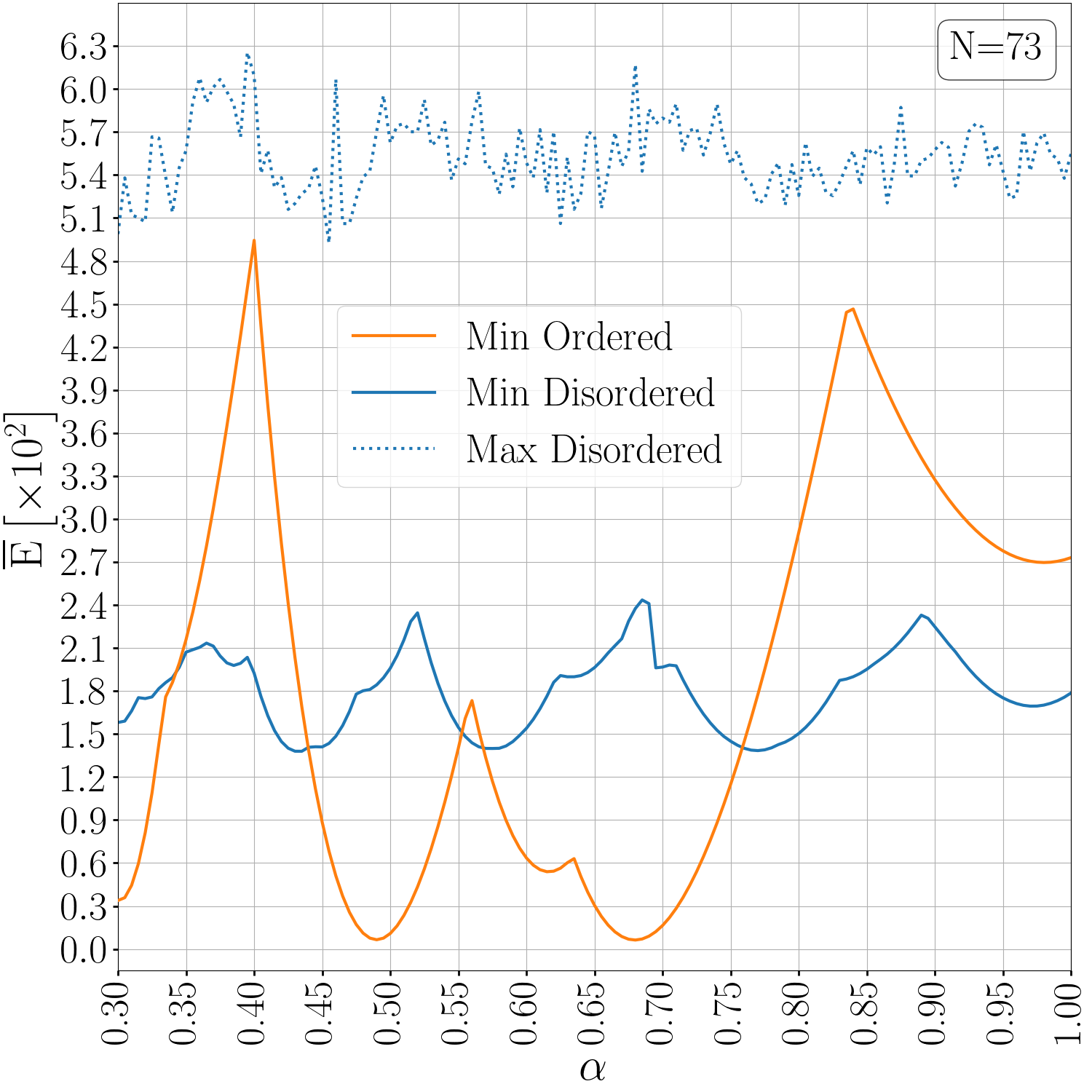} & 
    \includegraphics[width=.309\textwidth]{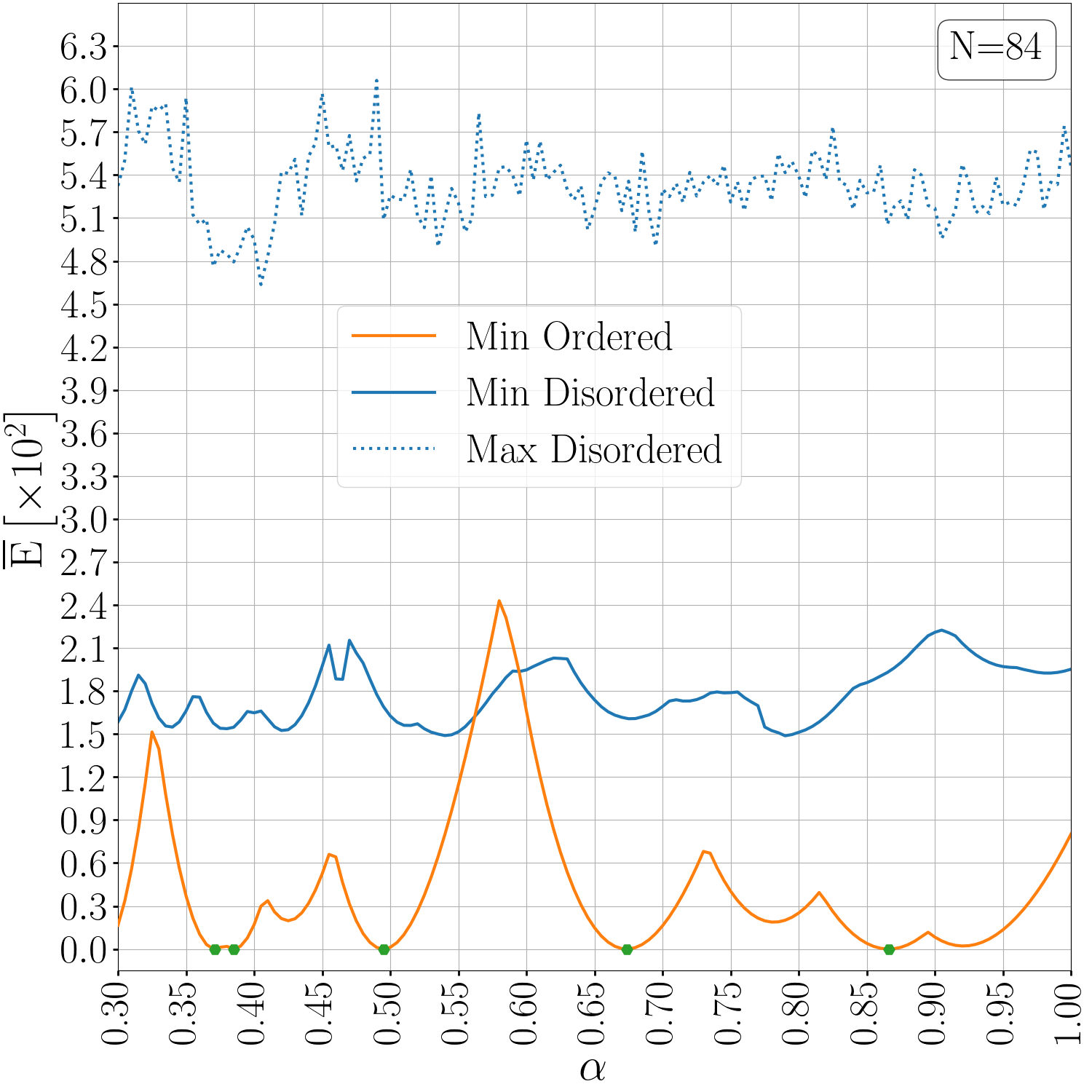}
    \end{tabular}
    \end{center}
    \vspace{-0.3in}
    \caption{The minimum volumetric excess energy for ordered (orange) and disordered (blue-solid) equilibria, and maximum $\VEE$ for stable disordered (blue-dotted) computational equilibria from $\cB(N,\alpha, 5000)$ for $N=67, 73, 84$ (left to right). The green dots on the $N=84$ minimum ordered $\VEE$ curve indicate the set $\cH(84)$ of regular hexagonal tilings. %\textcolor{blue}{Should we use 4 diagrams here, so we can keep the width of the diagrams consistent or replace 73 with 67 to keep it 2 diagrams?} 
    %from Proposition\,\ref{p:Hex-tiling}.
    }
    \label{f:VEE}
\end{figure}

The Hookean-Voronoi energy possesses a large collection of stable equilibria with a broad distribution of $\VEE$ whose structure depends sensitively upon $N$ and $\alpha.$ To investigate this set and their basins of attraction under the gradient flow we generate bins $\cB=\cB(N,\alpha,S)$ composed of $S$ computational equilibria of the Hookean-Voronoi gradient flow generated by initial sites $\bx(0)\in\mathbb R^{2N}$ that are randomly uniformly distributed in $\Omega_{\alpha,N}^N$. We denote the average of a quantity $\mrF$ over the bin by $\langle \mrF\rangle_\cB$.  We consider six  moderate values  $N =\{61, 67, 73, 81, 84, 100\}$ and vary the aspect ratio in increments of $0.01$ for $\alpha\in[0.3,1]$, forming $6\times71$ bins each with $S=5000$ simulations. Each bin is partitioned into two disjoint sub-bins,  one of ordered equilibria, $\cB_o$ for which all average radii of the associated Voronoi tessellation are equal to within the fixed tolerance $\delta_{\rm ord}:=10^{-8}$. The complement is the disordered bin, $\cB_d$. While the value of $\delta_{\rm ord}$ is small, every tessellation identified as disordered has at least one pair of defects.
For the disordered tessellations the number of Voronoi regions that are not 6-sided are identified -- this is the defect number $\mrD(\eqx)$ of the  computational equilibrium $\eqx.$ To each ordered sub-bin $\cB_o(N,\alpha,S)$ all associated single-string equilibria are added.  For each $N$ and $\alpha$ the minimum $\VEE$ of $\eqx\in\cB_o(N,\alpha,S)$ and the minimum and maximum $\VEE$ over $\cB_d$ are identified. These quantities, called the minimum ordered and minimum and maximum disordered $\VEE$ respectively, are presented in Figure\, \ref{f:VEE} as graphs over $\alpha\in[0.3,1]$ for $N= 67$, $N=73$, both primes, and $N=84=(2^2)\cdot(3\cdot7)$.  

The minimum and maximum disordered $\VEE$ are relatively constant with respect to $\alpha$ while the minimum ordered $\VEE$ is much more sensitive. For the six values of $N$ considered the maximum disordered $\VEE$ ranges between $0.048-0.063$ and the minimum disordered $\VEE$ is even flatter, residing in the range $0.015-0.024$. This lower value insensitive to the bin size, unchanged for $S$ between $500$ and $5000$. The maximum disordered $\VEE$ increases by about 5\% as the bin size $S$ increases from $500$ to $5000$ and stable disordered states with high $\VEE$ and small basins of attraction are realized.
For the six values of $N$ the minimum ordered $\VEE$ ranges from $0-0.0597$ with the highest value achieved at $\alpha=0.425$ and $N=67$. 
Figure\,\ref{f:Hex} (right) shows that the minimum ordered $\VEE$ is formed from families of single-string equilibria with cusp at values of $\alpha$ for which the role of ordered minimizer is exchanged. The range of the ordered minimum $\VEE$ shown in Figure\,\ref{f:VEE} is dramatic for the prime values $N=67$ with a sharp peak at $\alpha=0.425$ and $N=73$ with peaks at $\alpha=0.4$ and $\alpha=0.65$. Conversely, the more factorable value $N=84=2^2\cdot3\cdot7$ has a minimum ordered $\VEE$ with a more modest range with 5 values of $\alpha$ which support a regular hexagonal tiling with $\VEE=0$ for $\alpha\in(0.3,1]\cap \cH(84)=
\Bigl\{
%\frac{{1}{126}, \frac{1}{42}, \frac{2}{63}, \frac{1}{ 18}, \frac{1}{14}, \frac{2}{21}, \frac{1}{ 6}, 
\frac{3\sqrt{3}}{14}, \frac{2\sqrt{3}}{9}, \frac{2\sqrt{3}}{7}, \frac{7\sqrt{3}}{ 18}, \frac{\sqrt{3}}{2}\Bigr\},$ corresponding to regular hexagonal tessellations.  The minimum $\VEE$ remains low yielding only a single, minor peak of $0.024$ at $\alpha=0.58$. %\textcolor{red}{Figure\,\ref{f:VEE} seems to reprise Leo Tolstoy's opening sentence in the novel Anna Karenina -- disordered states are disordered in the same way, but ordered states are each ordered in their own way. However, much structure lurks within the disordered states as well.}
\begin{figure}[H]
    \begin{center}
    \begin{tabular}{ccc}
    \includegraphics[width=.4\textwidth]{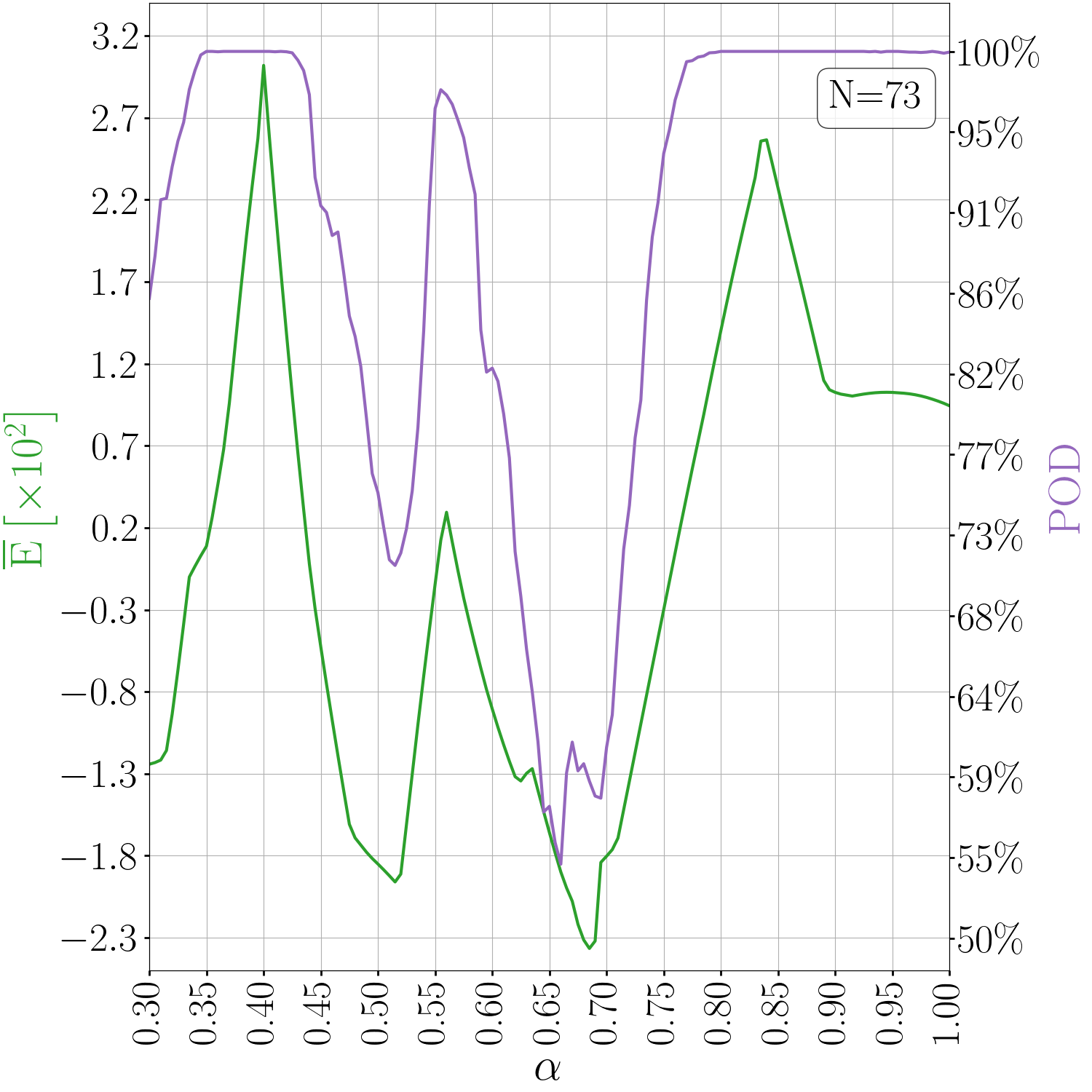} & 
    \includegraphics[width=.4\textwidth]{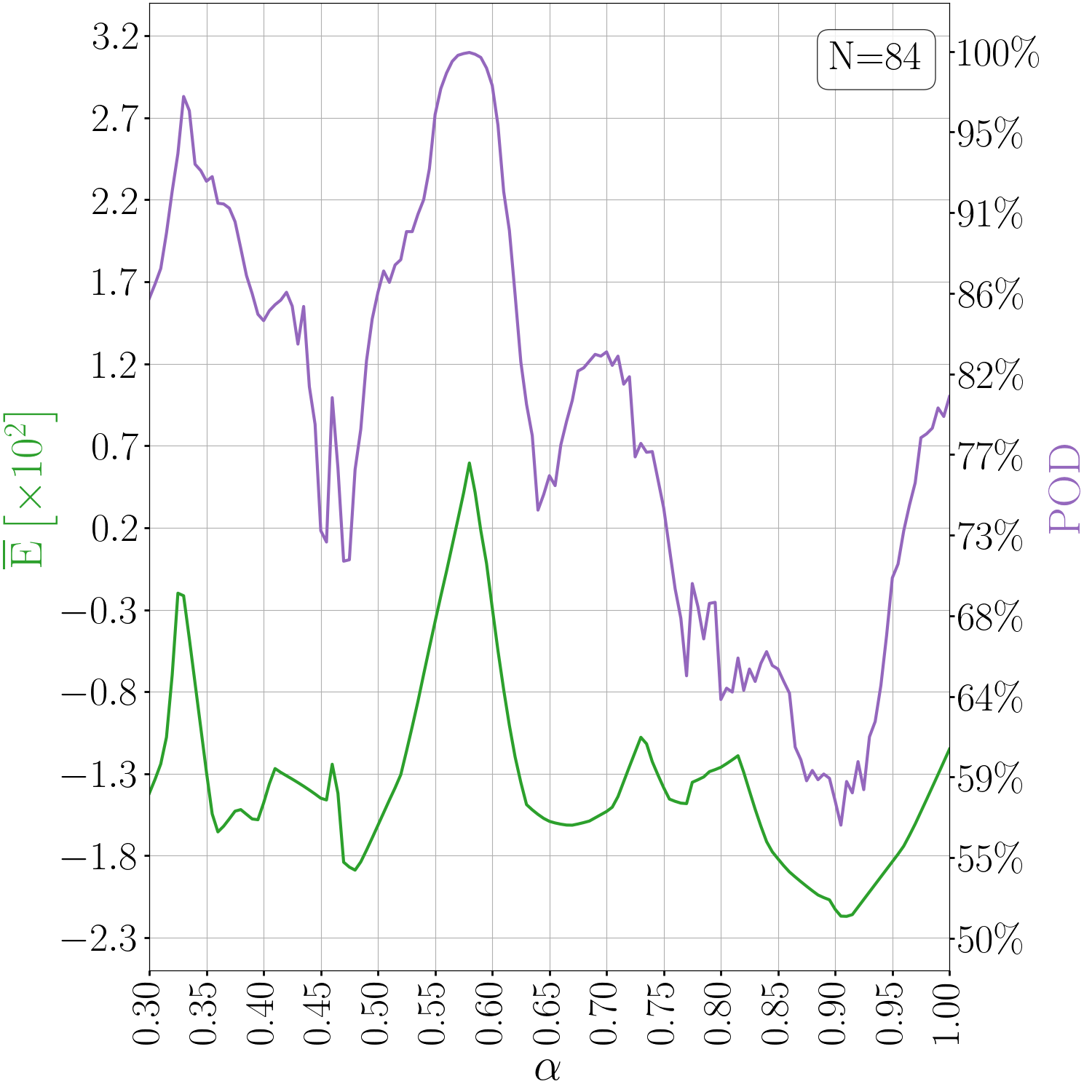} &
    \end{tabular}
    \includegraphics[width=.6\textwidth]{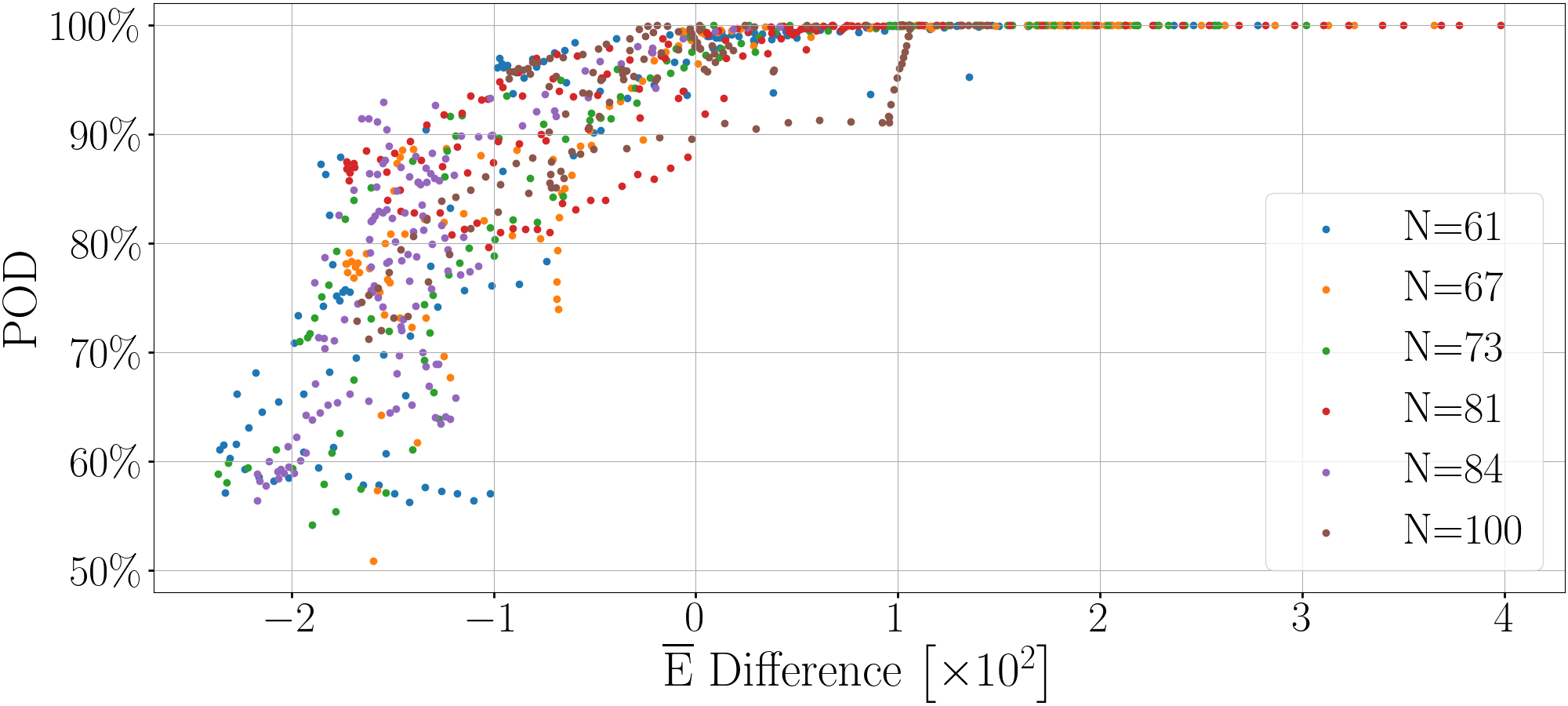}
    \end{center}
    \vspace{-0.25in}
    \caption{(top) The probability of disorder (purple curve right-axis) and difference between minimum $\VEE$ for ordered and disordered states (green curve left-axis) for the cases $N=73$ (left) and $N=84$ (center) gathered from $\cB(N,\alpha,5000)$ for each value of $\alpha$. (bottom) Scatter plot of the difference between minimum ordered $\VEE$ and minimum disordered $\VEE$ and the probability of disorder from $\cB(N,\alpha,5000)$ over each value of $\alpha\in[0.3,1]$ and $N$ as listed.}
    \label{f:POD}
\end{figure}
For a given $N$ and $\alpha$, the probability of disorder (POD) denotes the probability an initial data with randomly distributed sites converges to an equilibrium which has a disordered tessellation. This was tabulated $N=73$ and $N=84$, showing that its value correlates strongly to the difference between the minimum ordered and the minimum disordered $\VEE$. 
In Figure\,\ref{f:POD} (top), when the minimum ordered $\VEE$ exceeds the disordered $\VEE$ by more than $2\times 10^{-3}$ (green curve -- left axis), then the POD approaches $100\%$ (purple curve right-axis). 
This observation is further substantiated by the scatter plot of POD verses the difference in the minimum ordered and disordered $\VEE$ depicted in Figure\,\ref{f:POD} (bottom) for each $\alpha\in[0.3,1]$ and the selected values of $N.$ The basins of attraction of disordered equilibria are preponderant for these $N$ when  the difference in minimum $\VEE$  exceeds $10^{-2}.$ 
%\textcolor{blue}{Keith- Careful, $\VEE$ thresholds may not scale well with $N$.}

Each bin of stable computational equilibria contains a wide range of $\VEE$. This is illustrated through the ``anti-cumulative'' $\VEE$ distribution $\acDi:\bR \mapsto [0,1]$, defined as the percentage of stable equilibria whose $\VEE$ exceed a given value. The anti-cumulative $\VEE$ distribution is expressed in terms of a non-negative probability density $f_{\acDi}$, as
\beq\label{e:AC-VEE}
\acDi(s;N,\alpha):=\int_s^\infty f_{\acDi}(t;N,\alpha)\ \mrd t.
\eeq
The distribution $f_\acDi$ is approximated from the bins $\cB(N,1,5000)$ for each of the values of $N$ presented in Figure\,\ref{f:POD} (right). This reveals significant structure generated by low energy equilibria with broad basins of attraction that induce sharp drops in $f_\acDi$. Conversely, at higher values of $\VEE$  there is a continuous decline in $f_\acDi$ arising from a proliferation of equilibria with smaller basins of attraction. 
\begin{figure}[H]
    \begin{center}
    \begin{tabular}{cc}
        \includegraphics[width=.4\textwidth]{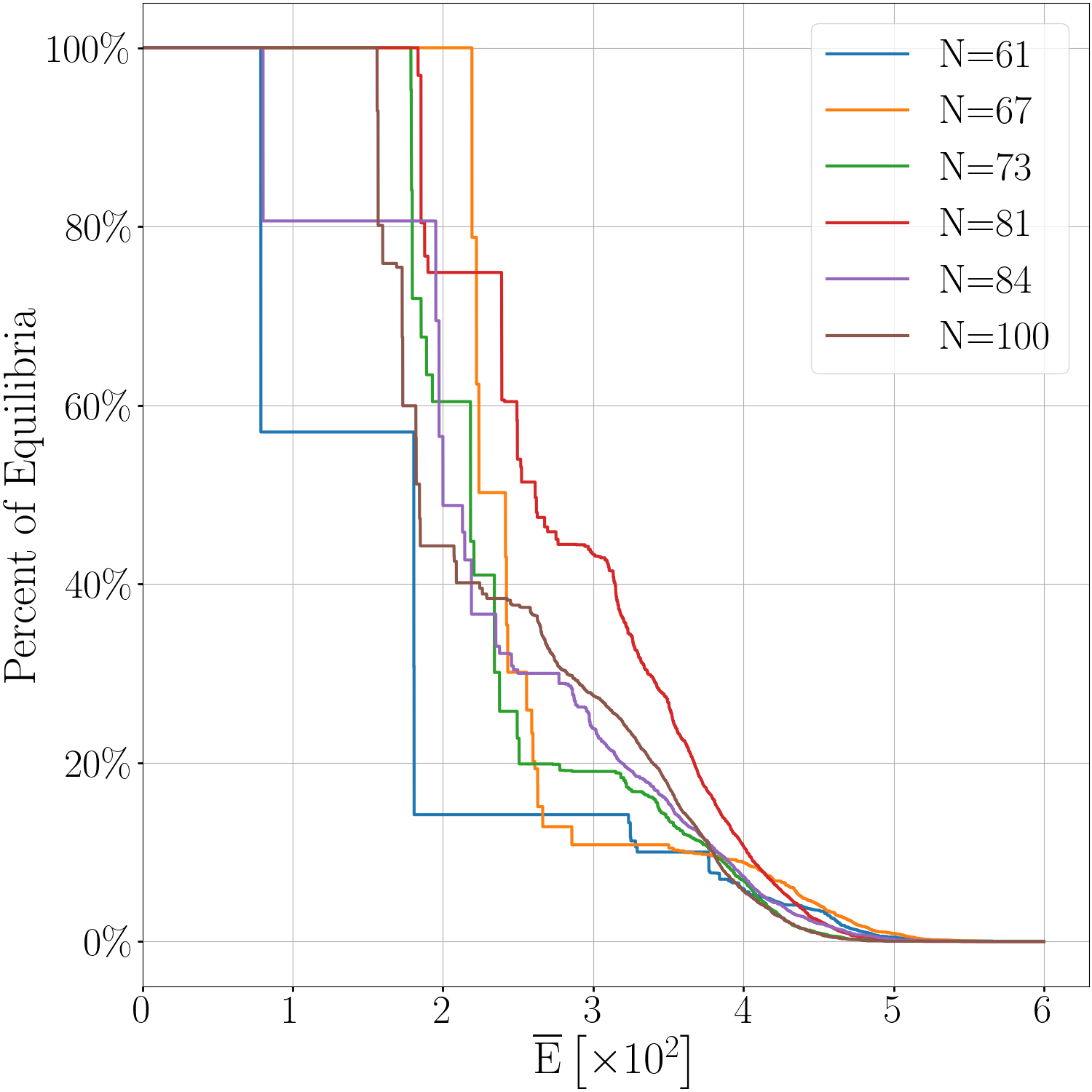}&
        \includegraphics[width=0.48\textwidth]{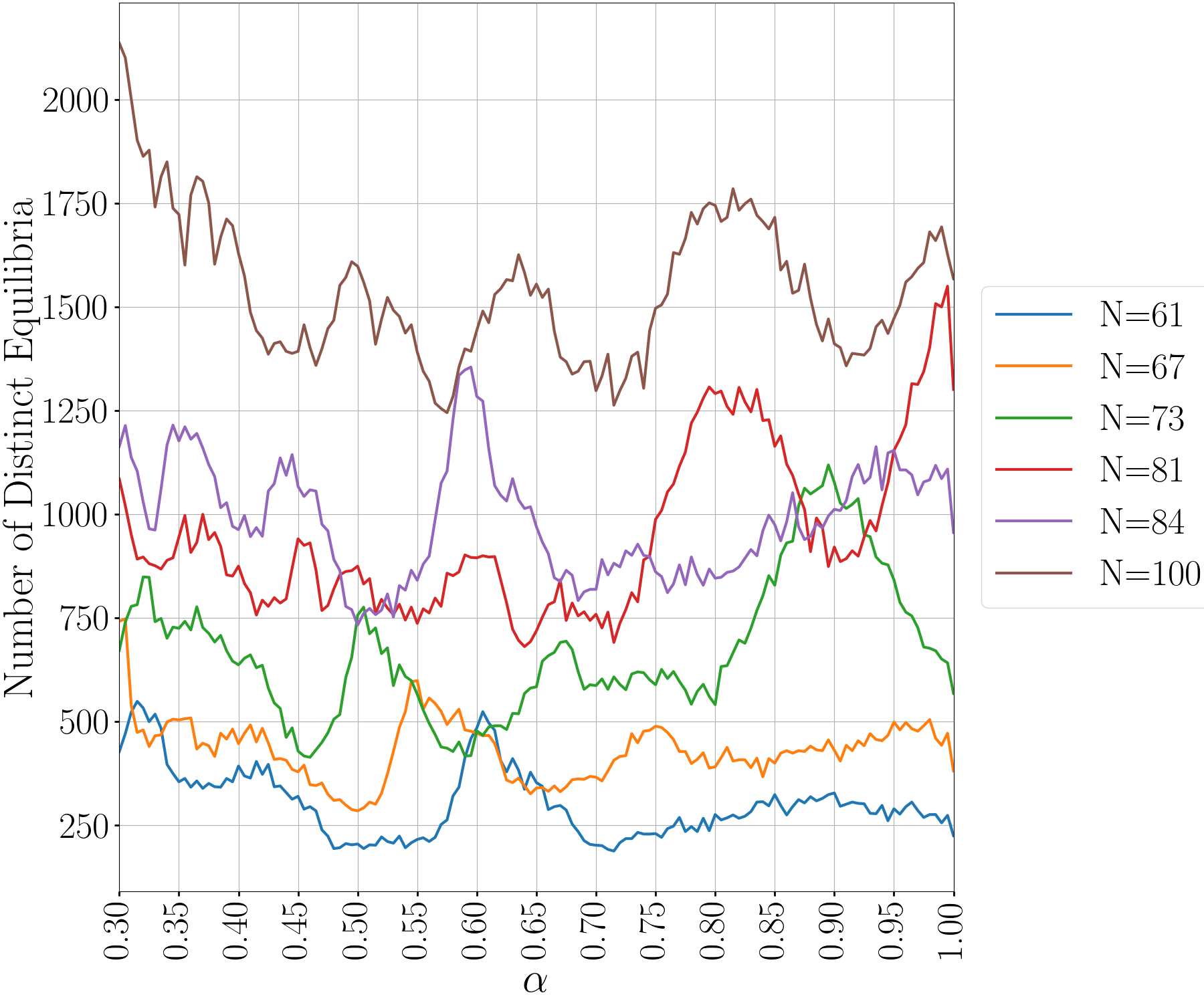} 
    \end{tabular}
    \end{center}
    \vspace{-0.3in}
    \caption{ (left)
    The anti-cumulative $\VEE$ distribution $f_\acDi$ for $\alpha=1$ and the six values of $N$, approximated from $\cB(N,1,5000).$ (right) The number of \emph{distinct} computational equilibria in $\cB(N,\alpha,5000)$ for the given values of $N.$}
    \label{f:acD-DOS}
\end{figure}
For $N=61$ the equilibria with large basins of attraction occur at two values of $\VEE$ below 0.02, collectively attract roughly 85\% of all initial data. These two jumps are followed by a flat region that manifests no basins of attraction, and then for $\VEE>0.03$ by a cascade of equilibria with distinct VEE, in excess of 200, that have marginal basins of attraction. The predominance of low energy equilibria as attractors of the flow, weakens as $N$ increases.  For all six values of $N$ the collective size of the basins of attraction of stable equilibria decay rapidly with $\VEE$, with $f_\acDi$ converges rapidly to zero for $\VEE$ greater than $0.05.$  The size of the basins of attraction of the dominant equilibria varies substantially with $N$. For $N=73$ the basins of attraction are highly fractured. Indeed, the dominant equilibria have $\VEE\sim 0.02$ and $0.025$, and collectively attract only 35\% of the simulations. The balance of roughly 600 distinct equilibria have $\VEE$ mostly larger then $0.025$ and individually attract less than 1\% of the simulations but collectively serve as end-states for over 60\% of the orbits. For $N=73$ and $\alpha=1$ the probability of disorder is approximately $100\%$, Figure\,\ref{f:POD} (left), so the fractured basins of attraction correspond to disordered equilibria. 

\begin{figure}[H]
    \begin{center}
    \begin{tabular}{cc}
        \includegraphics[width=.4\textwidth]{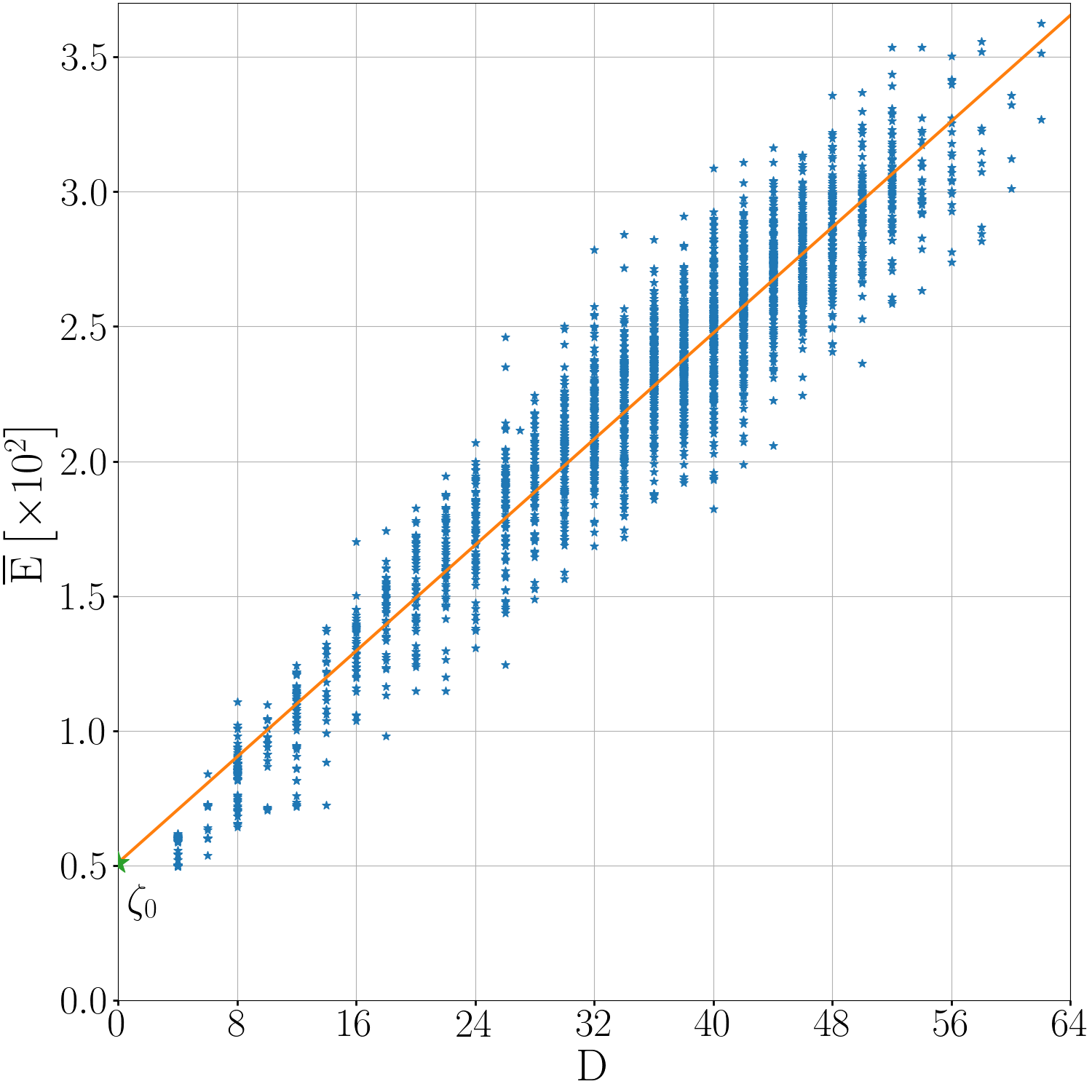}&
        \includegraphics[ width=.48\textwidth]{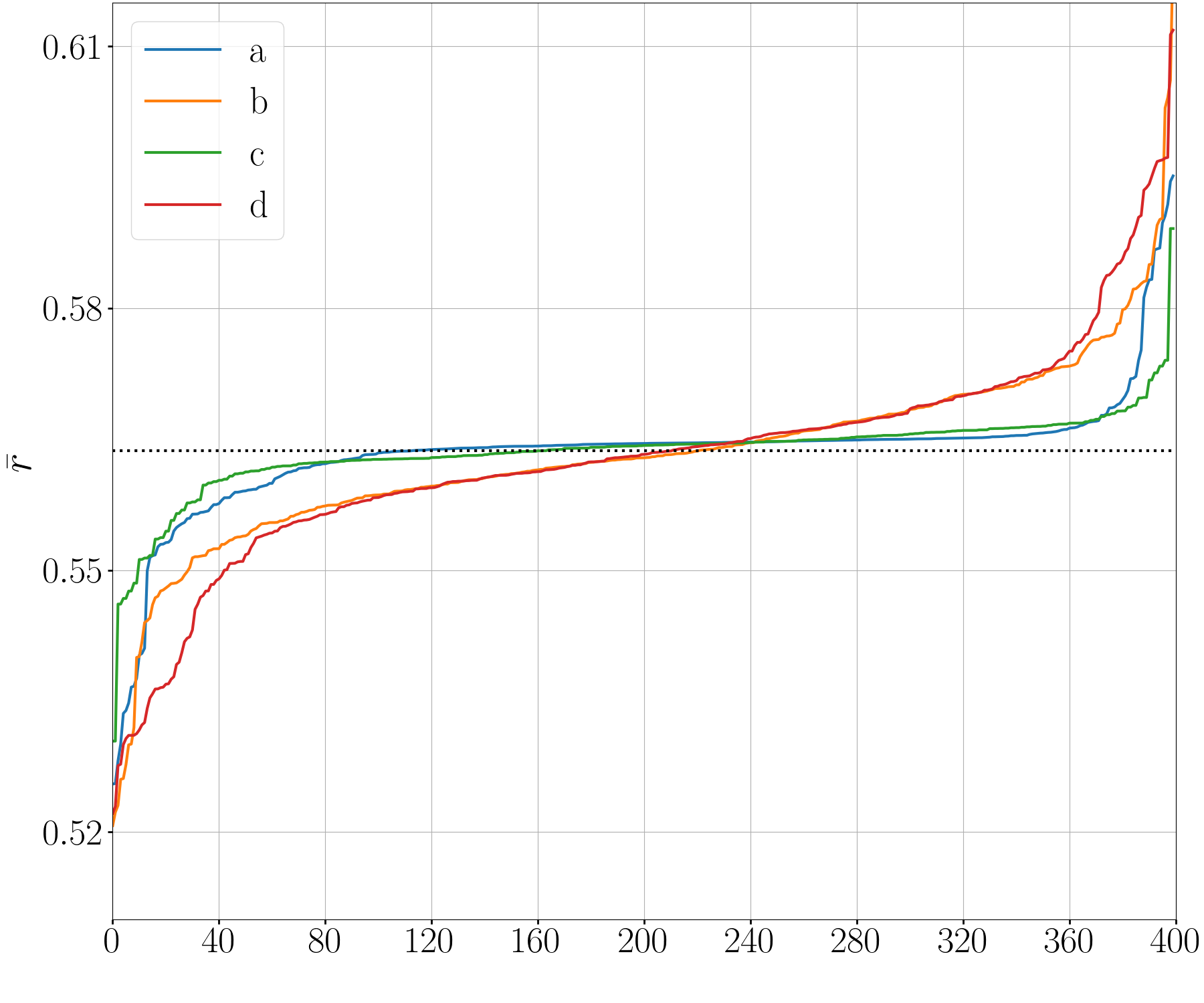}
    \end{tabular}
    \end{center}
    \vspace{-0.2in}
    \caption{  (left) Relation of $\VEE$ and defect number $\mrD$ determined from $\cB(400,1,2500)$. The linear fit determines the slope $\zeta(N,1)$ and the ground-state $\zeta_0(N,1)=0.502\times 10^{-2}.$
    (right) The ordered average radii $\br$ for the four $N=400$ tessellations shown in Figure\,\ref{f:N400a} (a-top left, b-top right) and (c-bottom left, d-bottom right). The dotted horizontal line is the average radius of a regular hexagon of unit area, $\bar{r}_6=0.5637.$
    }
    \label{f:EpD}
\end{figure}

The dependence of the number of \emph{distinct} computational equilibria contained in $\cB(N,\alpha,5000)$ is presented as a function of $\alpha\in[0.3,1]$ in Figure\,\ref{f:acD-DOS} (right). To prevent over counting equilibria that are translations or $\pi$ rotations of each-other, the $N$ average radii of the Voronoi tessellation of each computational equilibrium are ordered by increasing size to form the ordered average radii vector $\br\in\mathbb R^N$, see Figure\,\ref{f:EpD} (right).  These vectors are compared, and the associated equilibria are labeled as distinct if the $l^2$-norm of the difference of their ordered average radii vectors differ by more than $\delta_{\rm dist}=10^{-5}$. The vector $\br$ is a good proxy for distinctness as it determines the Hookean-Voronoi energy and is invariant under symmetric transformations of the underlying sites. The sensitivity of the basins of attraction of equilibria is highlighted by the fact that for each of the 6 values of $N$ the number of distinct computational equilibria within $\cB(N,\alpha, 5000)$  varies by more than a factor of $2$ over the range of $\alpha$.

%\textcolor{blue}{Kenny -- For $N=73$ lets make two new graphs. (a) For each $\alpha$ find the disordered equilibria with the largest number of defects that has lower energy than all stable single-strings, and plot that verses $\alpha$. (b) The percentage of disordered equilibria that have lower energy than all stable single-string solutions -- when this is 100\%, verify that there are no stable single-strings. }

Disordered Voronoi tessellations are composed of regions $\{\region{i}\}_{i=1}^N$ with a range of average radii. Many of these regions may have negative site energy $\VEE(\region{i})<0$ while, as guaranteed by Theorem\,\ref{t:PosVEE}, the tessellation energy $\VEE(\bx)\geq 0.$ Significantly, the  simulations show an absolute correlation between disorder and defects: all disordered tessellations  contained at least one pair of defective regions. Presented in Figure\,\ref{f:pair_collision}, the passage from a disordered but non-defective tessellation to a defective one requires a vertex collision. This places a clear threshold between ordered tessellations and defective ones, and no equilibria are found in this gap.  It is informative to correlate the number defects  $\mrD=\mrD(\eqx)$ with the $\VEE$ of the tessellation.  These values were tabulated for $\cB(400,1,2500)$ and presented as a scatter plot in Figure\,\ref{f:EpD} (left). While there is significant variation in $\VEE$ at a fixed number of defects, there is a strong, approximately linear, correlation between the $\mrD$ and $\VEE,$
\beq
\label{e:D-VEE}
\VEE(\eqx)\approx  \zeta_0(N,\alpha)+\zeta_1(N,\alpha) \mrD(\eqx).
\eeq
The slope $\zeta_1$ characterizes the average gain in $\VEE$ per defect, with a best linear fit given by the value $\zeta_1(400,1)\approx 0.0487\times10^{-2}.$ The value $\zeta_0(N,\alpha)$ is referred to as the ``ground-state'' $\VEE$ arising as the extrapolation of $\VEE$ from defect filled tessellations to the ``extrapolated energy'' of a zero-defect tessellation.
Remarkably the ground-state $\eta_0$ has little correlation with the actual minimum ordered $\VEE.$ 

\begin{figure}[H]
    \begin{center}
    \begin{tabular}{cc}
        \includegraphics[width=.35\textwidth]{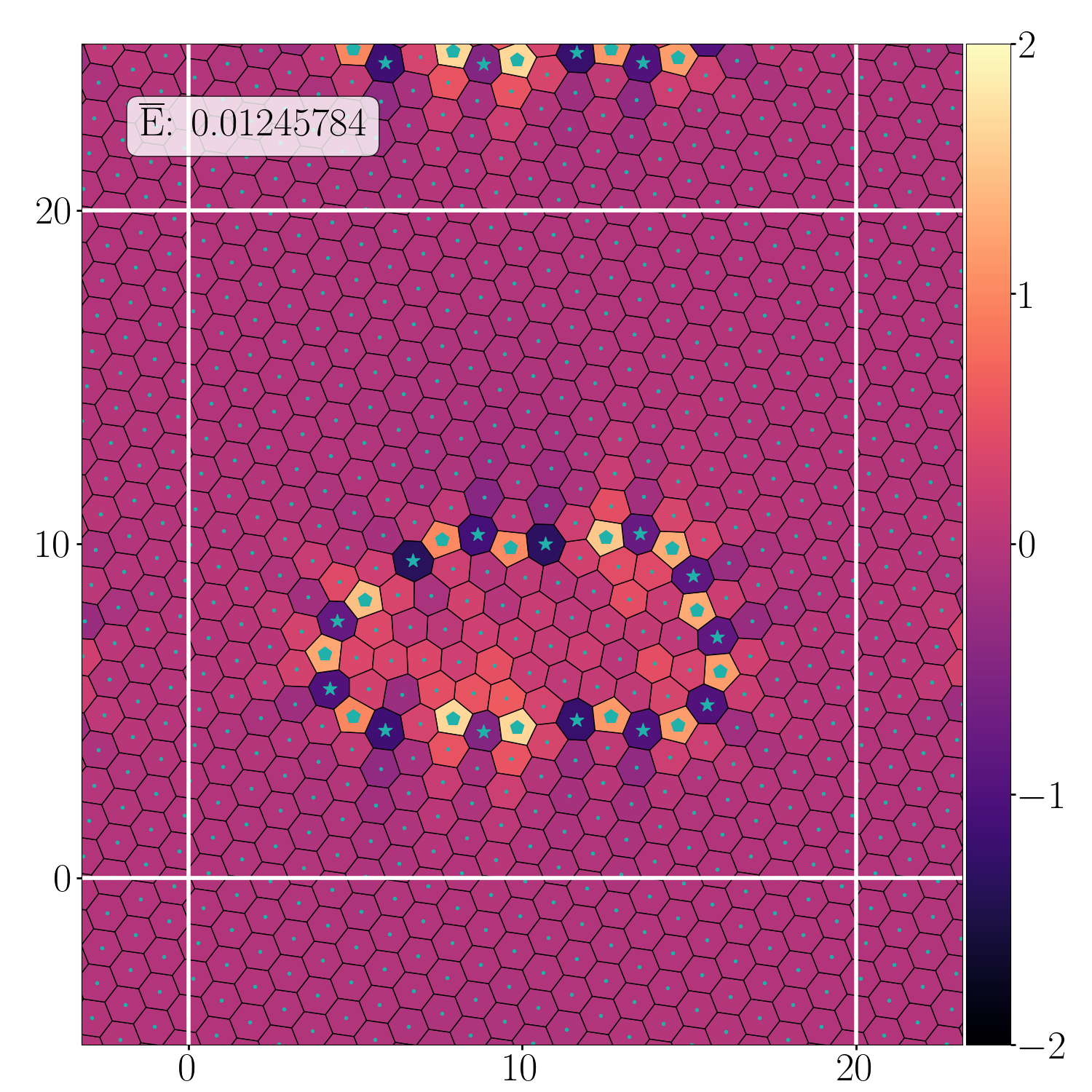} &
        \includegraphics[width=.35\textwidth]{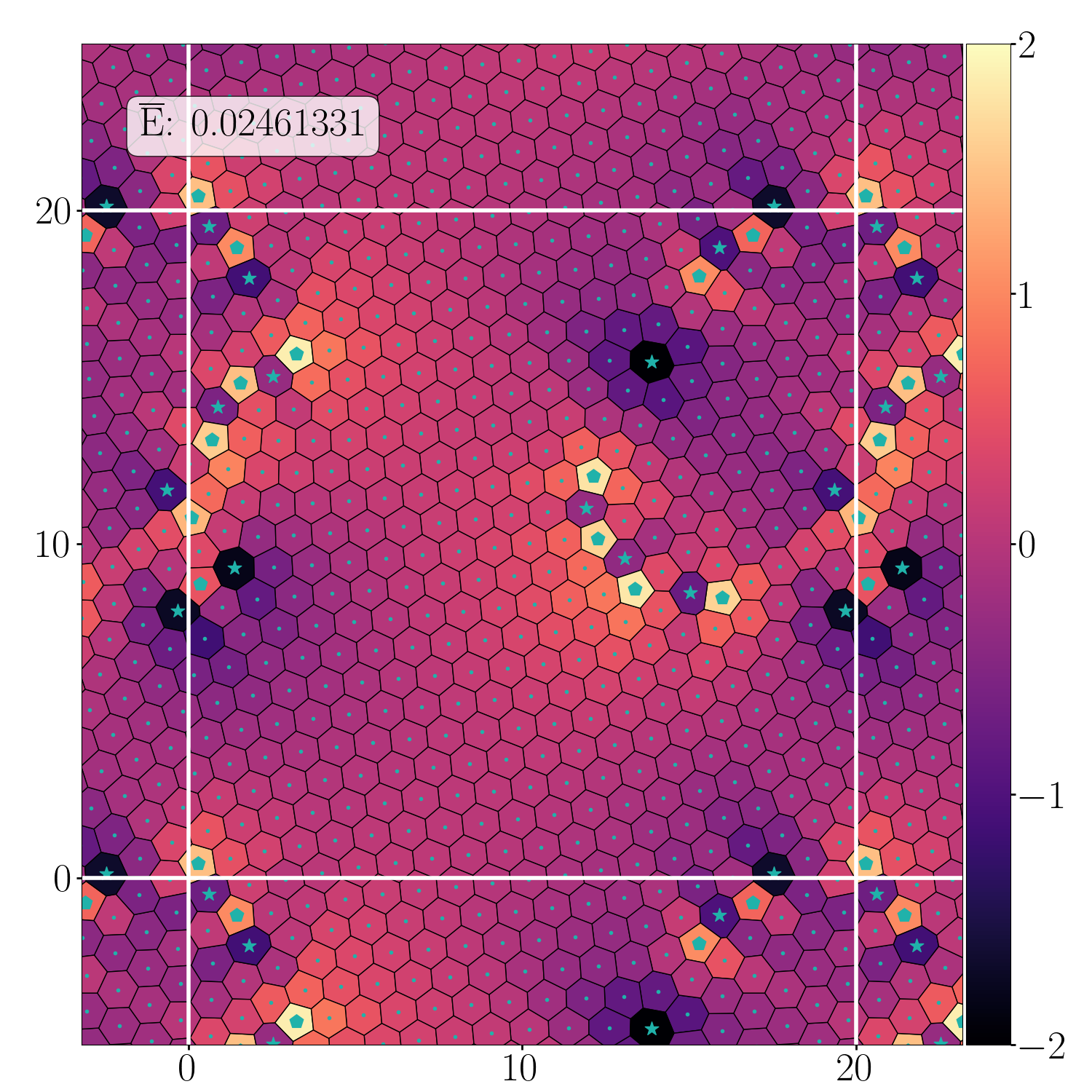} \\
        \includegraphics[width=.35\textwidth]{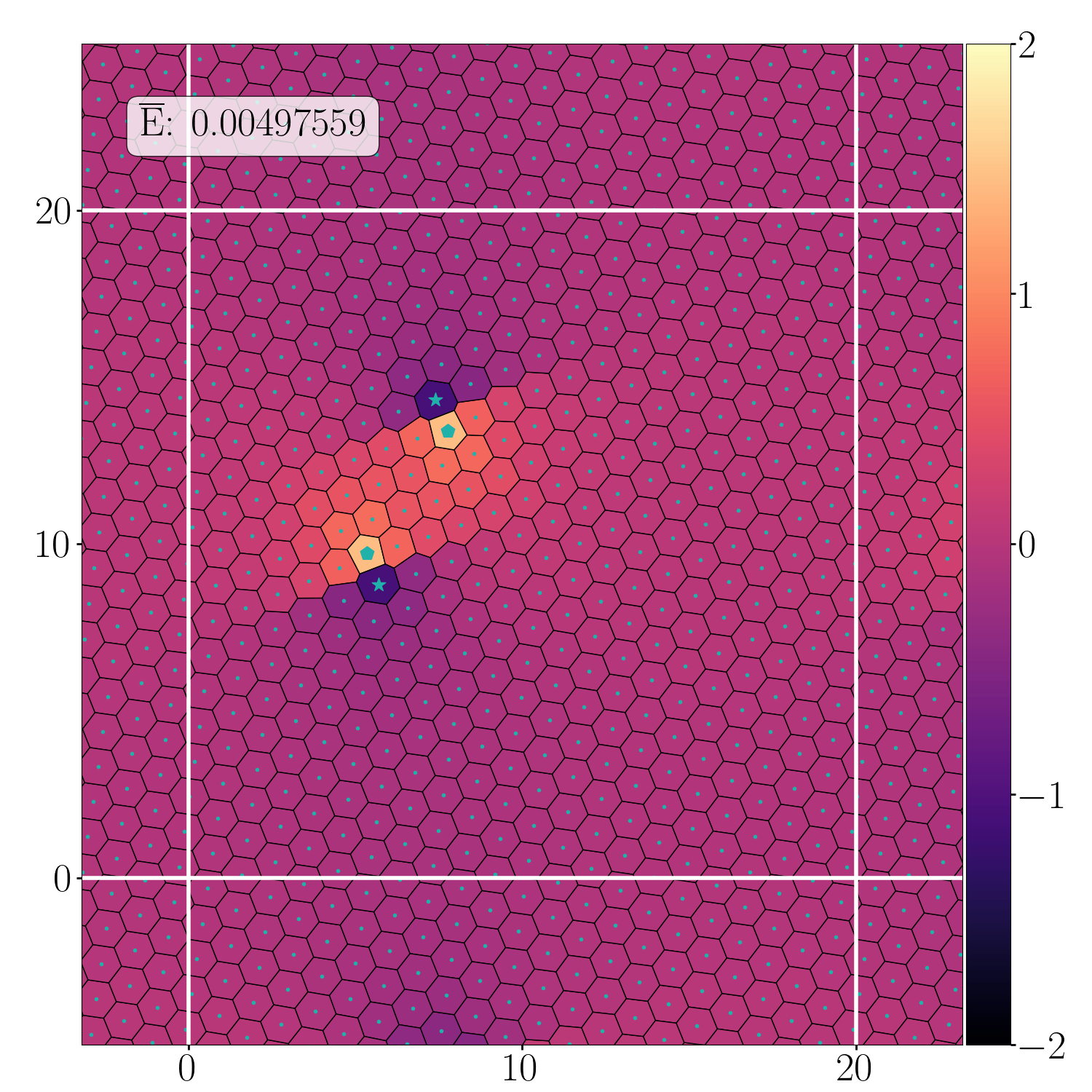} &
        \includegraphics[width=.35\textwidth]{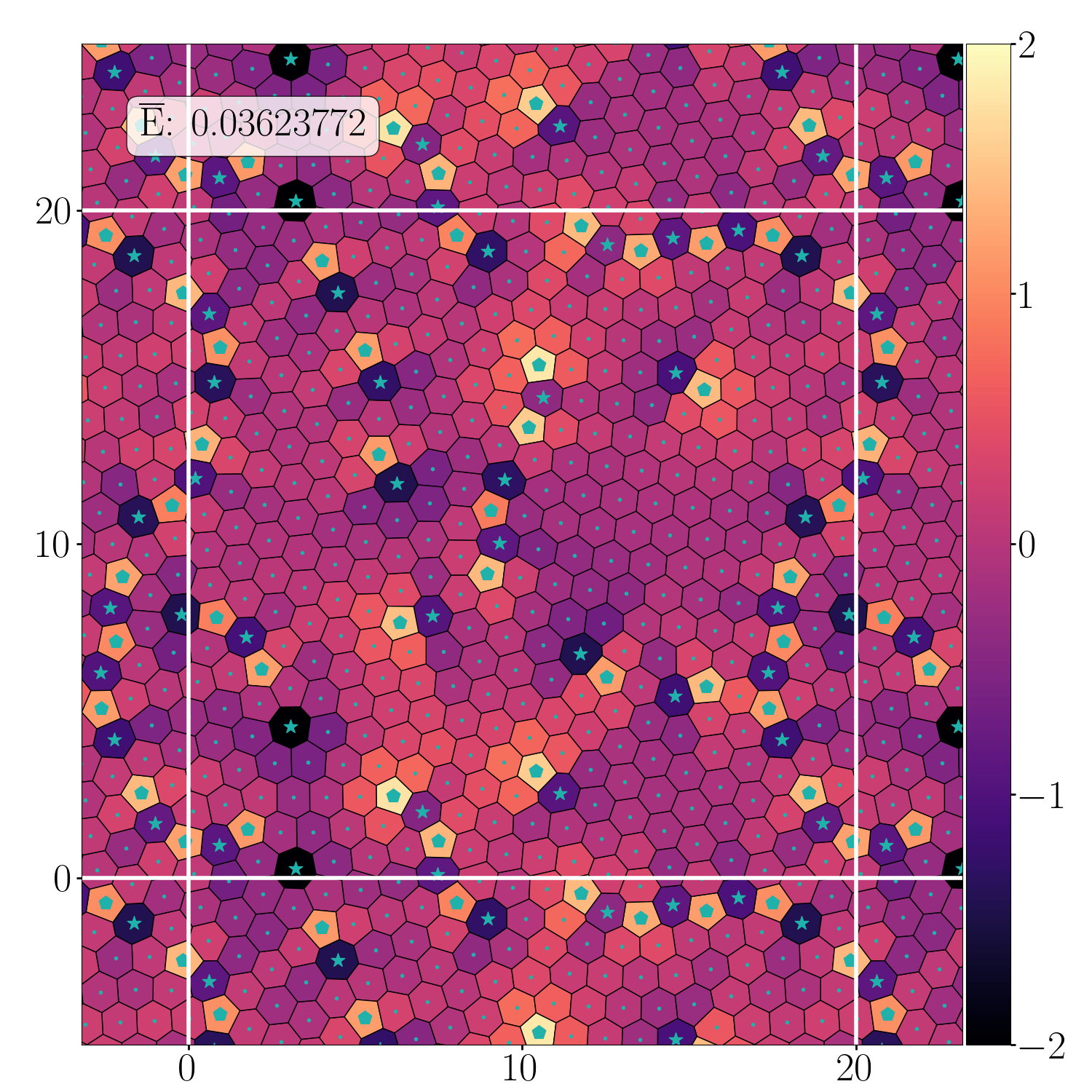}
    \end{tabular}
    \end{center}
    \vspace{-0.3in}
    \caption{(top) The bin $\cB(400,1,2500)$ yielded 78 tessellations with $\mrD=26$ defects. These two represent the lowest and the highest $\VEE$ among this set. (bottom) The two tessellations from the bin with (left) the lowest number, $\mrD=4$, and  (right) the highest number, $\mrD=62$, of defects.}
    \label{f:N400a}
\end{figure}

The relation \eqref{e:D-VEE} casts light on the minimum and maximum disordered $\VEE$ of Figure\,\ref{f:VEE}. For fixed $N$ the range of $\VEE$ of disordered tessellations is relatively constant, roughly from $0.015-0.06$. The minimum disordered $\VEE$ typically corresponds to equilibria with few defects, with the maximum disordered $\VEE$ having many defects. However order, that is the absence of all defects, can be energetically costly and can in some cases raise the minimum ordered $\VEE$ to be comparable to the maximum disordered $\VEE.$ The system can demand a high price for order, which can be relaxed by a modest injection of disorder. Examples of the relationship between order and energy in $\cB(400,1,2500)$  are presented in Figure\,\ref{f:N400a}. The top row shows the equilibria in that bin with $\mrD=26$ defects, while the bottom row shows the tessellations corresponding to the two equilibria with the lowest and the highest $\VEE$. Each Voronoi region is shaded according to its $\VEE$, with dark shading (lower $\VEE$) corresponding to larger regions with more edges and lighter shading (higher $\VEE$) corresponding to smaller regions with fewer edges. The equilibria with the lowest $\VEE$ from the $\mrD=26$ group has its defects arranged in an ellipse formed by alternating 5- and 7-sided defect pairs. The ellipse forms a boundary  dividing the domain into two parts populated with similar hexagons. The $\mrD=26$ equilibria with the highest $\VEE$ has a tessellation in which its defects are arranged in isolated pairs of 5-7 defects and one chain of 6 defects. Each of its defects reside in a background of hexagonal regions which have non-trivial variation in their $\VEE$. 

The bottom row of Figure\,\ref{f:N400a} presents the tessellations corresponding to the lowest, $\mrD=4$, and highest $\mrD=62$ number of defects in $\cB(400,1,2500)$.  Given the range in $\VEE$ between the equilibria with four defects and $\VEE=0.0049$ and the equilibrium with $62$ defects and $\VEE=0.0351$, it is surprising that the gradient flow is unable to combine the many defect pairs in the higher energy equilibrium. The equilibria with large numbers of defects possess an array of chains of $5-7$ defects of various lengths. With a large number of defects the combinatorial possibilities for chain lengths is considerable. However defect numbers $\mrD$ above 15\% of $N$ are seldom observed, most likely because the corresponding equilibria are unstable, and susceptible to vertex collisions that reduce defect numbers. This raises a fundamental question about the tolerance of defects in the Hookean-Voronoi gradient flow and the average energy that they contribute to its equilibria, in particular in the large $N$ limit.

%\begin{figure}[H]
%    \begin{center}
%    \begin{tabular}{c}
%    \includegraphics[width=.4\textwidth]{img/fig13/Near-Near-Neighbors-N400-D6+60.png} %    \includegraphics[width=.4\textwidth]{img/fig12/N400D62E3623 - 00046.png}
%    \end{tabular}
%    \end{center}
%    \vspace{-0.3in}
%    \caption{The probability density for the first- and second-near neighbors from $\cB(400,1,2500)$ over tessellations with defect number $\mrD=6$ (blue) and $\mrD=60$ (orange).}
%  \label{f:NN-nbr}
%\end{figure}

%The impact of defect number on structure of the tessellation can be examined through the first and second near-neighbor probability density. This quantity tabulates the probability that a near-neighbor site (first-near neighbor) or the near-neighbors of a near-neighbor site (second-near neighbors) appear at a particular distance from the given site.  These density functions are presented in Figure\,\ref{f:NN-nbr} for tessellations from $\cB(400,1,2500)$ with defects $\mrD=6$ and $\mrD=60.$ Each probability distribution has three principle peaks at distances $d=1.1, 1.8,$ and $2.2.$ However the quasi-ordered tessellations with $\mrD=6$ have much higher and narrower peaks than the more disordered tessellations with $\mrD=60$. Indeed the width-at-half-max is roughly 5 times greater in the second and third peaks of $\mrD=60$ compared to $\mrD=6$. This shows the long-range impact of defect number on disorder within the tessellation.

\subsection{Large \texorpdfstring{$N$}{N} limits: Universal behavior}

\begin{figure}[H]
    \begin{center}
    \begin{tabular}{cc}
    \includegraphics[width=0.309\textwidth]{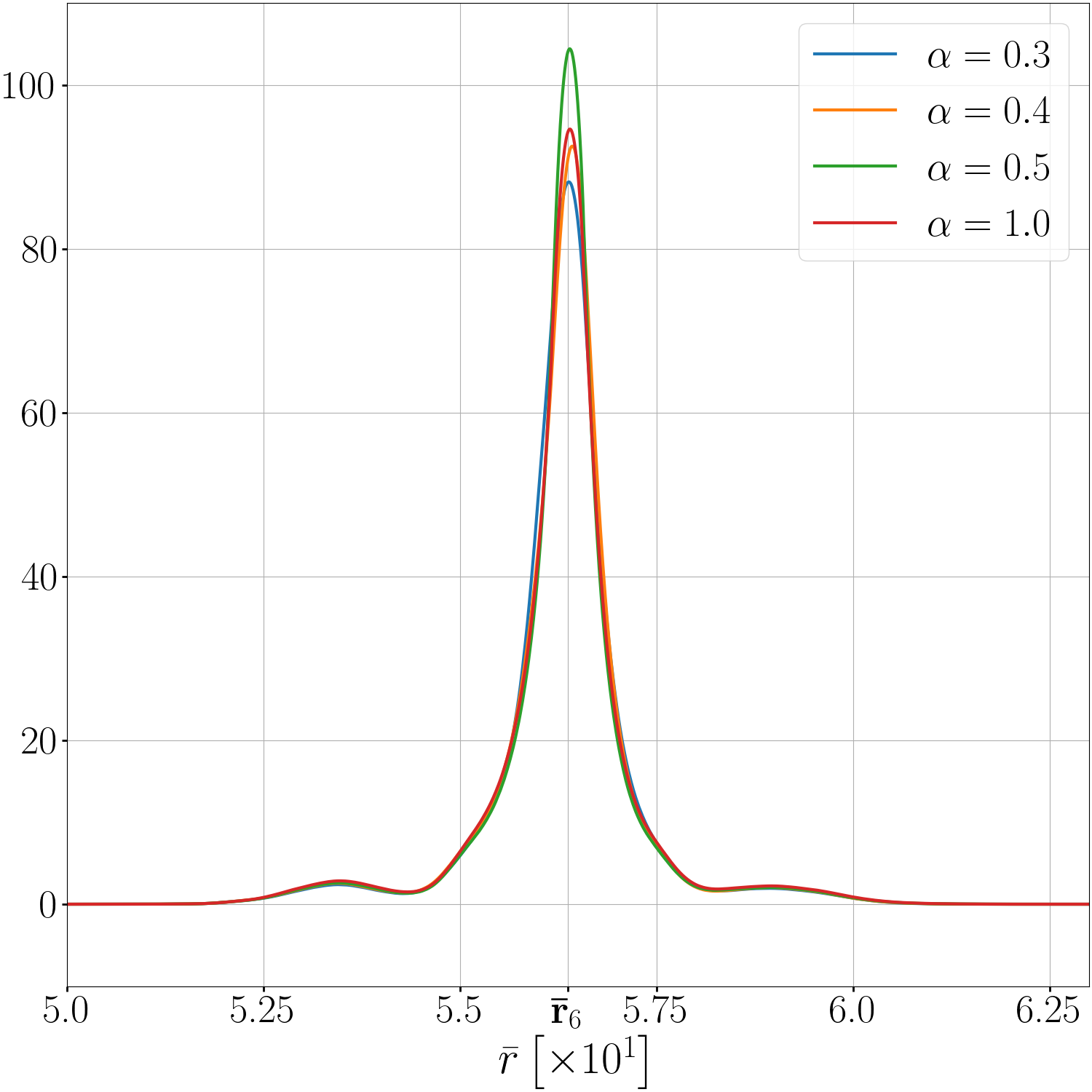}
    \includegraphics[width=0.309\textwidth]{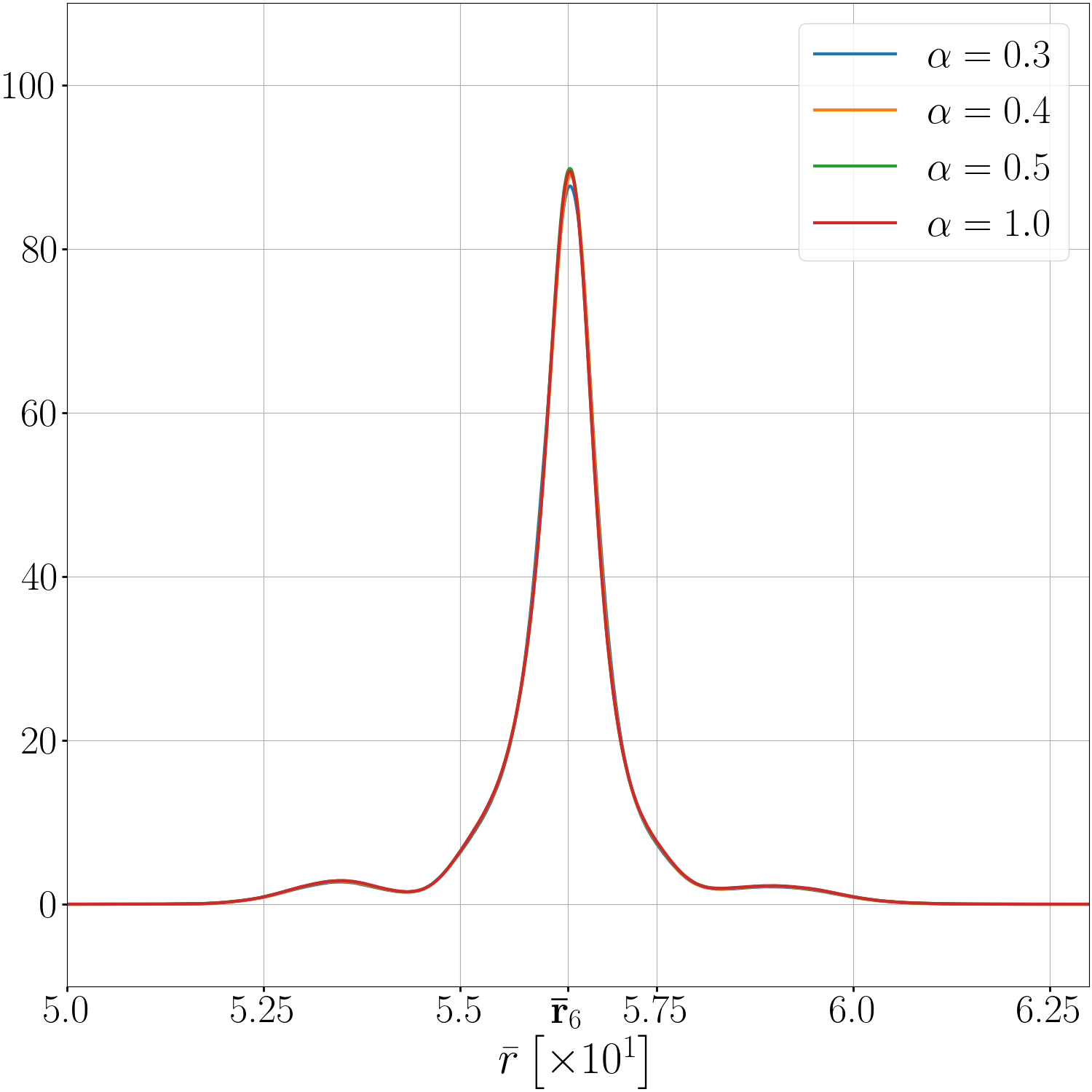}
    \includegraphics[width=0.309\textwidth]{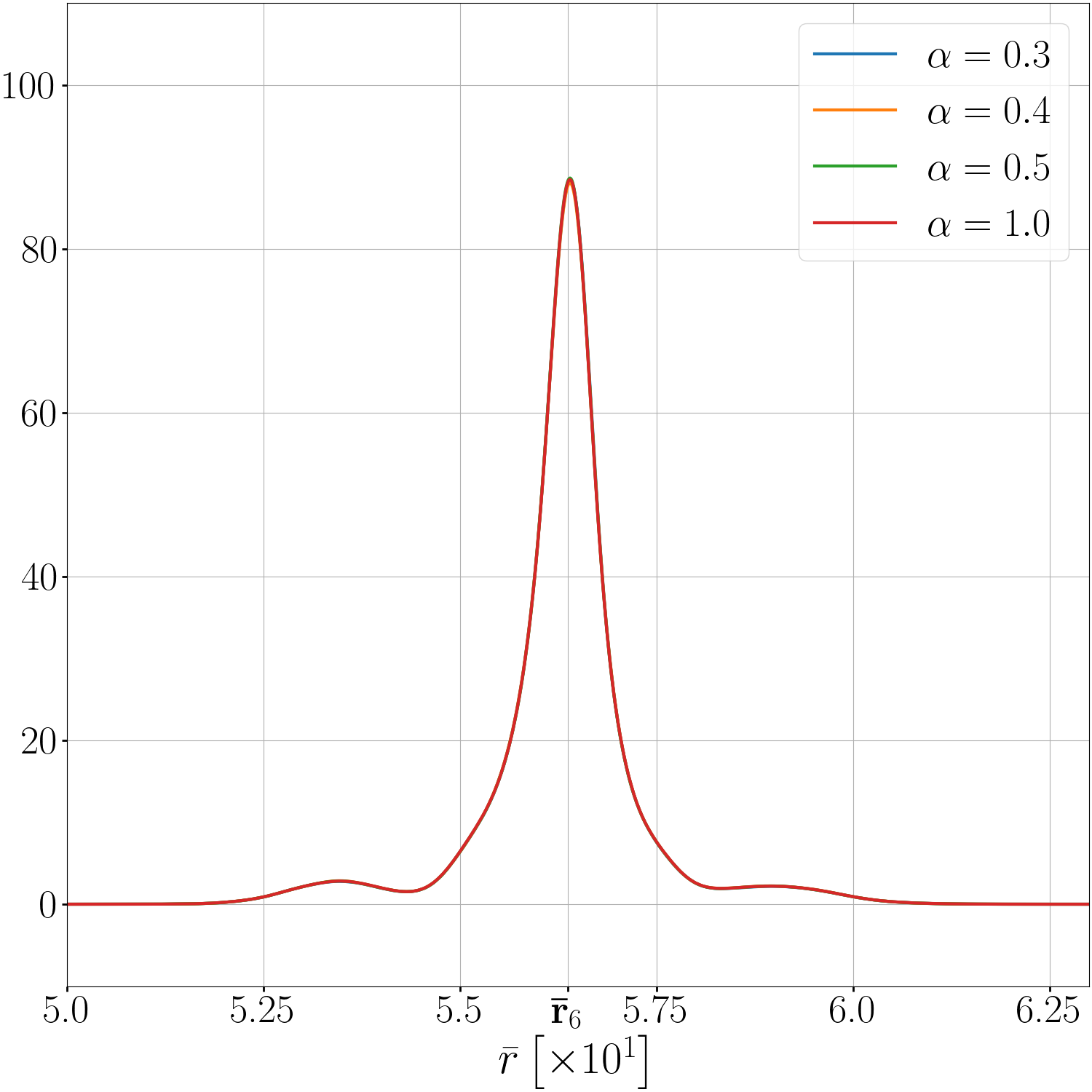} 
    \end{tabular}
    \end{center}
    \vspace{-0.3in}
    \caption{The distribution of average radii over the bins (left) $\cB(400,\alpha,2500)$ and (center) $\cB(1000,\alpha,2500)$ and (right) $\cB(2000,\alpha, 2500)$  for $\alpha=0.3, 0.4, 0.5$ and $1.0.$ The average radius of the unit-area regular hexagon, $\bar{r}_6$, is marked by a vertical line.}
  \label{f:NN-average-radii}
\end{figure}

\begin{figure}[H]
    \begin{center}
        \begin{tabular}{cc}
        \includegraphics[width=0.4\textwidth]{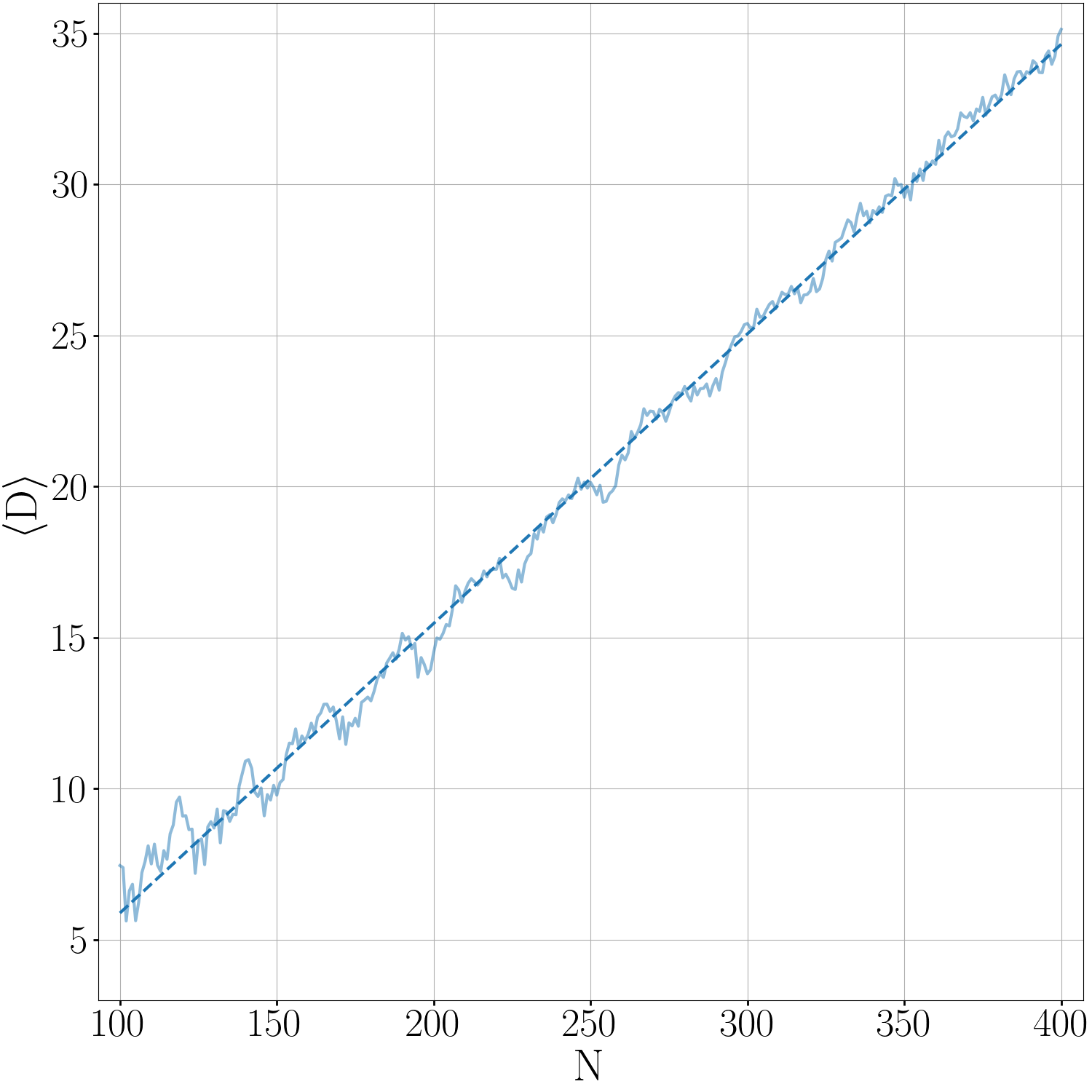} &
        \includegraphics[width=0.533\textwidth]{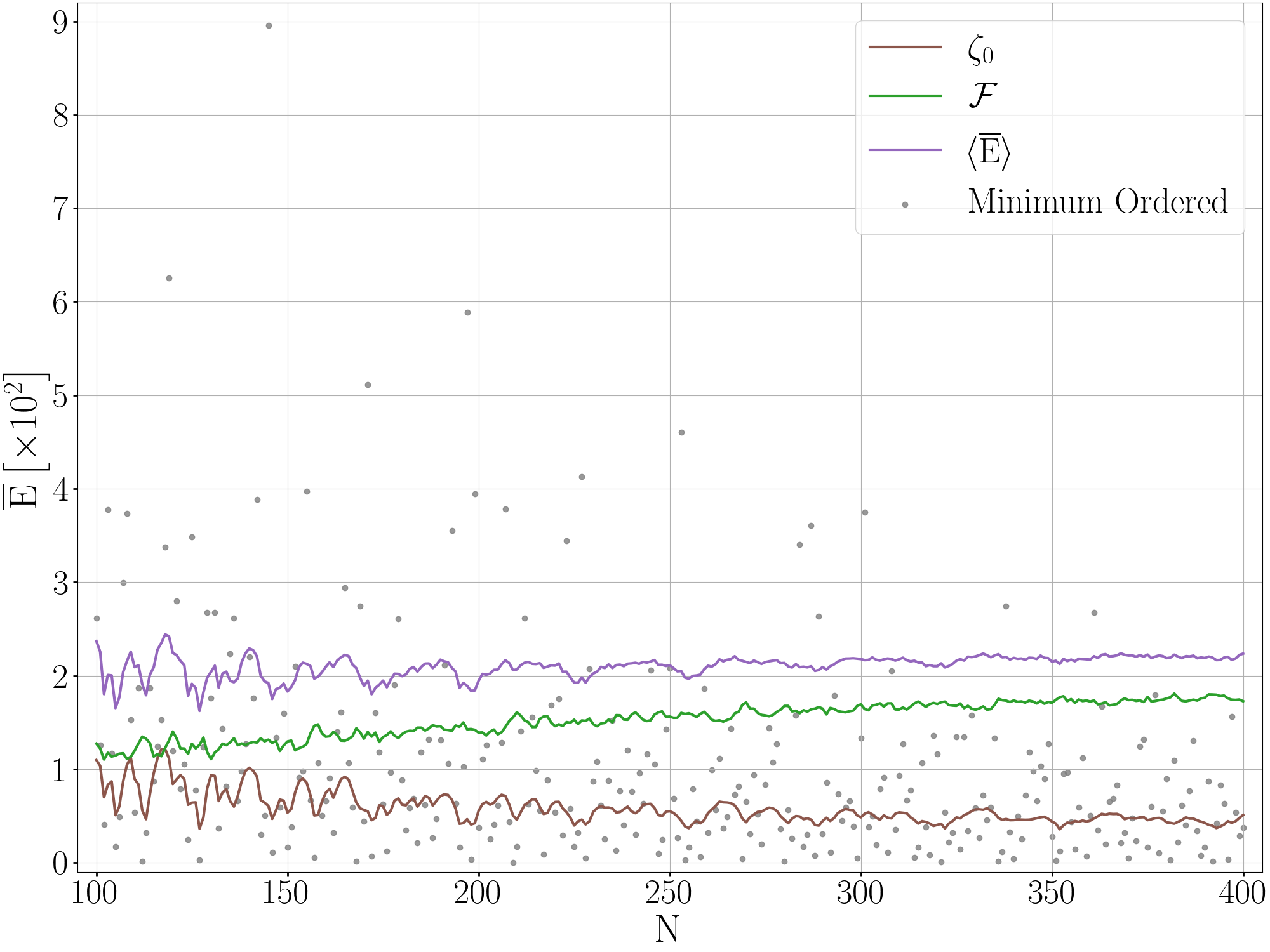} 
        \end{tabular}
    \end{center}
    \vspace{-0.2in}
    \caption{(left) For $\alpha=1$, the average defects $\langle\mrD\rangle_\cB$ computed from $\cB(N,1,2500)$ for $N=100,\ldots, 400$. The best linear fit to  
    $\langle\mrD\rangle_{\cB(N)}$ for $N\geq 100$ has slope
    $\eta_1(1)\approx 0.088.$ (right) The average $\langle\VEE\rangle_\cB$ decomposed into the ground-state $\zeta_0(N,1)$ and the frustration $\cF(N,1)$. The minimum ordered $\VEE$ is presented as disconnected dots.}
    \label{f:Large-N}
\end{figure}

\tcb{Simulations of gradient flows of the Hookean-Voronoi energy suggest a large $N$ convergence of the inherent states to universal or ``bulk'' distributions. We first focus on the large $N$ behaviour of individual Voronoi regions by examining the distribution of average radii within the inherent states and the dependence  of this distribution upon $N$ and $\alpha.$ 
Figure\,\ref{f:NN-average-radii} presents the probability distributions of the average radii computed from all the states in bins $\cB(N,\alpha, 2500)$ for $N=400, 1000$ and $2000,$ and aspect ratios $\alpha=0.3, 0.4, 0.5, $ and $1.$  For $N=400$ there is visible variation in the distribution with domain aspect ratio $\alpha$ which has largely been virtually eliminated for $N=2000.$ The distributions are almost symmetric about $\bar{r}=\bar{r}_6$, the average radius of the unit-area regular hexagon.  The spread of average radii is on the order of $10^{-1}$, with important bumps at $\bar{r}=5.3$ and $\bar{r}=5.8$ which correspond to 5 and 7 sided defects respectively. This suggests that the density of defects does not diminish with large $N$, and indeed approaches a universal form. Significantly the average energy of a tessellation from $\cB(N,\alpha,S)$ is controlled by the average of the average radii of the sites of the tessellation. This is measured in the asymmetry of the peaks in Figure,\ref{f:NN-average-radii} which, as guaranteed by Theorem\,\ref{t:PosVEE}, is slightly smaller than $\bar{r}_6.$ Indeed, as shown in Figure\,\ref{f:Large-N}, the average per-site energy of a tessellation is on the order of $10^{-2}$, one order of magnitude smaller than the average energy of a typical Voronoi region.}

\tcb{To quantify the ubiquity of defects, and their impact on the collective behavior of the  $\cB(400,1,2500)$ we introduce the \emph{frustration} $\cF$, 
of a bin $\cB(N,\alpha,S)$ of a tessellation} defined as the product of the energy per defect and the expected number of defects,
\beq\label{e:frust}
\cF(\cB(N,\alpha,S))=  \zeta_1(N,\alpha)\langle \mrD\rangle_{\cB}.
\eeq
The frustration represents the energy gap between the average value $\langle \VEE\rangle_\cB$ and the ground-state $\zeta_0$, defined in \eqref{e:D-VEE}. Like $\VEE$, frustration is averaged over $N$, representing a per-site contribution to $\langle\VEE\rangle_\cB$ from the defects that the gradient flow is unable to eliminate. In particular we have the relation
\beq
\langle \VEE\rangle_{\cB} = \zeta_0+ \cF.
\eeq
The asymptotic behaviour of these three quantities at large $N$ gives considerable insight into the role of defects.  The roughly linear relation between $\VEE$ and defect number $\mrD$ at fixed $N=400$ might seem to suggest that $\langle \mrD\rangle_{\cB}$ remains bounded with growing $N$. To determine this relation we formed the the bins $\cB(N,1,2500)$ for $N=100,\ldots 400$ and determined the average number of defects, $\langle \mrD\rangle_{\cB(N,1,2500)}.$
Presented in Figure\,\ref{f:Large-N} (left), this average defect count is well approximated by the linear relation 
$$\langle \mrD\rangle_{\cB(N,\alpha,S)} \sim \eta_1(\alpha) N-\eta_0,$$
for $N\gg 1.$ For $\alpha=1$ the best linear fit yields a proportionality $\eta_1(1)=0.088$ suggesting that between $8-9\%$ of Voronoi regions in a given equilibria are defects for $\alpha=1$ at large $N.$ 
%\begin{Remark}
%Given the sensitive dependence of ordered packings $\VEE$ upon $N$, it seems intuitive that the proportionality constant $\eta$ is likely largely independent of $\alpha$ since in the large $N$ limit the 'divisibility' issues that control defects will average out.
%\end{Remark} 
 Figure\,\ref{f:Large-N} (right), compares the ground-state energy $\zeta_0,$ the frustration $\cF$, their sum  $\langle \VEE\rangle_\cB$, and the minimum ordered $\VEE.$ The ground-state energy decreases with $N$ while the frustration grows. Their sum, the bin-averaged energy remains largely constant. This is consistent with a saturating frustration and suggests that in a large $N$ limit, taking expectation weighted by size of basin of attraction, the expected ratio of cells which are defects converges to a fixed value and moreover these defects  make a non-zero volumetric contribution to the expected system energy. This suggests that the ground state energy is not broadly accessible, having a vanishing basin of attraction. Yet more surprising, the minimum ordered $\VEE$ does not correlate with the ground-state $\zeta_0$. These simulations show that the minimum ordered $\VEE$ can exceed the average $\langle\VEE\rangle_\cB$ by up to a factor of 4, and can also be zero when a regular hexagonal tiling is admissible.  A wide range of minimal ordered $\VEE$ has been numerically observed out to $N=5000$.  If there is convergence in the large $N$ limit for minimum ordered energy, it is significantly slower than that observed for the disordered states. This suggests that care should be taken in extrapolating between special ordered equilibria and disordered states in these packing problems.  \tcb{It will require further study to determine if the sensitivity of the minimum ordered energy to aspect ratio $\alpha$ and site number $N$ is a feature of the rectangular domain or choice of periodic boundary conditions}.

\begin{figure}[H]
    \begin{center}
        \begin{tabular}{cc}
        \includegraphics[width=0.4\textwidth]{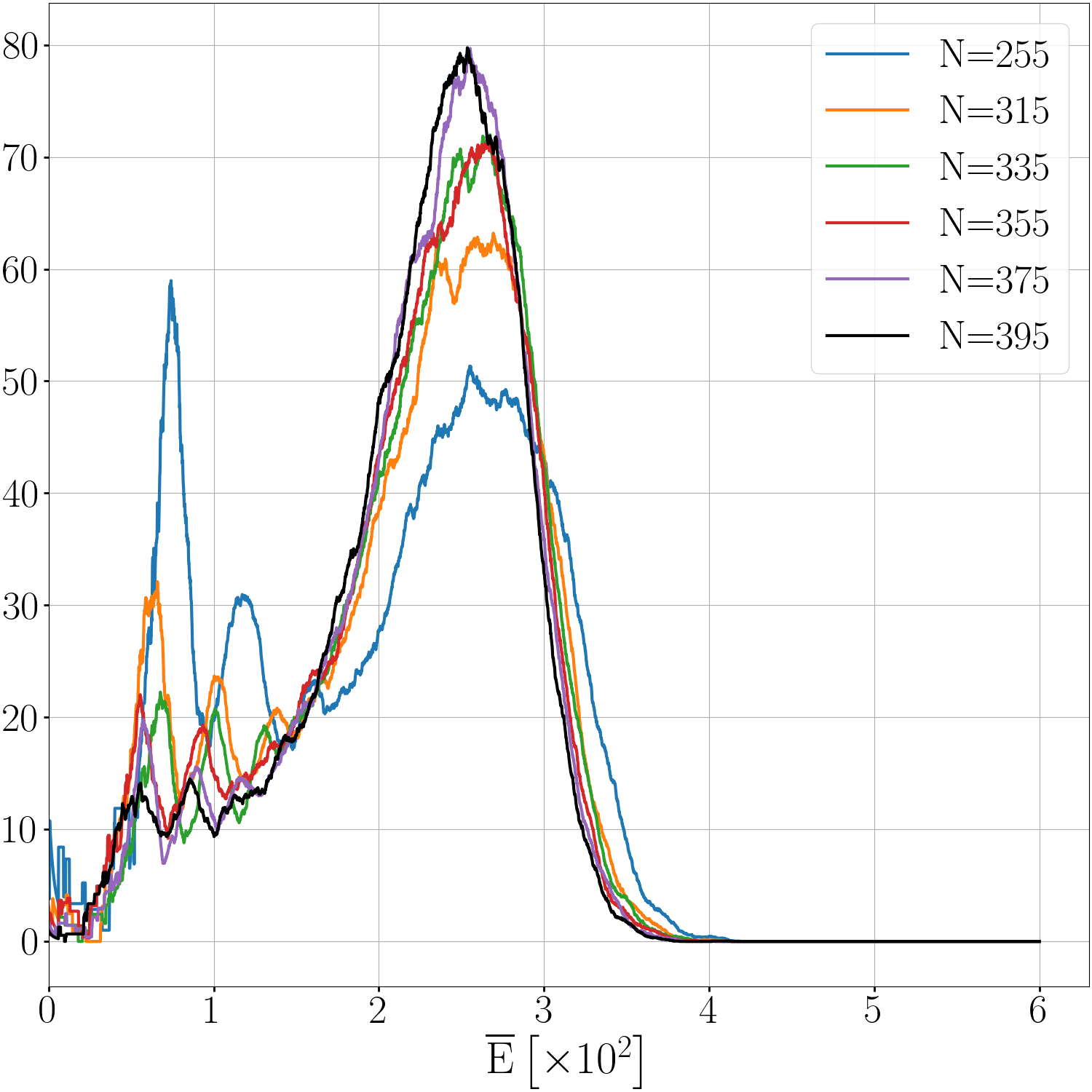} &
        \includegraphics[width=0.4\textwidth]{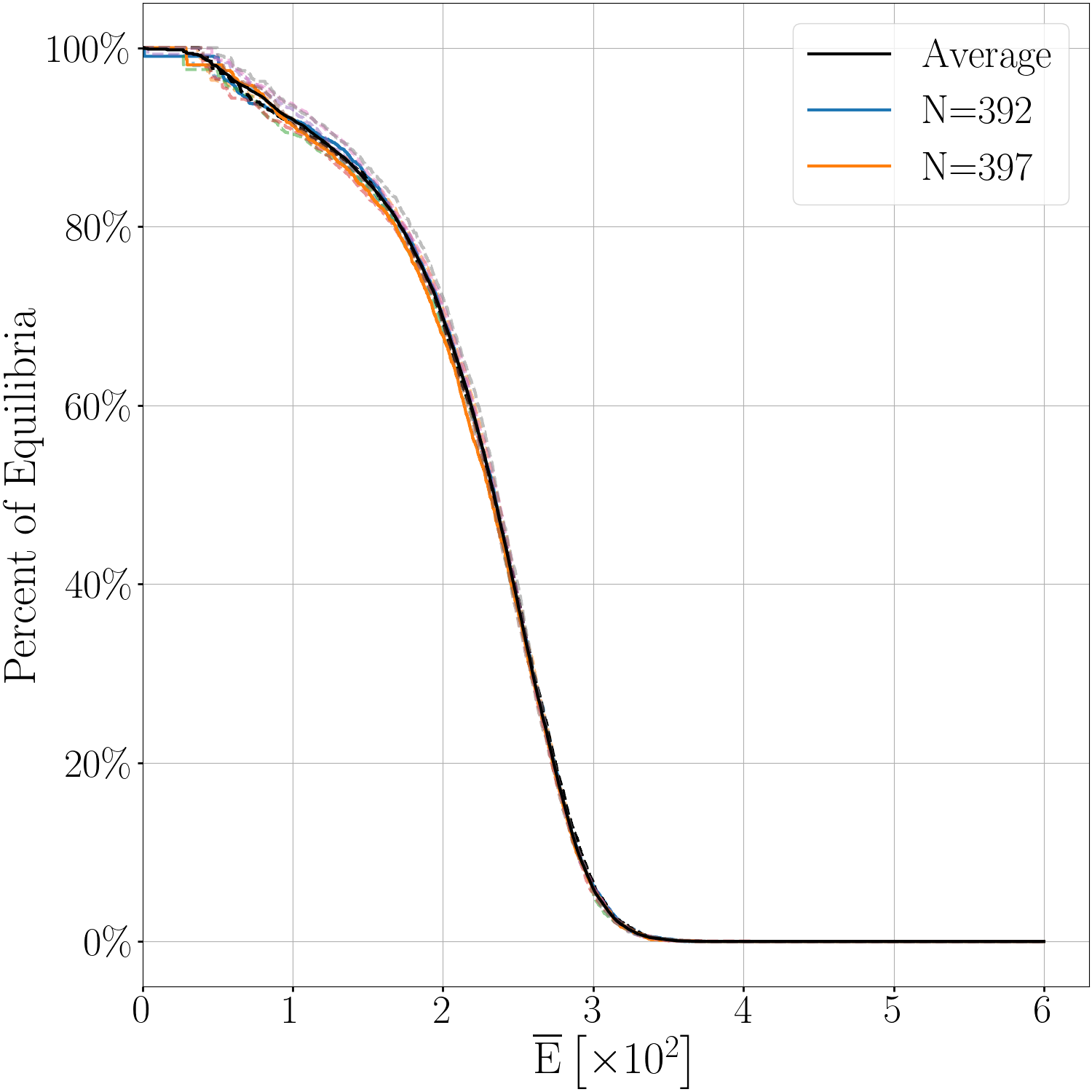}
        \end{tabular}
        \vspace{-0.15in}
    \end{center}
    \hspace{0.443in}
    \includegraphics[width=0.8355\textwidth]{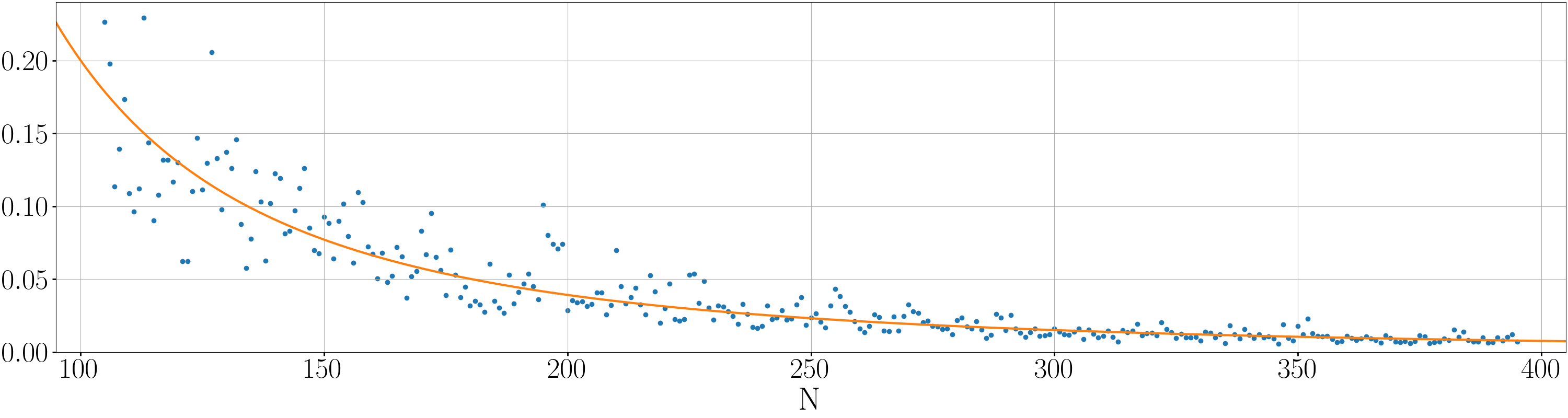} 
    \vspace{-0.1in}
    \caption{(top-left) The averaged tessellation probability densities $\langle f_\PD\rangle_{[N-5,N+5]}$  for values of $N$ as indicated, as determined from $\cB(N,1,2500)$ for $N=100, \ldots 400.$ 
    (top-right) The density for the averaged anti-cumulative tessellation 
    distribution $\langle f_\acDi\rangle_{[390,400]}$ (solid black) together with the densities for each of the anti-cumulative tessellation distributions  $f_\acDi(N),$ for $N=390,\ldots 400$ (dashed lines). 
    The densities for distributions $\acDi(392)$ and $\acDi(397)$ are indicated via solid, colored lines for comparison. 
    (bottom) The $l^2$ relative error between the sliding average $\langle f_\acDi\rangle_{[N-5,N+5]}$ and $f_\acDi(N)$ for $N=105\ldots 395$. The graph $y=10120\cdot N^{-2.35}$ (red line) is provided for reference.}
    \label{f:Universal}
\end{figure}
To quantify the convergence to a bulk-like state in the large $N$ limit we form the probability densities $f_\PD(N,\alpha)$ associated to the probability  $\PD(N,\alpha,S)$ that a random initial data converges to an equilibria of prescribed $\VEE.$  The densities $f_\PD(N,1)$ are approximated from the bins $\cB(N,1, 2500)$ for $N=100, \ldots 400.$ To observe convergence, for a given value $N=N_0$ we construct the sliding average of the probability densities over the window of 11 different values of $N= N_0-5, \ldots, N_0+5$. The associated probability measure is denoted $\langle \PD\rangle_{[N_0-5,N_0+5]}$ and its density is referred to as the averaged probability density and denoted $\langle f_\PD\rangle_{[N_0-5,N_0+5]}$. As presented in Figure\,\ref{f:Universal}, the averaged probability densities tend towards a uni-modal distribution as $N$ increases from $250$ to $395$. For $N\leq 355$ the averaged probability density has a broader top on its main peak, whose central value decreases slightly with increasing $N$. Significantly, for values of $N<320$, in particular represented by $N=255$, the averaged densities show oscillations with strong peaks for $\VEE$ in the range $0.2-1.5\times 10^{-2}.$ These correspond to attractivity of low-energy, low-defect count equilibria. The attractivity of these low energy equilibria is \emph{suppressed} for larger values of $N$. This corresponds to a shift of probability density away from equilibria with lower $\VEE$, and lower defect numbers, towards  equilibria with higher defect numbers and with $\VEE$ centered around $2.5\times10^{-2}$. This suggests that for larger values of $N$ the higher defect equilibria are preponderant, forming a basin of attraction maze that is increasingly difficult for the gradient flow to navigate and find the low defect (semi-ordered) equilibria. The proliferation of equilibria is reflected in the fact that 2461 out of the 2500 equilibria in $\cB(400,1,2500)$ received a single hit, with the most frequently visited equilibria receiving $4$ hits. The higher-defect equilibria shield the low-defect equilibria, and the basins of attraction of the low-defect equilibria shrink as a proportion of the total state space. 

\begin{figure}[H]
    \begin{center}
    \begin{tabular}{cc}
    \includegraphics[width=0.309\textwidth]{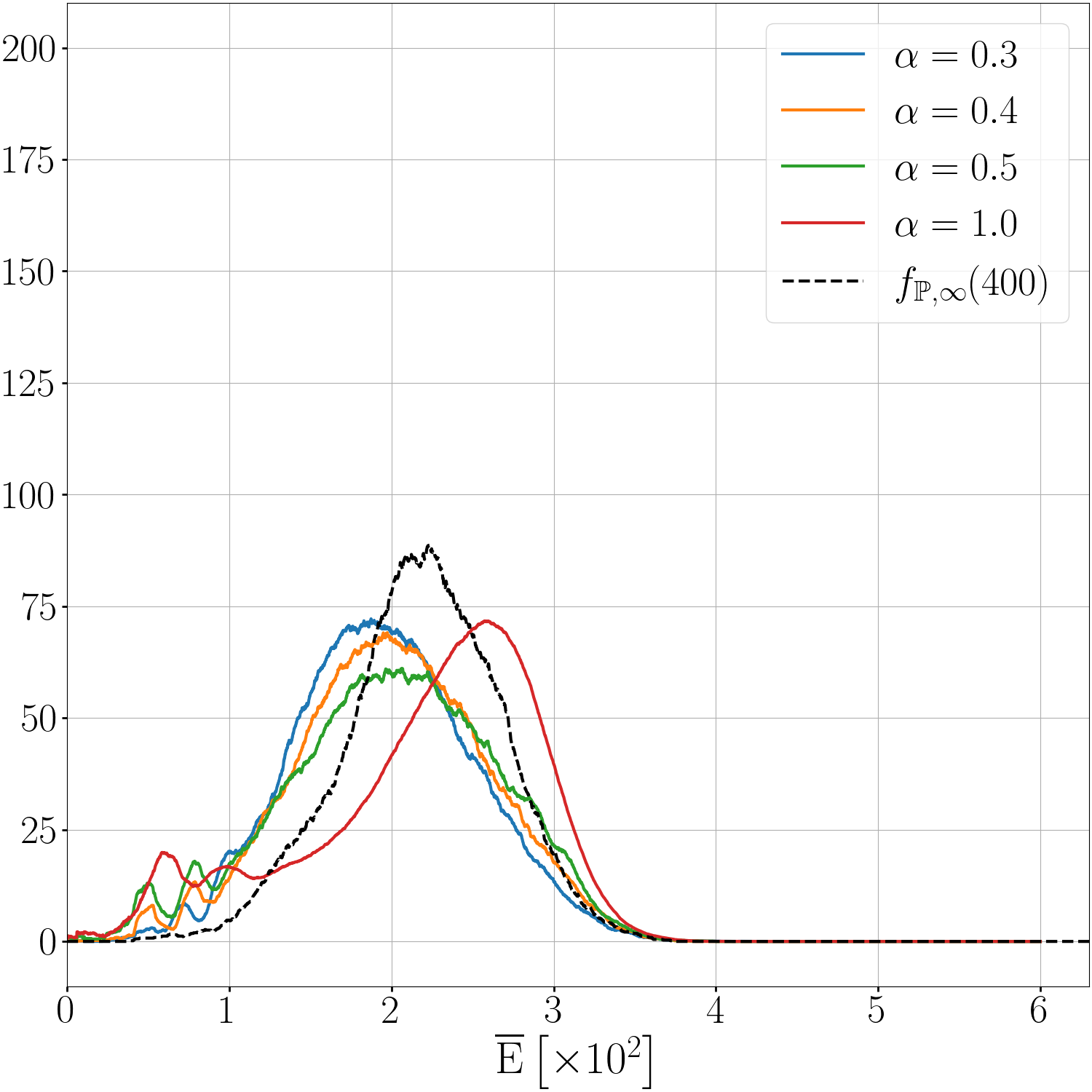}
    \includegraphics[width=0.309\textwidth]{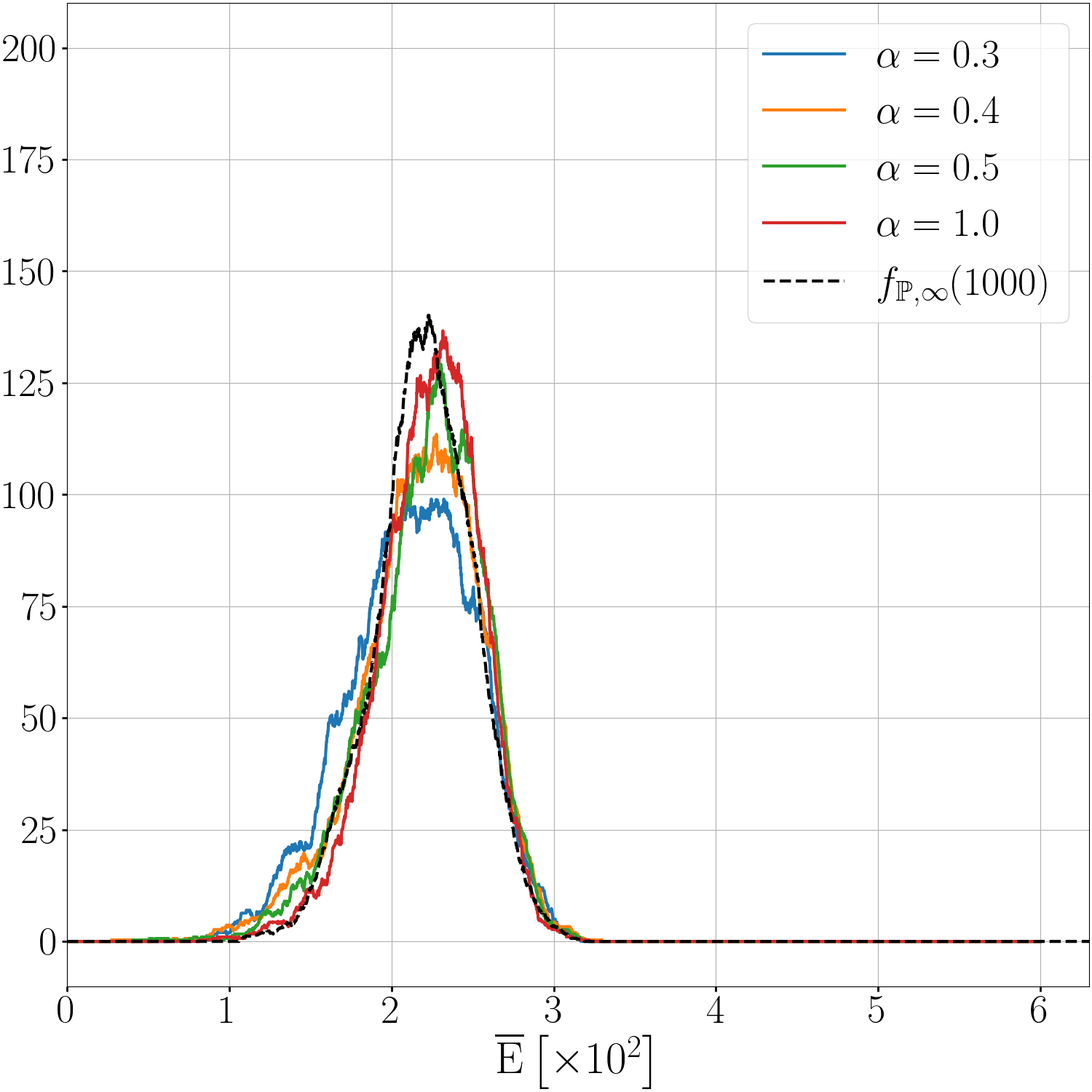}
    \includegraphics[width=0.309\textwidth]{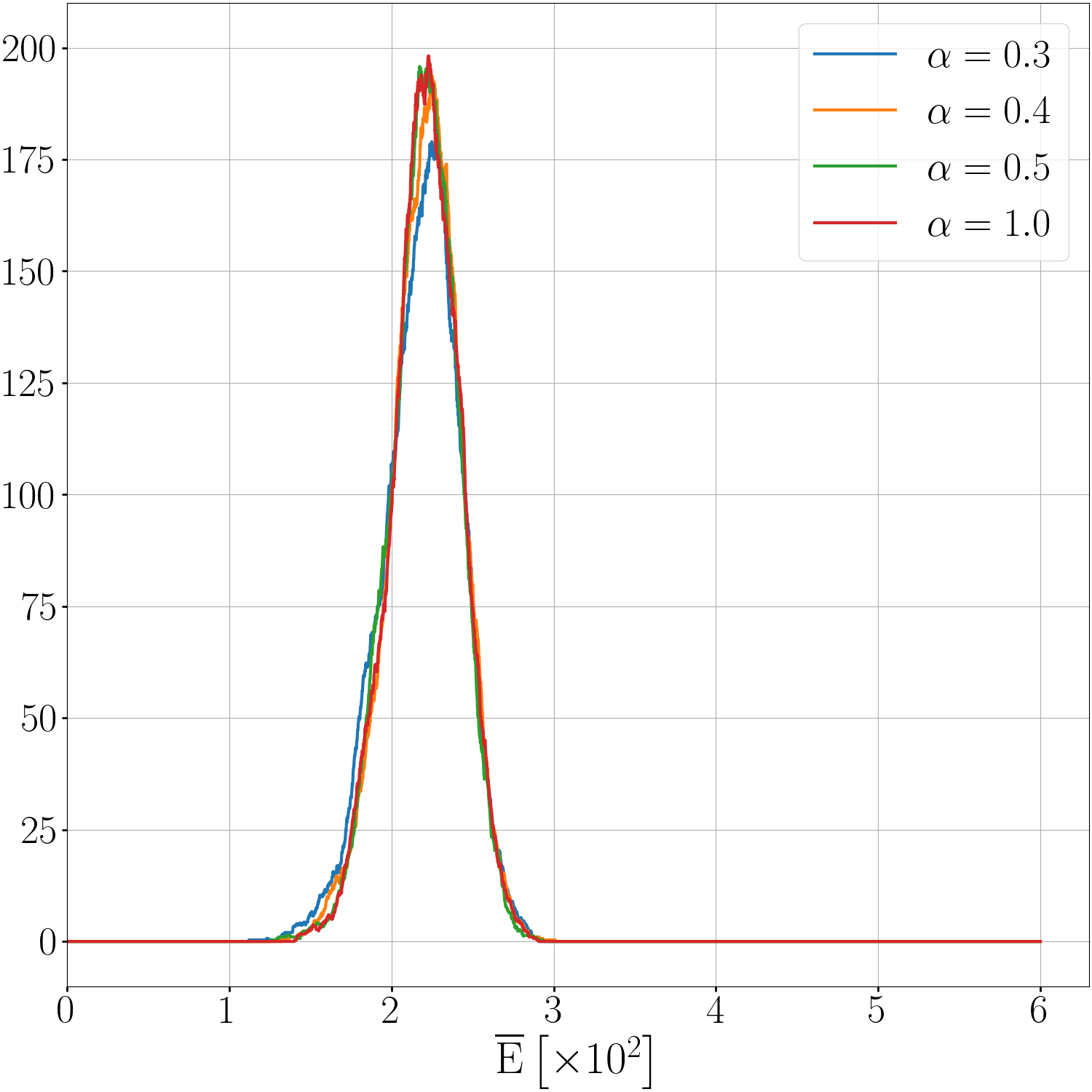}
    \end{tabular}
    \end{center}
    \vspace{-0.3in}
    \caption{
    The probability densities for the averaged probability distributions $f_\PD(N,\alpha)$ generated from $\cB(N,\alpha,2500)$ for each of the indicated values of $\alpha$ and for $N=400$ (left), $N=1000$ (center), and $N=2000,$ (right). The dotted line graph depicts the asymptotic density $f_{\PD,\infty}(\cdot, N)$, defined in \eqref{e:Conj} with $f_{\PD,*}$ and $\bar{E}_*$ approximated from $f_\PD(2000,1)$ and scaled to $N=400$ (left) and $1000$ (center).}
    %\caption{Anti-cumulative distribution density $f_V(\mrD)$ over Voronoi regions, binned by the number of defects $D=4\ldots 62$ (dashed) from $\cB(400,1,2500)$. The lowest, $\mrD=4$, and the highest, $\mrD=62$, defect distributions are solid and colored. The distribution $f_V$ for the  entire bin is solid black. (right)}
    \label{f:Cell-VEE}
\end{figure}

In a second, competing effect, the maximum value of $\VEE$ observed for a stable equilibria at a given $N$ and $\alpha$ becomes lower at larger $N$. The shift away from both low energy and higher energy equilibria squeezes the averaged probability density from below and above, pushing it towards a uni-modal shape. Comparing Figure\, \ref{f:acD-DOS} (left) and Figure\,\ref{f:Universal} (top-left) shows that the highest observed energy of 
a stable equilibrium drops from $\VEE\approx 5.0\times 10^{-2}$ for $N<100$ to values below $4\times 10^{-2}$ for $N>300$. The trend is monotonic with increases in $N$. Indeed in Figure\,\ref{f:Universal} (top-left) each of the averaged probability densities decays rapidly to zero for $\VEE>3.5\times10^{-2}.$ The 
extent of the convergence at large $N$ is examined through the anti-cumulative distribution $\acDi(N,1)$ and the associated sliding averages, $\langle f_\acDi \rangle_{[N-5,N+5]}$ of its density. In Figure\,\ref{f:Universal} (top-right) the averaged anticummulant densities for $N=390, \ldots, 400$ are compared to $f_\acDi(N)$. For these values of $N$ the  density $f_\acDi$ is relatively smooth, the drops found for the $N<100$ in Figure\,\ref{f:acD-DOS} (left) have been replaced with a continuous decline. For the larger values of $\VEE$ the density $f_\acDi$ is relatively 
independent of $N$, with only a $1\%$ difference between $f_\acDi$ for any of the values of $N$ and its average over the window. This is confirmed in Figure\,\ref{f:Universal} (right), which shows that the $L^2$ relative error between $f_\acDi(N)$ and its sliding average decays to less than $1\%$ for $N>350.$ A log-log linear fit of this decay suggests that the anti-cumulative distribution and its sliding average satisfy
\beq
 \|f_\acDi(N)-\langle f_\acDi\rangle_{[N-5, N+5]}\|_{L^2} \leq C N^{-\beta},
 \eeq
 where the value $\beta=2.35,$ yields the best fit.

\tcb{We investigate the large $N$ behavior of the probability distributions $f_{\PD}(N,\alpha)$ and their sensitivity to $\alpha.$ We form bins $\cB(N,\alpha,2500)$ for domain aspect ratios $\alpha=0.3, 0.4, 0.5, 1.0$ and for $N=400, 1000, 2000.$  For larger values of $N$ the majority of sites are anticipated to be sufficiently far from $\partial\Omega_{\alpha,N}$ to arrive at a ``bulk'' state. However for small aspect ratios, the distance to boundary is generically smaller, and the convergence is anticipated to be slower in $N.$ Figure\,\ref{f:Cell-VEE} (right) shows the probability densities for the $\PD$ generated from the bin $\cB(400,\alpha, 2500)$ for each of $\alpha =0.3, 0.4, 0.5,$ and $1.0.$   For $\alpha\leq 0.5$ the central peak is shifted left and is broader when compared to the peak for $\alpha=1$. Moreover the low energy equilibria are suppressed. The upper bound on $\VEE$ of stable equilibria is robustly independent of $\alpha.$ This suggests that the impact of the boundary is still felt for $N=400.$ For a given aspect ratio $\alpha$ the average distance to the boundary is proportional to $d_\partial:=\sqrt{N\alpha}$. For $\alpha=0.3$ and $N=400$, the value  $d_\partial\approx 10.9$ is comparable to that for $\alpha=1$ and $N=120$. From Figure\,\ref{f:Large-N} (right) the $\alpha=1$ and $N=120$ bin does not display convergence to a bulk limit. The situation changes dramatically for $N=1000$ and $N=2000$. For these values of $N$ the distributions become more localized. The dependence upon $\alpha$ is reduced, with smaller $\alpha$ corresponding to a slightly lower, broader peak. This motivates the conjecture that the large $N$ bulk distribution is independent of $\alpha$ and is self-similar in $N$ about a fixed energy $\bar{E}_*$. Specifically we conjecture the existence of a limiting bulk distribution which takes the form
\beq\label{e:Conj}
f_{\PD,\infty}(\bar{E};N) = \sqrt{N} f_{\PD,*}\left( \frac{\bar{E}-\bar{E}_*}{\sqrt{N}}\right)
.
\eeq
The form of $f_{\PD,*}$ and value of $\bar{E}_*$ determine the bulk distribution the energy of the inherent states. To compare to simulations we  approximate $f_{\PD,*}$ from $f_{\PD}(2000,1)$ with $\bar{E}_*=2.227\times10^{-2}$, and plot the scaled distributions $f_{\PD,\infty}$ in Figure\,\ref{f:Cell-VEE} (left)-(center) as dotted-black curves for $N=400$ and $N=1000$ respectively. A consequence of this conjecture is the limit
\beq\label{e:Conj-2}
\lim\limits_{N\to\infty}f_\PD(N,\alpha) = \delta_{\bar{E}_*}.
\eeq
In the bulk limit the distribution of energies of invariant states is a $\delta$-function at a prescribed frustration level $\bar{E}_*.$}

\section{Discussion and Conjectures}

This work establishes that the Hookean-Voronoi energy generates a complex landscape with a broad range of inherent states. We have examined the stable equilibria  determining the density of defects and the sizes of the basins of attraction of the inherent states when grouped according to their energy level. For smaller values of site number $N$, particularly $N<150$, these distributions are very sensitive to both the number of sites $N$ and to the aspect ratio $\alpha$ of the rectangular domain. For these value the basins of attraction are generically dominated by several low-energy equilibria which attract a significant portion of the total phase space. As $N$ increases a growing collection of stable, moderate-defect-number equilibria form small basins of attraction that increasingly dominate, and fracture, the phase space. These moderate energy equilibria exploit their numerical superiority to crowd out the basins of attraction of the lower energy, more ordered equilibria, decreasing the share of the total phase space that they attract and raising the system frustration. On the other hand, an upper limit in $\VEE$ appears above which all equilibria seem to be unstable. This upper limit decreases with increasing $N$, approaching a limit that establishes a sharp cut-off on the density of stable equilibria at higher $\VEE$ for larger $N$. \tcb{These two effects push the probability density of the equilibrium energy distribution into a uni-modal curve which seems to define the `bulk state' of the system. We conjecture that these two effects generate a self-similar distribution of energies of inherent states that converges to a $\delta$ function localized at a prescribed bulk-energy $\VEE_*.$} 

%\tcb{For $N$ near 400, the probability density for the equilibrium energies are comparable for $\alpha\geq 0.7$ but differ significantly for domain aspect ratios $\alpha<0.7$. These lower aspect ratio domains retain a largely unimodal equilibrium energy distribution, however their peak is both broader and shifted to lower energy. This suggests that the average volumetric excess energy $\langle \VEE\rangle_{\cB(400,\alpha,S)}$ is lower for smaller aspect ratios. This seems counter-intuitive as the lower aspect ratio can be seen as a constraint, lowering the packing options available to the system. This is a second example of the more bulk-like system, where domains with higher aspect ratio experience a higher degree of frustration.}

It is natural to consider the impact of thermal annealing -- the temporally limited addition of white noise to the gradient flow to eject the system from local minima so that it can find lower energy minima. This is a standard method to quench a system's frustration. As can be inferred from Figure\,\ref{f:Equilibium} (bottom), the minimum energy to destabilize an equilibrium, is approximately $\frac12 \nu(\eqx)\|\bx(0)-\eqx\|_2^2\approx 10^{-1}.$ This value is one order of magnitude larger than the variation between the $\VEE$ of the disordered equilibria, which is roughly $2\times 10^{-2}$. This rough estimate of the energy barrier between minima is in agreement with numerical simulations which find ``escape energies'' of $\VEE\approx 2\times 10^{-1}$ for $N=400$ equilibria with $\VEE\leq 3\times 10^{-2}$. The escape energy decreases slightly for equilibria with $\VEE$ near the upper limit of $4.5\times 10^{-2}.$ This suggests that annealing would require insertion of energy at a level that lifts the system above the energy of the maximum stable states. Subsequent removal of the  annealing perturbations would merely lead the system to relax back onto the energy landscape dominated by disordered equilibrium. In this sense the Hookean Voronoi system may resemble a spin glass, a frustrated system in which energy barriers also dominate variation between local minima, \cite{Spin-Glass, Spin-Glass2}.

There are several fundamental questions to be addressed. The first is the observation that an equilibrium can be disordered only if it has defects. In the more than $10^6$ simulations conducted in this study, defect free equilibria only arose from ordered tessellations, indeed only from single-string equilibria. This motivates the twin conjecture that ordered equilibria only arise from the single string tessellations and that stable equilibrium are either ordered or have defects.  Ordered tessellations can be continuously perturbed into disordered ones, but introducing a defect into a defect-free tessellation requires pushing the system through a vertex collision. This introduces a gap between ordered and disordered equilibrium. This gap may be related to the lack of correlation between the ground state energy $\zeta_0$ and the minimum energy of the ordered (single string) packings. 
%The ground state energy reflects the extrapolation of equilibria energy back to a zero defect state. The actual minimum $\VEE$ available to the system in a zero defect state may be significantly lower, even zero, or it can be many times higher. This shows that 
This emphasizes the observations that ordered states are not good proxies for system behavior, especially at large values of $N$. 
%For ordered states the tessellation energy equals the region energy, while for disordered states these two quantities generically differ by two orders of magnitude.

\tcb{A second question is the role of boundary conditions. Changes in boundary conditions to a rigid wall or to oblique periodic domains may have significant impact on the bulk statistics, in particular if these conditions nucleate regular hexagonal tilings. These changes could lead to arctic circle type phenomena observed in domino tilings in which an ordered layer at the boundary breaks into a disordered region at a sufficient distance, see \cite{ACT-95, ACT-01}. }
%A second question is to determine if the $\VEE$ is always positive. This is equivalent to asking if the average energy of the regions in a tessellation is bounded below by the energy of the regular hexagon of unit area. In all simulations this is the case, with zero $\VEE$ attained only for tessellations by regular hexagons. Nonetheless, the distributions of $\VEE$ present in any disordered tessellation always includes regions with substantially negative $\VEE$. It is only in the average over the whole tessellation that the $\VEE$ is positive. 
%Establishing this conjecture requires eliminating the existence of defect-filled equilibria with negative $\VEE$, a substantially more challenging task than showing that ordered tessellations have non-negative energy.
\tcb{It is important to quantify the structural stability of disorder in two-dimensional packing problems, and if this disorder is qualitatively different in three-dimensional setting. Quasi-ordered structure in the inherent states has been observed in other packing models in two space dimensions. This included particle-particle interaction models such as those based upon a Lenard-Jones type potential, \cite{Weber-93} and the $k$-space overlap potential the builds excluded volume based upon the area of overlap of circles centered at near neighbors \cite{Tor11}. The Lenard-Jones models produces an relatively rough inherent state energy landscape that supports a first-order phase transition with a relatively large melting entropy between ordered and disordered states. Conversely the $k$-space overlap model generates a large $N$ energy landscape with frustration characterized by inherent states with mild variations in energy, as is found in the Hookean-Voronoi energy presented here.}

\tcb{The most fundamental question raised by this study concerns the large $N$ or ``bulk'' limit of the Hookean-Voronoi energy. The numerical studies presented suggests that $\PD(N,\alpha)$ approaches a universal large $N$ limit, and that this limit supports a non-zero frustration $\VEE_*$.  If the asymptotic limit \eqref{e:Conj} holds, what determines the structure of $f_{\PD,*}$ and the value of $\VEE_*$? The simulations presented here cannot rule out that for yet larger values of $N$ the average of the distribution $f_{\PD}(N,\alpha)$ tends to zero. The limiting bulk behavior seems to arise from a combinatorial domination of moderate-defect inherent states over low-defect, quasi-ordered ones, and a mechanism that drives instability in higher energy equilibria.   A rigorous justification of a bulk limit in the Hookean Voronoi system may require novel combinatorial and probabilistic arguments.}

\section{Acknowledgement}
The second author recognizes support from the NSF through grants DMS 1813203 and DMS 2205553. All authors thank Yuan Chen, Sulin Wang, and  Zhao Wei for helpful discussions during the preliminary stage of the research.

\bibliographystyle{siam}
\bibliography{main.bib}
\end{document}